\documentclass[aps,prl,reprint,superscriptaddress,amsmath,amssymb]{revtex4-2}
\usepackage{amsfonts}
\usepackage{times}
\usepackage[cmyk]{xcolor}
\usepackage{pgfplotstable}
\usepackage{url}
\usepackage{hyperref}
\usepackage{amsthm}
\usepackage{amsmath}
\usepackage{tabularx}
\usepackage{graphicx}
\usepackage{array}
\usepackage{cleveref}
\usepackage{makecell}
\usepackage{multirow}
\newtheorem{thm}{Theorem}\crefname{thm}{Theorem}{Theorems}
\crefname{lem}{Lemma}{Lemmas}
\crefname{prp}{Proposition}{Propositions}
\crefname{cor}{Corollary}{Corollaries}
\crefname{dfn}{Definition}{Definitions}
\crefname{section}{Section}{Sections}
\crefname{appendix}{Appendix}{Appendices}

\newcommand{\ket}[1]{\ensuremath{\left\vert{#1}\right\rangle}}
\newcommand{\SQiSW}{\mathrm{SQ\lowercase{i}SW} }
\newcommand{\iSWAP}{\mathrm{\lowercase{i}SWAP} }
\newcommand{\CNOT}{\mathrm{CNOT} }
\newcommand{\CPhase}{\mathrm{CPHASE} }
\newcommand{\SWAP}{\mathrm{SWAP} }

\date{\today}

\begin{document}
\title{Quantum Instruction Set Design for Performance}

\author{Cupjin Huang}
\thanks{These two authors contributed equally}

\affiliation{Alibaba Quantum Laboratory, Alibaba Group USA, Bellevue, Washington 98004, USA}

\author{Tenghui Wang}
\thanks{These two authors contributed equally}

\affiliation{Alibaba Quantum Laboratory, Alibaba Group, Hangzhou, Zhejiang 311121, P.R.China}

\author{Feng Wu}
\affiliation{Alibaba Quantum Laboratory, Alibaba Group, Hangzhou, Zhejiang 311121, P.R.China}

\author{Dawei Ding}
\affiliation{Alibaba Quantum Laboratory, Alibaba Group USA, Sunnyvale, California 94085, USA}

\author{Qi Ye}
\affiliation{Alibaba Quantum Laboratory, Alibaba Group, Beijing 100102, P.R.China}
\affiliation{Institute for Interdisciplinary Information Sciences, Tsinghua University, Beijing 100084, People's Republic of China}

\author{Linghang Kong}
\affiliation{Center for Theoretical Physics, Massachusetts Institute of Technology,Cambridge, MA 02139, USA}

\author{Fang Zhang}
\affiliation{Alibaba Quantum Laboratory, Alibaba Group USA, Bellevue, Washington 98004, USA}

\author{Xiaotong Ni}
\affiliation{Alibaba Quantum Laboratory, Alibaba Group, Hangzhou, Zhejiang 311121, P.R.China}

\author{Zhijun Song}
\affiliation{Alibaba Quantum Laboratory, Alibaba Group, Hangzhou, Zhejiang 311121, P.R.China}

\author{Yaoyun Shi}
\affiliation{Alibaba Quantum Laboratory, Alibaba Group USA, Bellevue, Washington 98004, USA}

\author{Hui-Hai Zhao}
\affiliation{Alibaba Quantum Laboratory, Alibaba Group, Beijing 100102, P.R.China}

\author{Chunqing Deng}
\email{chunqing.cd@alibaba-inc.com}
\affiliation{Alibaba Quantum Laboratory, Alibaba Group, Hangzhou, Zhejiang 311121, P.R.China}

\author{Jianxin Chen}
\email{liangjian.cjx@alibaba-inc.com}
\affiliation{Alibaba Quantum Laboratory, Alibaba Group USA, Bellevue, Washington 98004, USA}

\begin{abstract}
    A quantum instruction set is where quantum hardware and software meet. We develop new characterization and compilation techniques for non-Clifford gates to accurately evaluate different quantum instruction set designs. We specifically apply them to our fluxonium processor that supports mainstream instruction $\iSWAP$ by calibrating and characterizing its square root $\SQiSW{}$. We measure a gate fidelity of up to $99.72\%$ with an average of $99.31\%$ and realize Haar random two-qubit gates using $\SQiSW{}$ with an average fidelity of $96.38\%$. This is an average error reduction of $41\%$ for the former and a $50\%$ reduction for the latter compared to using $\iSWAP$ on the same processor. This shows designing the quantum instruction set consisting of $\SQiSW{}$ and single-qubit gates on such platforms leads to a performance boost at almost no cost.
\end{abstract}

\maketitle

A quantum instruction set is the interface that quantum hardware vendors provide to software. This has a subtle difference from a native gate set although they are often used interchangeably. A native gate set contains quantum operations that are physically realizable on a quantum computing platform. In contrast, a quantum instruction set is a carefully chosen set of gates with consideration for cost or performance of the processor. Although one can always include more quantum gates to an instruction set to more efficiently implement algorithms, there are additional considerations. First of all, the added instruction must be of sufficient accuracy. In addition, since we need to calibrate each instruction, it is preferable to limit the number of gates~\cite{Lao+21}. This is reminiscent of the debate between RISC (Reduced Instruction Set Computer) and CISC (Complex Instruction Set Computer) in classical computing~\cite{BMV+15}.

Now, any quantum instruction set capable of universal quantum computing contains at least one entangling two-qubit gate. Given that errors of single-qubit gates are usually much less than that of two-qubit gates, the design of a quantum instruction set is mainly about the selection of two-qubit gates. However, evaluating quantum instruction sets with different two-qubit gates has the following challenges.
\begin{enumerate}
    \item The set of native two-qubit gates and how accurately each gate can be realized is highly dependent on hardware implementation. Therefore, it is difficult to make a direct comparison between quantum instruction sets at a software level;
    \item It is a non-trivial task to faithfully characterize a general quantum gate. In particular, the conventional approach of using Clifford-based randomized benchmarking (RB)~\cite{emerson2005scalable} cannot be used to benchmark non-Clifford gates.
    \item The compilation of quantum algorithms into quantum instructions is only extensively studied for $\CNOT$~\cite{JST+20,AAM18,VW04}. Although there are studies of compiling into general gates~\cite{davis2019heuristics,Lao+21}, these approaches use heuristics and so cannot give reliable data for designing an instruction set.
\end{enumerate}

With these challenges in mind, in this letter we re-evaluate the design of the mainstream quantum instruction set for superconducting circuits that is based on $\CNOT{}$, $\iSWAP{}$, or gates that differ from either only by single-qubit gates~\footnote{Note that some gate schemes such as those of cross-resonance gate~\cite{krantz2019quantum} or the Mølmer–Sørensen gate~\cite{bruzewicz19} for ion-traps can realize the gate with continuous single-qubit phase degrees of freedom. However, from a two-qubit perspective, these gate sets are equivalent to the singleton set $\{\CNOT{}\}$.}. Given all the focus on maximizing the fidelities of these two specific types of gates~\cite{krantz2019quantum, bruzewicz19}, it is worthwhile to critically analyze these particular choices of quantum instruction sets. In particular, we consider the instruction set with $\iSWAP$ and single-qubit gates. We numerically compute and experimentally demonstrate that for a quantum processor with such an instruction set, 
designing an instruction by replacing $\iSWAP$ with its matrix square root $\SQiSW{}$ can both reduce the gate error and improve compilation capabilities substantially. This 
new design can therefore significantly improve the quantum instruction set.

More specifically, the square root of the $\iSWAP$ gate, abbreviated as $\SQiSW{}$, whose matrix form can be written as $$\SQiSW\equiv\begin{bmatrix}1&0&0&0\\0&\frac1{\sqrt{2}}&\frac{i}{\sqrt{2}}&0\\0&\frac{i}{\sqrt{2}}&\frac1{\sqrt{2}}&0\\0&0&0&1\end{bmatrix},$$ is in the $\iSWAP$ family and has been proposed and experimentally realized in earlier works~\cite{arute2020observation, google2020hartree,bialczak2010quantum,mi2021information, abrams2020implementation, Peterson2020fixeddepthtwoqubit, PRXQuantum.2.040309, goerz2017charting}. The most common technique to realize $\iSWAP{}$ is to bring two transversally coupled qubits into resonance, and $\SQiSW{}$ is realized by halving the gate time for $\iSWAP{}$. Thus, $\SQiSW{}$ is expected to be less susceptible to errors from decoherence and stray interactions. Despite hints of promise, few works have systematically studied the $\SQiSW{}$ from a holistic perspective, which we aim to accomplish in this work. In particular, we establish its control accuracy and compilation efficiency as follows.

\begin{itemize}
    \item Control accuracy: Taking into account decoherence, stray coupling between qubits, and instrumental limitations on experimental control parameters, we conduct a numerical simulation with experimentally achievable circuit parameters and decoherence times and estimate that we can realize $\SQiSW{}$ with an error of about $5\times 10^{-4}$,  which is only half of the error of $\iSWAP{}$ on the same system. 
    \item Compilation efficiency:  We prove that an arbitrary two-qubit gate can be compiled using at most three uses of the $\SQiSW{}$ gate interleaved with arbitrary single-qubit gates and give an analytical scheme to compile a two-qubit gate with the optimal number of $\SQiSW{}$ gates. Moreover, we prove that 79\% of two-qubit gates (under the Haar measure) can be generated with only two uses of the $\SQiSW{}$  gate. In comparison, only a zero-measure set of two-qubit gates can be generated using two uses of common two-qubit gates such as $\CPhase$ family gates, $\SWAP$ family gates or the $\iSWAP$ gate. As a result, using $\SQiSW{}$ to compile two-qubit Haar random or random Clifford gates leads to error reductions of about 63\% or 35\% compared to using $\iSWAP{}$, assuming that the gate error of $\SQiSW{}$ is half of that of $\iSWAP{}$ and that single-qubit gate errors are negligible. 
\end{itemize} 

Furthermore, we experimentally demonstrate these advantages of $\SQiSW$ on a superconducting quantum processor consisting of two capacitatively-coupled fluxonium qubits. The calibration of $\SQiSW{}$ can be performed by following an efficient procedure for general excitation number conserving gates in~\cite{arute2020observation}. The detailed calibration on our specific platform can be found in the Supplementary Material. In order to characterize $\SQiSW$, a non-Clifford gate, we develop a framework for benchmarking a general quantum gate based which we call {\em interleaved fully randomized benchmarking (iFRB)}. Fully randomized benchmarking (FRB) is similar to RB but instead of using Clifford gates, we go back to the original proposal of using Haar randomgates~\cite{emerson2005scalable}. iFRB is the corresponding interleaved variant and does not require the target gate to lie in the Clifford group or exhibit any group structure (another example is the $\CNOT$-dihedral group~\cite{garion2021experimental}). It is applicable whenever an efficient compilation scheme for an arbitrary gate is available, which we establish through our explicit compilation scheme for $\SQiSW{}$. We experimentally benchmark $\SQiSW{}$ using iFRB and compare the results to $\iSWAP{}$ in the same experimental run. We observe a $\SQiSW{}$ fidelity of up to $99.72\%$ with average $99.31\%$. The latter is a $41\%$ error reduction compared to the corresponding $\iSWAP$ fidelity. We also realize Haar random two-qubit gates using $\SQiSW{}$ and measure an average fidelity of $96.38\%$, the error reduction being $50\%$ compared to that of $\iSWAP$. This additional reduction demonstrates the superior compilation capabilities of $\SQiSW{}$.

\emph{The $\SQiSW{}$ gate scheme.}---Two-qubit gates in the $\iSWAP$ family can be implemented by
turning on a transverse coupling between qubits for a given time $t$~\cite{krantz2019quantum}. In the weak coupling limit where the coupling energy is much smaller than the qubit energy, we can neglect the transition between $\ket{00}$ and $\ket{11}$ by performing the rotating wave approximation (RWA) therefore the interaction Hamiltonian takes the well-known form of the $XY$ spin interaction: $g(\sigma_x \sigma_x + \sigma_y \sigma_y)$ where $g$ is the $XY$ coupling strength between the two qubits.  During the interaction, the population exchange of the two-qubit system can be described by the unitary
\begin{align*}
    U(t) & 
    = \left[\begin{matrix}
            1 & 0              & 0              & 0 \\
            0 & \cos(\theta)   & -i\sin(\theta) & 0 \\
            0 & -i\sin(\theta) & \cos(\theta)   & 0 \\
            0 & 0              & 0              & 1
        \end{matrix}\right] ,
\end{align*}
where $\theta\equiv tg_{}/4$. When $g_{} >0$, we can implement $\SQiSW{}^\dagger$ by $U(t=\pi/g_{})$ and $\iSWAP^\dagger$ by $U(t=2\pi/g_{})$. When $g_{}<0$, we instead implement $U(t=-\pi/g_{})=\SQiSW$. Note that $\SQiSW{}$  is equivalent to $\SQiSW^\dag$ under local unitaries: we have $\SQiSW^\dag = (Z\otimes I)\SQiSW(Z\otimes I)$. The properties of these two gates are essentially identical other than minor sign differences in compilation. In the subsequent sections we will consider the $\SQiSW{}$  gate even though $\SQiSW^\dag$ is more common in physical implementations.

In realistic systems, this gate scheme is complicated by possible stray longitudinal coupling ($ZZ$ interaction), inaccurate flux tuning (to bring qubits into resonance) and decoherence~\cite{barends2019,ganzhorn2020,PhysRevX.11.021058}. Since $\SQiSW{}$ takes roughly half of the gate time of $\iSWAP$, the corresponding contributions from the major error sources are approximately halved as well. A more detailed analysis of the error contributions, together with numerical simulations, can be found in Section~III of the Supplemental Material. Considering the recent progress on fluxonium fabrication~\cite{nguyen2019} and the precision of the arbitrary wave generator, according to our simulation we can expect an infidelity as small as $5\times 10^{-4}$ for the $\SQiSW{}$ gate. 

Since $\SQiSW^2 = \iSWAP$, replacing $\iSWAP$ in a quantum instruction set with $\SQiSW$ in principle results in negligible fidelity loss, possibly nonzero due to imprecise control on real devices. Assuming this, we next show that replacing $\iSWAP$ with $\SQiSW$ yields strict performance gain by proving that $\SQiSW$ is superior at certain compilation tasks.

\emph{Compilation capabilities of $\SQiSW$.}---We study how effective $\SQiSW{}$ can compile quantum circuits, in particular arbitrary two-qubit gates assuming single-qubit gates are free resources. We compare this with those of other widely used two-qubit gates. We give a more detailed analysis of the ability of $\SQiSW{}$ to compile multiqubit quantum circuits, such as $W$ state preparation, in Section~II of the Supplemental Material.

To minimize the two-qubit gate count when compiling circuits, there is a convenient tool classifying the equivalence classes of two-qubit gates modulo single-qubit gates:

\begin{thm}[KAK decomposition~\cite{zhang2003geometric}]
For an arbitrary $U\in SU(4)$, there exists a unique $\Vec{\eta}=(x,y,z), \frac{\pi}4\geq x\geq y\geq |z|$, single qubit rotations $A_0,A_1,B_0,B_1\in SU(2)$ and a global phase $g\in\{1,i\}$ such that
$$U = g\cdot\left(A_1\otimes A_2\right)\exp\{i\Vec{\eta}\cdot\Vec{\Sigma}\}\left(B_1\otimes B_2\right),$$
where $\Vec{\Sigma} \equiv [\sigma_X\otimes \sigma_X, \sigma_Y\otimes \sigma_Y,\sigma_Z\otimes \sigma_Z]$. The tuple $(g,\Vec{\eta}, A_0, A_1, B_0, B_1)$ is called the \emph{KAK decomposition} of the unitary $U$.
\end{thm}

The equivalence class of a two-qubit unitary $U$ modulo local, or single-qubit, unitaries is characterized by the \emph{interaction coefficients} $\eta(U)$, which lives in a 3-dimensional tetrahedron called the \emph{Weyl chamber}~\footnote{Note that there are many ways one can choose the canonicalization. Our convention follows the one in~\cite{cirq_developers_2021_4586899} and~\cite{cross2019validating}.} $$W\equiv\{\frac{\pi}{4}\geq x\geq y\geq |z|\ \textrm{and}\ z\geq 0 \text{ if }x=\frac{\pi}{4}\}\subset \mathbb{R}^3.$$ We say that two unitaries $U,V\in SU(4)$ are locally equivalent, or $U\sim V$, if $\eta(U)=\eta(V)$, that is, they can be transformed into each other with only single-qubit gates.

We prove that other than trivial compilations of gates that are locally equivalent to $I$ or $\SQiSW{}$, a two-qubit gate can be realized with 2 applications of $\SQiSW{}$ if and only if its interaction coefficients $(x,y,z)$ satisfy $x-y\geq |z|$. All other two-qubit gates can be realized with 3 applications of $\SQiSW{}$.

An explicit compilation scheme for all two-qubit gates minimizing $\SQiSW{}$ gate count, the corresponding proofs, and specific compilation schemes for special families of two-qubit gates can be found in Section~I of the Supplemental Material. To complete the compilation scheme, arbitrary single-qubit gates can be implemented in hardware using for example the PMW-3 scheme in~\cite{chen2021compiling}.

We also compare the ability of $\SQiSW{}$ to compile two-qubit gates with that of other two qubit gates. The average two-qubit gate count required to generate Haar random two-qubit gates $\mathcal{L}_H$ and random Clifford gates $\mathcal{L}_C$, together with other properties, are summarized in~\Cref{tab:2qgates}. It can be seen that, although only being half of the $\iSWAP$ gate, the $\SQiSW{}$ gate is actually much more efficient in compiling Haar random two-qubit gates than the $\iSWAP$ gate, while not being much worse in terms of compiling random Clifford gates. If we assume that $\SQiSW{}$ has half the error of that of $\iSWAP$, by using $\SQiSW{}$ we expect a $63\%$ reduction in the average infidelity for Haar random gates and a $35\%$ reduction for random Clifford gates. However, note that other two-qubit gates have been proposed in the literature, such as the $B$ gate~\cite{zhang2004minimum} and the $\iSWAP^{\frac 3 4}$ gate~\cite{Peterson2020fixeddepthtwoqubit}. However the former does not have a high-fidelity experimental gate scheme while the latter does not have an explicit compilation scheme. $\SQiSW{}$ is the first gate to our knowledge that outperforms $\iSWAP{}$ and $\CNOT{}$ in compilation abilities \emph{while still admitting a high-fidelity gate scheme and an explicit compilation scheme.}

\begin{table}[ht]
    \centering
    \begin{tabular}{|c||c|c|c|}
           \hline
          & $\SQiSW{}$ & $\CNOT$~\cite{ye2004super} & $\iSWAP$~\cite{ye2004super} \\
          \hline
          $\mathcal{L}_H$&2.21&3&3\\
          \hline
          $\mathcal{L}_C$&1.95&1.5&1.5\\
          \hline
    \end{tabular}
    \caption{Comparison between $\SQiSW{}$ and conventional $\CNOT$/$\iSWAP$ in terms of compiling Haar random two-qubit gates and random Clifford gates.}
    \label{tab:2qgates}
\end{table}

\emph{Experiment.}---We experimentally realize $\SQiSW{}$ and $\iSWAP{}$ on a fluxonium quantum processor. Fluxonium is an attractive candidate for the next generation of superconducting qubits because of its large anharmonicity and long coherence times. The anharmonicity is an order of magnitude larger compared to that of transmon which allows for performing low-leakage gates without the need for complex pulse shaping. With proper calibration, the fidelities of quantum gates are mainly limited by decoherence. Therefore, fluxonium is particularly suitable for demonstrating the advantages of $\SQiSW{}$. 

The experiment setup and quantum processor used in this experiment are the same as that of~\cite{Bao2021}, but this experiment is performed in a different cool-down cycle. For individual single-qubit gates, the two qubits $Q_A$ and $Q_B$ are fixed at their sweet spots corresponding to 1.09~GHz and 1.33~GHz, respectively, and a resonant microwave pulse generated by an arbitrary waveform generator (AWG) is applied to each qubit through their own charge line.

The two fluxonium qubits in our processor are capacitively coupled, with a transverse coupling strength $g$. To realize an iSWAP-like gate, we modulate the external flux of $Q_A$ to bring it in resonance with $Q_B$. The flux modulation pulse has a simple error-function shape and is also generated by an AWG and applied to $Q_A$ through its individual flux line. We use the same method as in Ref.~\cite{Bao2021} to calibrate the flux amplitude $\Phi_{\textrm{amp}}$ that brings the two qubits in resonance. With a coupling strength of $g/2\pi\approx11.2$~MHz, the gate times of $\iSWAP$ and $\SQiSW{}$ are 48.5~ns and 24.6~ns, respectively. The detailed calibration of the SQiSW gate is summarized in Section~IV of the Supplementary Material. 

We apply a variant of \emph{interleaved randomized benchmarking} (iRB)~\cite{magesan2012efficient} to characterize the fidelity of $\SQiSW{}$ and compare it with $\iSWAP{}$. The iRB protocol characterizes the fidelity of a particular \emph{target} gate by measuring the difference between the fidelity of a random Clifford gate, and a random Clifford gate appended by the target gate, both through the standard RB protocol~\cite{emerson2007symmetrized}. The iRB protocol requires that the target gate be chosen from the Clifford group as the reference gate set (the random gates) are all from the Clifford group, which is not directly applicable to the $\SQiSW{}$. Instead, in our experiments we choose the random gates from the Haar random distribution on all two-qubit rotations $SU(4)$. We call randomized benchmarking based on the whole (special) unitary group \emph{fully randomized benchmarking} (FRB), and the corresponding interleaved version \emph{interleaved fully randomized benchmarking} (iFRB). The reason we apply FRB and iFRB in our experiments is two-fold: First, benchmarking arbitrary gates requires sampling the gates from an entire special unitary group; second, Clifford-group RB cannot directly benchmark non-Clifford gates such as $\SQiSW{}$. FRB and iFRB applicable for characterizing $\SQiSW{}$ because we have an explicit compilation scheme and the required gates are all realized with high fidelities.

\begin{figure}
  \includegraphics[width=8.3cm]{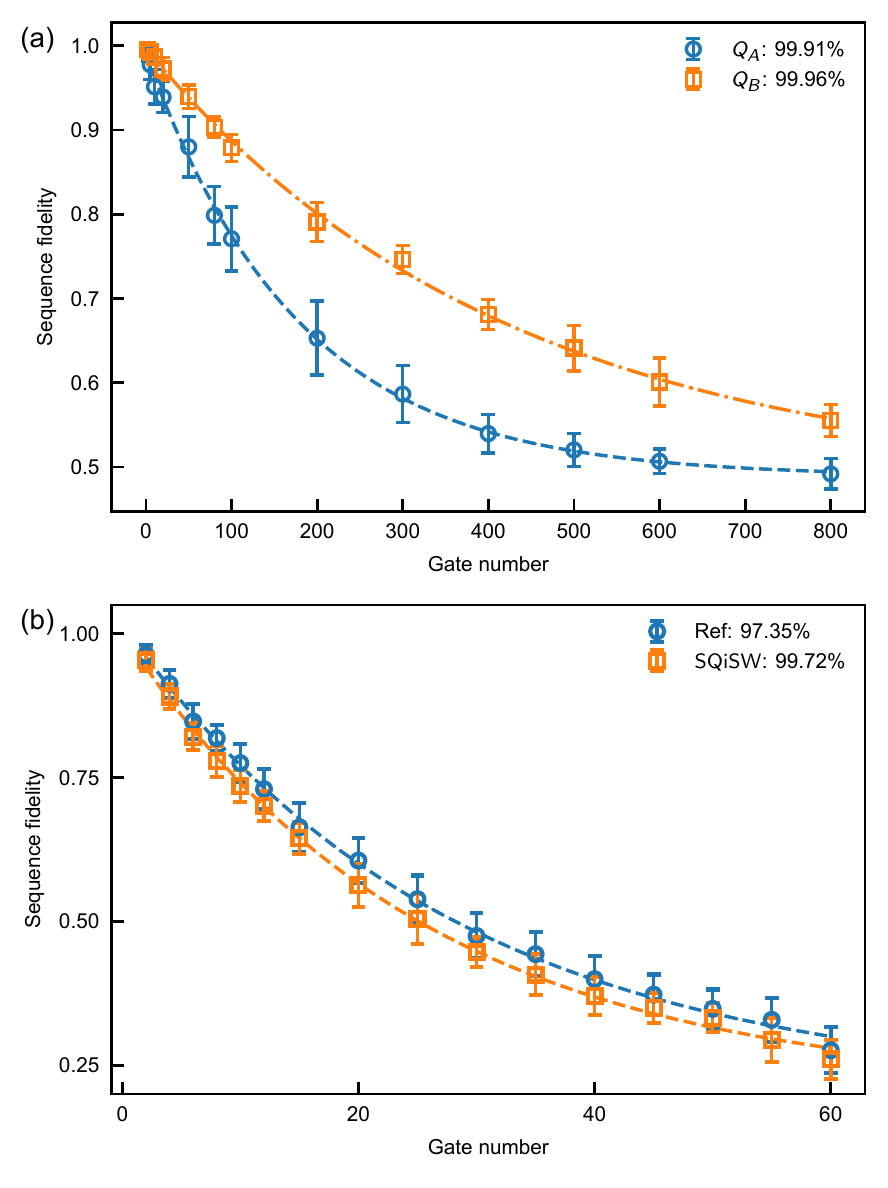}
  \caption{\label{fig:frb} (a) Simultaneous single-qubit FRB. The legend presents the average fidelity of a single Pauli rotation. (b) Sequence fidelity of FRB and iFRB for $\SQiSW{}$ for the run with highest fidelity. Each data point is averaged from 20 random sequences.}
 \end{figure}

In order to compile arbitrary two-qubit gates, we first need to compile arbitrary single-qubit gates. To do this, we employ the results in~\cite{chen2021compiling} to compile an arbitrary single-qubit gate using 3 phase-shifted microwave (PMW) pulses. In particular, we use the PMW-3 scheme. In~\Cref{fig:frb}(a), we show the results of simultaneous single-qubit FRB using the PMW-3 scheme. An arbitrary single-qubit gate is complied with 3 primary Pauli rotations with rotation angles $\pi/2$ or $\pi$. The extracted average fidelity of the Pauli rotations of $Q_{A}$ ($Q_{B}$) is 99.91\% (99.96\%).
We also include data for simultaneous single-qubit RB in Section~IV of the Supplementary material.
The Clifford RB provides a standard baseline for single-qubit gate fidelities, and we see the results are comparable to the high-fidelity implementations on superconducting qubits~\cite{krantz2019quantum}. We also see that FRB can achieve comparable fidelities of more than 99.9\% per Pauli rotation. This in particular demonstrates that the PMW-3 scheme can achieve state-of-the-art fidelities. Each microwave pulse is $10~\text{ns}$ long and is calibrated using known techniques developed for example in~\cite{chen2018metrology,chen2016measuring,sheldon2016characterizing} and detailed in~\cite{Bao2021}.

\begin{figure}
    \centering
    \includegraphics[width=8.3cm]{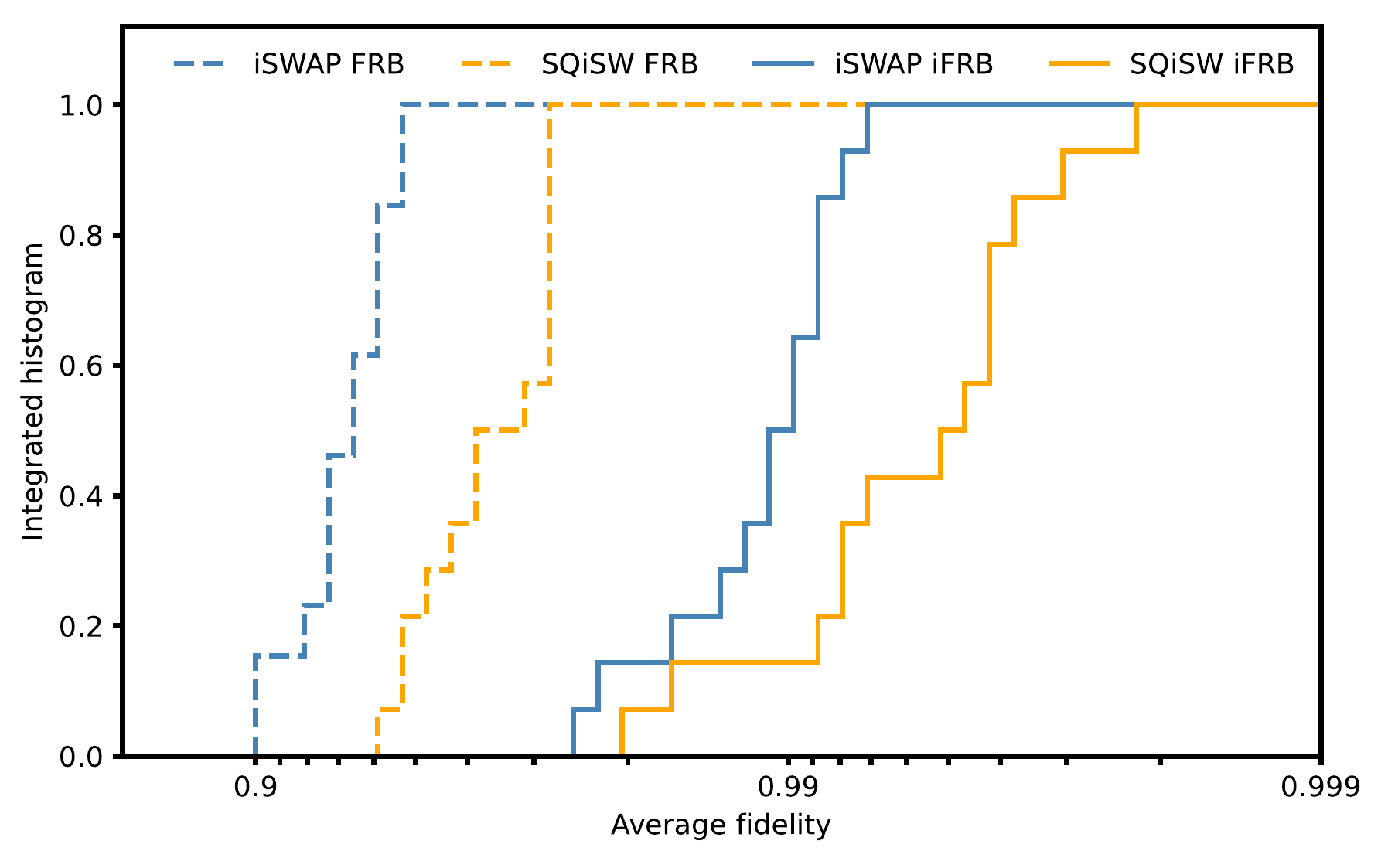}
\caption{Cumulative histogram of two-qubit FRB and iFRB fidelities of $\iSWAP{}$ and $\SQiSW{}$ from 14 experiments.\footnote{Although we measure a $\SQiSW$ iFRB of $99.80\%$ in one of the runs, a closer look into the data reveals a poor fitting due to highly fluctuating data. We include this data point here for sake of completeness, but do not report the corresponding fidelity value as our highest.} }
\label{fig:FRB_iFRB}
\end{figure}

Theoretically, iFRB and iRB should be measuring the same quantity~\cite{kong2021framework}. We run iFRB for $\SQiSW{}$ and $\iSWAP{}$ and compare the results in~\Cref{fig:FRB_iFRB}. We show the sequence fidelity exponential decays for the highest measured $\SQiSW{}$ fidelity in~\Cref{fig:frb}(b). We also give average values and the standard deviation for each metric in~\Cref{tab:frb_ifrb_data}. We include in the table our measurements of iRB for $\iSWAP{}$, which, as expected, we find very similar to the iFRB~\footnote{We also measured an average $\iSWAP{}$ RB value of $97.26\%$. This is significantly different from the FRB value, but this is because we use on average 1.5 $\iSWAP{}$ gates to compile Cliffords while we use 3 to compile arbitrary two-qubit gates. }. We also find that the average iFRB infidelity of $\SQiSW{}$ is reduced by 41\% compared to $\iSWAP{}$, while the FRB is reduced by $50\%$. This is similar to what is expected from the theoretical predictions (50\% and 63\% reduction, respectively). In particular, the additional reduction for FRB is testament to $\SQiSW{}$'s superior compilation abilities. More details on our benchmarking schemes are provided in Section~V of the Supplementary Material. In Section~IV of the Supplementary Material we also discuss a fluctuation we observed in coherence times of our qubits, which we can possibly attribute to undesired coupling with a drifting two-level system. 

\begin{table}[hpt]
    \centering
    \begin{tabular}{|c|c||c|c|}
    \hline
    & &  Fidelity & Infidelity\\
    \hline
    \multirow{2}{*}{$\SQiSW{}$} & iFRB &  $99.31 \pm 0.50 \%$ & $6.851 \times 10^{-3}$\\
    \cline{2-4}
    & FRB & $96.38 \pm 0.98\%$ & $3.615 \times 10^{-2}$\\
    \hline
    \multirow{3}{*}{$\iSWAP{}$} & iFRB &  $98.84 \pm 0.51\%$ & $11.63 \times 10^{-3}$ \\
    \cline{2-4}
    & iRB & $98.88 \pm 0.44\%$ & $11.15\times 10^{-3}$ \\
    \cline{2-4}
    & FRB & $92.74 \pm 2.17\%$ & $7.265 \times 10^{-2}$ \\
    \hline
    \end{tabular}
    \caption{Average values and standard deviations of different benchmarking metrics for $\SQiSW{}$ and $\iSWAP{}$ over 14 experiments. FRB is the fidelity of arbitrary two-qubit gates, while iFRB and iRB is the fidelity of the target two-qubit gate using the FRB and RB schemes, respectively. We also give the corresponding infidelity to make the error reduction more transparent.}
    \label{tab:frb_ifrb_data}
\end{table}

\emph{Conclusions.}---In this letter we re-evaluate the design of the quantum instruction set for superconducting qubits or any quantum computing platform that offers $\iSWAP$ as a native gate. By taking only roughly half of the time of $\iSWAP$, the $\SQiSW{}$ gate is expected to be implemented with much higher fidelity. Moreover, it has superior compilation capabilities than $\iSWAP{}$ in the task of compiling arbitrary two-qubit gates. An iFRB experiment, which can benchmark non-Clifford gates, on our capacitatively coupled fluxonium quantum processor shows the gate error is reduced by $41\%$ and the Haar random two-qubit gate error is reduced by $50\%$ compared to $\iSWAP{}$ on the same chip .

Our work indicates that while two-qubit gates such as $\CNOT{}$ and $\iSWAP{}$ are usually chosen to be the two-qubit gates in quantum instruction sets, it is actually beneficial to investigate other gates that are natively supported on the platform. On our fluxonium processor, choosing $\SQiSW{}$ improves our system performance substantially with the average error rate of a Haar random two-qubit gate decreased by $50\%$. A series of works \cite{Lao+21, Peterson2022optimalsynthesis,abrams2020implementation} have studied how one can make full use of alternative two-qubit gates for compiling quantum circuits. We therefore advocate for a systematic study of hardware native gates as possible candidates to include as quantum instructions. 

Additionally, the FRB and iFRB schemes are general schemes for benchmarking arbitrary gates, assuming that a Haar random unitary can be implemented efficiently. Although such schemes face the same scaling issues of ordinary Clifford-based RB, it is still useful for characterizing hardware native gates, which are usually either single-qubit or two-qubit. Our work is the first to demonstrate the applicability of the iFRB scheme as a general sol~tion for benchmarking a non-Clifford gate. A recent work \cite{kong2021framework} established a framework for analyzing matrix exponential decay behaviors of RB schemes that involve a continuous gate set under gate-dependent noise. We leave the theoretical interpretation of the iFRB decay exponent in terms of the target gate fidelity for future work.
\\~\\
We would like to thank Casey Duckering, Eric Peterson and Jun Zhang for insightful discussions. We thank all those Alibaba Quantum Laboratory members who contributed to the development of the quantum hardware. DD would like to thank God for all of His provisions. LK is supported by NSF grant CCF-1452616.

\bibliographystyle{unsrt}
\bibliography{ref}

\begin{thebibliography}{10}

\bibitem{Lao+21}
Lingling Lao, Prakash Murali, Margaret Martonosi, and Dan Browne.
\newblock Designing calibration and expressivity-efficient instruction sets for
  quantum computing.
\newblock In {\em 2021 ACM/IEEE 48th Annual International Symposium on Computer
  Architecture (ISCA)}, pages 846--859, 2021.

\bibitem{BMV+15}
Emily Blem, Jaikrishnan Menon, Thiruvengadam Vijayaraghavan, and Karthikeyan
  Sankaralingam.
\newblock Isa wars: Understanding the relevance of isa being risc or cisc to
  performance, power, and energy on modern architectures.
\newblock {\em ACM Trans. Comput. Syst.}, 33(1), mar 2015.

\bibitem{emerson2005scalable}
Joseph Emerson, Robert Alicki, and Karol {\.Z}yczkowski.
\newblock Scalable noise estimation with random unitary operators.
\newblock {\em Journal of Optics B: Quantum and Semiclassical Optics},
  7(10):S347, 2005.

\bibitem{JST+20}
Jiaqing Jiang, Xiaoming Sun, Shang-Hua Teng, Bujiao Wu, Kewen Wu, and Jialin
  Zhang.
\newblock Optimal space-depth trade-off of cnot circuits in quantum logic
  synthesis.
\newblock In {\em Proceedings of the Fourteenth Annual ACM-SIAM Symposium on
  Discrete Algorithms}, pages 213--229. SIAM, 2020.

\bibitem{AAM18}
Matthew Amy, Parsiad Azimzadeh, and Michele Mosca.
\newblock On the controlled-not complexity of controlled-not--phase circuits.
\newblock {\em Quantum Science and Technology}, 4(1):015002, 2018.

\bibitem{VW04}
Farrokh Vatan and Colin Williams.
\newblock Optimal quantum circuits for general two-qubit gates.
\newblock {\em Phys Rev. A}, 69:032315, Mar 2004.

\bibitem{davis2019heuristics}
Marc~Grau Davis, Ethan Smith, Ana Tudor, Koushik Sen, Irfan Siddiqi, and Costin
  Iancu.
\newblock Heuristics for quantum compiling with a continuous gate set.
\newblock {\em arXiv preprint arXiv:1912.02727}, 2019.

\bibitem{Note1}
Note that some gate schemes such as those of cross-resonance gate~\cite
  {krantz2019quantum} or the Mølmer–Sørensen gate~\cite {bruzewicz19} for
  ion-traps can realize the gate with continuous single-qubit phase degrees of
  freedom. However, from a two-qubit perspective, these gate sets are
  equivalent to the singleton set $\{\protect \mathrm {CNOT} {}\}$.

\bibitem{krantz2019quantum}
Philip Krantz, Morten Kjaergaard, Fei Yan, Terry~P Orlando, Simon Gustavsson,
  and William~D Oliver.
\newblock A quantum engineer's guide to superconducting qubits.
\newblock {\em Applied Physics Reviews}, 6(2):021318, 2019.

\bibitem{bruzewicz19}
Colin~D. Bruzewicz, John Chiaverini, Robert McConnell, and Jeremy~M. Sage.
\newblock Trapped-ion quantum computing: {Progress} and challenges.
\newblock {\em Applied Physics Reviews}, 6(2):021314, June 2019.

\bibitem{arute2020observation}
Frank Arute, Kunal Arya, Ryan Babbush, Dave Bacon, Joseph~C Bardin, Rami
  Barends, Andreas Bengtsson, Sergio Boixo, Michael Broughton, Bob~B Buckley,
  et~al.
\newblock Observation of separated dynamics of charge and spin in the
  {Fermi-Hubbard} model.
\newblock {\em arXiv preprint arXiv:2010.07965}, 2020.

\bibitem{google2020hartree}
Google~AI Quantum, Collaborators*†, Frank Arute, Kunal Arya, Ryan Babbush,
  Dave Bacon, Joseph~C Bardin, Rami Barends, Sergio Boixo, Michael Broughton,
  Bob~B Buckley, et~al.
\newblock Hartree-fock on a superconducting qubit quantum computer.
\newblock {\em Science}, 369(6507):1084--1089, 2020.

\bibitem{bialczak2010quantum}
Radoslaw~C Bialczak, Markus Ansmann, Max Hofheinz, Erik Lucero, Matthew Neeley,
  AD~O’Connell, Daniel Sank, Haohua Wang, James Wenner, Matthias Steffen,
  et~al.
\newblock Quantum process tomography of a universal entangling gate implemented
  with {J}osephson phase qubits.
\newblock {\em Nature Physics}, 6(6):409--413, 2010.

\bibitem{mi2021information}
Xiao Mi, Pedram Roushan, Chris Quintana, Salvatore Mandra, Jeffrey Marshall,
  Charles Neill, Frank Arute, Kunal Arya, Juan Atalaya, Ryan Babbush, et~al.
\newblock Information scrambling in quantum circuits.
\newblock {\em Science}, 374(6574):1479--1483, 2021.

\bibitem{abrams2020implementation}
Deanna~M Abrams, Nicolas Didier, Blake~R Johnson, Marcus~P da~Silva, and Colm~A
  Ryan.
\newblock Implementation of {XY} entangling gates with a single calibrated
  pulse.
\newblock {\em Nature Electronics}, pages 1--7, 2020.

\bibitem{Peterson2020fixeddepthtwoqubit}
Eric~C. Peterson, Gavin~E. Crooks, and Robert~S. Smith.
\newblock Fixed-{D}epth {T}wo-{Q}ubit {C}ircuits and the {M}onodromy
  {P}olytope.
\newblock {\em {Quantum}}, 4:247, March 2020.

\bibitem{PRXQuantum.2.040309}
Tobias Haug, Kishor Bharti, and M.S. Kim.
\newblock Capacity and quantum geometry of parametrized quantum circuits.
\newblock {\em PRX Quantum}, 2:040309, Oct 2021.

\bibitem{goerz2017charting}
Michael~H Goerz, Felix Motzoi, K~Birgitta Whaley, and Christiane~P Koch.
\newblock Charting the circuit qed design landscape using optimal control
  theory.
\newblock {\em npj Quantum Information}, 3(1):1--10, 2017.

\bibitem{garion2021experimental}
Shelly Garion, Naoki Kanazawa, Haggai Landa, David~C McKay, Sarah Sheldon,
  Andrew~W Cross, and Christopher~J Wood.
\newblock Experimental implementation of non-{C}lifford interleaved randomized
  benchmarking with a controlled-{S} gate.
\newblock {\em Physical Review Research}, 3(1):013204, 2021.

\bibitem{barends2019}
R.~Barends, C.~M. Quintana, A.~G. Petukhov, Yu~Chen, D.~Kafri, K.~Kechedzhi,
  R.~Collins, O.~Naaman, S.~Boixo, F.~Arute, K.~Arya, D.~Buell, B.~Burkett,
  Z.~Chen, B.~Chiaro, A.~Dunsworth, B.~Foxen, A.~Fowler, C.~Gidney,
  M.~Giustina, R.~Graff, T.~Huang, E.~Jeffrey, J.~Kelly, P.~V. Klimov,
  F.~Kostritsa, D.~Landhuis, E.~Lucero, M.~McEwen, A.~Megrant, X.~Mi, J.~Mutus,
  M.~Neeley, C.~Neill, E.~Ostby, P.~Roushan, D.~Sank, K.~J. Satzinger,
  A.~Vainsencher, T.~White, J.~Yao, P.~Yeh, A.~Zalcman, H.~Neven, V.~N.
  Smelyanskiy, and John~M. Martinis.
\newblock Diabatic gates for frequency-tunable superconducting qubits.
\newblock {\em Phys. Rev. Lett.}, 123:210501, Nov 2019.

\bibitem{ganzhorn2020}
M.~Ganzhorn, G.~Salis, D.~J. Egger, A.~Fuhrer, M.~Mergenthaler, C.~M\"uller,
  P.~M\"uller, S.~Paredes, M.~Pechal, M.~Werninghaus, and S.~Filipp.
\newblock Benchmarking the noise sensitivity of different parametric two-qubit
  gates in a single superconducting quantum computing platform.
\newblock {\em Phys. Rev. Research}, 2:033447, Sep 2020.

\bibitem{PhysRevX.11.021058}
Youngkyu Sung, Leon Ding, Jochen Braum\"uller, Antti Veps\"al\"ainen, Bharath
  Kannan, Morten Kjaergaard, Ami Greene, Gabriel~O. Samach, Chris McNally,
  David Kim, Alexander Melville, Bethany~M. Niedzielski, Mollie~E. Schwartz,
  Jonilyn~L. Yoder, Terry~P. Orlando, Simon Gustavsson, and William~D. Oliver.
\newblock Realization of high-fidelity cz and $zz$-free iswap gates with a
  tunable coupler.
\newblock {\em Phys. Rev. X}, 11:021058, Jun 2021.

\bibitem{nguyen2019}
Long~B. Nguyen, Yen-Hsiang Lin, Aaron Somoroff, Raymond Mencia, Nicholas
  Grabon, and Vladimir~E. Manucharyan.
\newblock High-coherence fluxonium qubit.
\newblock {\em Phys. Rev. X}, 9:041041, Nov 2019.

\bibitem{zhang2003geometric}
Jun Zhang, Jiri Vala, Shankar Sastry, and K~Birgitta Whaley.
\newblock Geometric theory of nonlocal two-qubit operations.
\newblock {\em Phys. Rev. A}, 67(4):042313, 2003.

\bibitem{Note2}
Note that there are many ways one can choose the canonicalization. Our
  convention follows the one in~\cite {cirq_developers_2021_4586899} and~\cite
  {cross2019validating}.

\bibitem{chen2021compiling}
Jianxin Chen, Dawei Ding, Cupjin Huang, and Qi~Ye.
\newblock Compiling arbitrary single-qubit gates via the phase-shifts of
  microwave pulses.
\newblock {\em arXiv preprint arXiv:2105.02398}, 2021.

\bibitem{zhang2004minimum}
Jun Zhang, Jiri Vala, Shankar Sastry, and K~Birgitta Whaley.
\newblock Minimum construction of two-qubit quantum operations.
\newblock {\em Phys. Rev. Lett.}, 93(2):020502, 2004.

\bibitem{ye2004super}
Ming-Yong Ye, Yong-Sheng Zhang, and Guang-Can Guo.
\newblock Super controlled gates and controlled gates in two-qubit gate
  simulations.
\newblock {\em arXiv preprint quant-ph/0407108}, 2004.

\bibitem{Bao2021}
Feng Bao, Hao Deng, Dawei Ding, Ran Gao, Xun Gao, Cupjin Huang, Xun Jiang,
  Hsiang-Sheng Ku, Zhisheng Li, Xizheng Ma, Xiaotong Ni, Jin Qin, Zhijun Song,
  Hantao Sun, Chengchun Tang, Tenghui Wang, Feng Wu, Tian Xia, Wenlong Yu, Fang
  Zhang, Gengyan Zhang, Xiaohang Zhang, Jingwei Zhou, Xing Zhu, Yaoyun Shi,
  Jianxin Chen, Hui-Hai Zhao, and Chunqing Deng.
\newblock Fluxonium: an alternative qubit platform for high-fidelity
  operations.
\newblock {\em arXiv:2111.13504 [quant-ph]}, December 2021.

\bibitem{magesan2012efficient}
Easwar Magesan, Jay~M Gambetta, Blake~R Johnson, Colm~A Ryan, Jerry~M Chow,
  Seth~T Merkel, Marcus~P Da~Silva, George~A Keefe, Mary~B Rothwell, Thomas~A
  Ohki, et~al.
\newblock Efficient measurement of quantum gate error by interleaved randomized
  benchmarking.
\newblock {\em Phys. Rev. Lett.}, 109(8):080505, 2012.

\bibitem{emerson2007symmetrized}
Joseph Emerson, Marcus Silva, Osama Moussa, Colm Ryan, Martin Laforest,
  Jonathan Baugh, David~G Cory, and Raymond Laflamme.
\newblock Symmetrized characterization of noisy quantum processes.
\newblock {\em Science}, 317(5846):1893--1896, 2007.

\bibitem{chen2018metrology}
Zijun Chen.
\newblock {\em Metrology of quantum control and measurement in superconducting
  qubits}.
\newblock PhD thesis, UC Santa Barbara, 2018.

\bibitem{chen2016measuring}
Zijun Chen, Julian Kelly, Chris Quintana, R~Barends, B~Campbell, Yu~Chen,
  B~Chiaro, A~Dunsworth, AG~Fowler, E~Lucero, et~al.
\newblock Measuring and suppressing quantum state leakage in a superconducting
  qubit.
\newblock {\em Phys. Rev. Lett.}, 116(2):020501, 2016.

\bibitem{sheldon2016characterizing}
Sarah Sheldon, Lev~S Bishop, Easwar Magesan, Stefan Filipp, Jerry~M Chow, and
  Jay~M Gambetta.
\newblock Characterizing errors on qubit operations via iterative randomized
  benchmarking.
\newblock {\em Phys. Rev. A}, 93(1):012301, 2016.

\bibitem{kong2021framework}
Linghang Kong.
\newblock A framework for randomized benchmarking over compact groups.
\newblock {\em arXiv preprint arXiv:2111.10357}, 2021.

\bibitem{Note3}
We also measured an average $\protect \mathrm {\lowercase {i}SWAP} {}$ RB value
  of $97.26\%$. This is significantly different from the FRB value, but this is
  because we use on average 1.5 $\protect \mathrm {\lowercase {i}SWAP} {}$
  gates to compile Cliffords while we use 3 to compile arbitrary two-qubit
  gates.

\bibitem{Peterson2022optimalsynthesis}
Eric~C. Peterson, Lev~S. Bishop, and Ali Javadi-Abhari.
\newblock Optimal synthesis into fixed {XX} interactions.
\newblock {\em {Quantum}}, 6:696, April 2022.

\bibitem{cirq_developers_2021_4586899}
Cirq Developers.
\newblock Cirq, March 2021.
\newblock {See full list of authors on Github: https://github
  .com/quantumlib/Cirq/graphs/contributors}.

\bibitem{cross2019validating}
Andrew~W Cross, Lev~S Bishop, Sarah Sheldon, Paul~D Nation, and Jay~M Gambetta.
\newblock Validating quantum computers using randomized model circuits.
\newblock {\em Phys. Rev. A}, 100(3):032328, 2019.

\end{thebibliography}


\begin{thebibliography}{10}

\bibitem{zhang2003geometric}
Jun Zhang, Jiri Vala, Shankar Sastry, and K~Birgitta Whaley.
\newblock Geometric theory of nonlocal two-qubit operations.
\newblock {\em Phys. Rev. A}, 67(4):042313, 2003.

\bibitem{tucci2005introduction}
Robert~R Tucci.
\newblock {An introduction to Cartan's KAK decomposition for QC programmers}.
\newblock {\em arXiv preprint quant-ph/0507171}, 2005.

\bibitem{zhang2004minimum}
Jun Zhang, Jiri Vala, Shankar Sastry, and K~Birgitta Whaley.
\newblock Minimum construction of two-qubit quantum operations.
\newblock {\em Phys. Rev. Lett.}, 93(2):020502, 2004.

\bibitem{makhlin2002nonlocal}
Yuriy Makhlin.
\newblock Nonlocal properties of two-qubit gates and mixed states, and the
  optimization of quantum computations.
\newblock {\em Quantum Information Processing}, 1(4):243--252, 2002.

\bibitem{watts2013metric}
Paul Watts, Maurice O'Connor, and Ji{\v{r}}{\'\i} Vala.
\newblock Metric structure of the space of two-qubit gates, perfect entanglers
  and quantum control.
\newblock {\em Entropy}, 15(6):1963--1984, 2013.

\bibitem{cross2019validating}
Andrew~W Cross, Lev~S Bishop, Sarah Sheldon, Paul~D Nation, and Jay~M Gambetta.
\newblock Validating quantum computers using randomized model circuits.
\newblock {\em Phys. Rev. A}, 100(3):032328, 2019.

\bibitem{ye2004super}
Ming-Yong Ye, Yong-Sheng Zhang, and Guang-Can Guo.
\newblock Super controlled gates and controlled gates in two-qubit gate
  simulations.
\newblock {\em arXiv preprint quant-ph/0407108}, 2004.

\bibitem{zhang2004optimal}
Jun Zhang, Jiri Vala, Shankar Sastry, and K~Birgitta Whaley.
\newblock Optimal quantum circuit synthesis from controlled-unitary gates.
\newblock {\em Phys. Rev. A}, 69(4):042309, 2004.

\bibitem{Peterson2020fixeddepthtwoqubit}
Eric~C. Peterson, Gavin~E. Crooks, and Robert~S. Smith.
\newblock Fixed-{D}epth {T}wo-{Q}ubit {C}ircuits and the {M}onodromy
  {P}olytope.
\newblock {\em {Quantum}}, 4:247, March 2020.

\bibitem{koponen2006discrete}
Laura Koponen, Ville Bergholm, and Martti~M Salomaa.
\newblock A discrete local invariant for quantum gates.
\newblock {\em Quantum Information \& Computation}, 6(1):58--66, 2006.

\bibitem{dur2001multipartite}
W~D{\"u}r.
\newblock Multipartite entanglement that is robust against disposal of
  particles.
\newblock {\em Phys. Rev. A}, 63(2):020303, 2001.

\bibitem{barends2019}
R.~Barends, C.~M. Quintana, A.~G. Petukhov, Yu~Chen, D.~Kafri, K.~Kechedzhi,
  R.~Collins, O.~Naaman, S.~Boixo, F.~Arute, K.~Arya, D.~Buell, B.~Burkett,
  Z.~Chen, B.~Chiaro, A.~Dunsworth, B.~Foxen, A.~Fowler, C.~Gidney,
  M.~Giustina, R.~Graff, T.~Huang, E.~Jeffrey, J.~Kelly, P.~V. Klimov,
  F.~Kostritsa, D.~Landhuis, E.~Lucero, M.~McEwen, A.~Megrant, X.~Mi, J.~Mutus,
  M.~Neeley, C.~Neill, E.~Ostby, P.~Roushan, D.~Sank, K.~J. Satzinger,
  A.~Vainsencher, T.~White, J.~Yao, P.~Yeh, A.~Zalcman, H.~Neven, V.~N.
  Smelyanskiy, and John~M. Martinis.
\newblock Diabatic gates for frequency-tunable superconducting qubits.
\newblock {\em Phys. Rev. Lett.}, 123:210501, Nov 2019.

\bibitem{ganzhorn2020}
M.~Ganzhorn, G.~Salis, D.~J. Egger, A.~Fuhrer, M.~Mergenthaler, C.~M\"uller,
  P.~M\"uller, S.~Paredes, M.~Pechal, M.~Werninghaus, and S.~Filipp.
\newblock Benchmarking the noise sensitivity of different parametric two-qubit
  gates in a single superconducting quantum computing platform.
\newblock {\em Phys. Rev. Research}, 2:033447, Sep 2020.

\bibitem{PhysRevX.11.021058}
Youngkyu Sung, Leon Ding, Jochen Braum\"uller, Antti Veps\"al\"ainen, Bharath
  Kannan, Morten Kjaergaard, Ami Greene, Gabriel~O. Samach, Chris McNally,
  David Kim, Alexander Melville, Bethany~M. Niedzielski, Mollie~E. Schwartz,
  Jonilyn~L. Yoder, Terry~P. Orlando, Simon Gustavsson, and William~D. Oliver.
\newblock Realization of high-fidelity cz and $zz$-free iswap gates with a
  tunable coupler.
\newblock {\em Phys. Rev. X}, 11:021058, Jun 2021.

\bibitem{krantz2019quantum}
Philip Krantz, Morten Kjaergaard, Fei Yan, Terry~P Orlando, Simon Gustavsson,
  and William~D Oliver.
\newblock A quantum engineer's guide to superconducting qubits.
\newblock {\em Applied Physics Reviews}, 6(2):021318, 2019.

\bibitem{nielsen2002simple}
Michael~A Nielsen.
\newblock A simple formula for the average gate fidelity of a quantum dynamical
  operation.
\newblock {\em Physics Letters A}, 303(4):249--252, 2002.

\bibitem{omalley2015}
P.~J.~J. O'Malley, J.~Kelly, R.~Barends, B.~Campbell, Y.~Chen, Z.~Chen,
  B.~Chiaro, A.~Dunsworth, A.~G. Fowler, I.-C. Hoi, E.~Jeffrey, A.~Megrant,
  J.~Mutus, C.~Neill, C.~Quintana, P.~Roushan, D.~Sank, A.~Vainsencher,
  J.~Wenner, T.~C. White, A.~N. Korotkov, A.~N. Cleland, and John~M. Martinis.
\newblock Qubit metrology of ultralow phase noise using randomized
  benchmarking.
\newblock {\em Phys. Rev. Applied}, 3:044009, Apr 2015.

\bibitem{nguyen2019}
Long~B. Nguyen, Yen-Hsiang Lin, Aaron Somoroff, Raymond Mencia, Nicholas
  Grabon, and Vladimir~E. Manucharyan.
\newblock High-coherence fluxonium qubit.
\newblock {\em Phys. Rev. X}, 9:041041, Nov 2019.

\bibitem{Bao2021}
Feng Bao, Hao Deng, Dawei Ding, Ran Gao, Xun Gao, Cupjin Huang, Xun Jiang,
  Hsiang-Sheng Ku, Zhisheng Li, Xizheng Ma, Xiaotong Ni, Jin Qin, Zhijun Song,
  Hantao Sun, Chengchun Tang, Tenghui Wang, Feng Wu, Tian Xia, Wenlong Yu, Fang
  Zhang, Gengyan Zhang, Xiaohang Zhang, Jingwei Zhou, Xing Zhu, Yaoyun Shi,
  Jianxin Chen, Hui-Hai Zhao, and Chunqing Deng.
\newblock Fluxonium: an alternative qubit platform for high-fidelity
  operations.
\newblock {\em arXiv:2111.13504 [quant-ph]}, December 2021.

\bibitem{arute2020observation}
Frank Arute, Kunal Arya, Ryan Babbush, Dave Bacon, Joseph~C Bardin, Rami
  Barends, Andreas Bengtsson, Sergio Boixo, Michael Broughton, Bob~B Buckley,
  et~al.
\newblock Observation of separated dynamics of charge and spin in the
  {Fermi-Hubbard} model.
\newblock {\em arXiv preprint arXiv:2010.07965}, 2020.

\bibitem{magesan2012efficient}
Easwar Magesan, Jay~M Gambetta, Blake~R Johnson, Colm~A Ryan, Jerry~M Chow,
  Seth~T Merkel, Marcus~P Da~Silva, George~A Keefe, Mary~B Rothwell, Thomas~A
  Ohki, et~al.
\newblock Efficient measurement of quantum gate error by interleaved randomized
  benchmarking.
\newblock {\em Phys. Rev. Lett.}, 109(8):080505, 2012.

\bibitem{helsen2020matchgate}
Jonas Helsen, Sepehr Nezami, Matthew Reagor, and Michael Walter.
\newblock Matchgate benchmarking: Scalable benchmarking of a continuous family
  of many-qubit gates.
\newblock {\em arXiv preprint arXiv:2011.13048}, 2020.

\bibitem{emerson2005scalable}
Joseph Emerson, Robert Alicki, and Karol {\.Z}yczkowski.
\newblock Scalable noise estimation with random unitary operators.
\newblock {\em Journal of Optics B: Quantum and Semiclassical Optics},
  7(10):S347, 2005.

\bibitem{knill2008randomized}
Emanuel Knill, Dietrich Leibfried, Rolf Reichle, Joe Britton, R~Brad Blakestad,
  John~D Jost, Chris Langer, Roee Ozeri, Signe Seidelin, and David~J Wineland.
\newblock Randomized benchmarking of quantum gates.
\newblock {\em Phys. Rev. A}, 77(1):012307, 2008.

\bibitem{emerson2007symmetrized}
Joseph Emerson, Marcus Silva, Osama Moussa, Colm Ryan, Martin Laforest,
  Jonathan Baugh, David~G Cory, and Raymond Laflamme.
\newblock Symmetrized characterization of noisy quantum processes.
\newblock {\em Science}, 317(5846):1893--1896, 2007.

\bibitem{gambetta2012characterization}
Jay~M Gambetta, Antonio~D C{\'o}rcoles, Seth~T Merkel, Blake~R Johnson, John~A
  Smolin, Jerry~M Chow, Colm~A Ryan, Chad Rigetti, S~Poletto, Thomas~A Ohki,
  et~al.
\newblock Characterization of addressability by simultaneous randomized
  benchmarking.
\newblock {\em Phys. Rev. Lett.}, 109(24):240504, 2012.

\bibitem{wallman2015robust}
Joel~J Wallman, Marie Barnhill, and Joseph Emerson.
\newblock Robust characterization of loss rates.
\newblock {\em Phys. Rev. Lett.}, 115(6):060501, 2015.

\bibitem{wallman2016robust}
Joel~J Wallman, Marie Barnhill, and Joseph Emerson.
\newblock Robust characterization of leakage errors.
\newblock {\em New Journal of Physics}, 18(4):043021, 2016.

\bibitem{garion2021experimental}
Shelly Garion, Naoki Kanazawa, Haggai Landa, David~C McKay, Sarah Sheldon,
  Andrew~W Cross, and Christopher~J Wood.
\newblock Experimental implementation of non-{C}lifford interleaved randomized
  benchmarking with a controlled-{S} gate.
\newblock {\em Physical Review Research}, 3(1):013204, 2021.

\bibitem{gottesman1998heisenberg}
Daniel Gottesman.
\newblock The {H}eisenberg representation of quantum computers.
\newblock {\em arXiv preprint quant-ph/9807006}, 1998.

\bibitem{arute2019quantum}
Frank Arute, Kunal Arya, Ryan Babbush, Dave Bacon, Joseph~C Bardin, Rami
  Barends, Rupak Biswas, Sergio Boixo, Fernando~GSL Brandao, David~A Buell,
  et~al.
\newblock Quantum supremacy using a programmable superconducting processor.
\newblock {\em Nature}, 574(7779):505--510, 2019.

\bibitem{reagor2018demonstration}
Matthew Reagor, Christopher~B Osborn, Nikolas Tezak, Alexa Staley, Guenevere
  Prawiroatmodjo, Michael Scheer, Nasser Alidoust, Eyob~A Sete, Nicolas Didier,
  Marcus~P da~Silva, et~al.
\newblock Demonstration of universal parametric entangling gates on a
  multi-qubit lattice.
\newblock {\em Science advances}, 4(2):eaao3603, 2018.

\bibitem{knill1995approximation}
Emanuel Knill.
\newblock Approximation by quantum circuits.
\newblock {\em arXiv preprint quant-ph/9508006}, 1995.

\bibitem{mckay2017efficient}
David~C McKay, Christopher~J Wood, Sarah Sheldon, Jerry~M Chow, and Jay~M
  Gambetta.
\newblock Efficient {Z} gates for quantum computing.
\newblock {\em Phys. Rev. A}, 96(2):022330, 2017.

\bibitem{kong2021framework}
Linghang Kong.
\newblock A framework for randomized benchmarking over compact groups.
\newblock {\em arXiv preprint arXiv:2111.10357}, 2021.

\bibitem{cirq_developers_2021_4586899}
Cirq Developers.
\newblock Cirq, March 2021.
\newblock {See full list of authors on Github: https://github
  .com/quantumlib/Cirq/graphs/contributors}.

\end{thebibliography}

\end{document}


\title{Supplemental Material for `Quantum Instruction Set Design for Performance'}

\maketitle
\tableofcontents
\section{Compilation: Two-qubit gates}
\label{app:compilation}
In this section, we first show some basic mathematical properties of the $\SQiSW$  gate and then study the information processing capabilities of the $\SQiSW$  gate, in particular its ability to compile two-qubit and higher operations. Specifically, we prove that an arbitrary two-qubit gate can be decomposed into at most three applications of the $\SQiSW$  gate interleaved by single qubit rotations and give explicit decompositions for certain families of two-qubit rotations. 
The $\CNOT$  gate and the $\iSWAP$ gate also generate all two-qubit gates with three applications; however we prove that a majority ($\sim79\%$) of two-qubit gates, under the Haar measure, can be generated using only two uses of the $\SQiSW$  gate, whereas gates generated by two uses of the $\CNOT$  gate or the $\iSWAP$ gate only spans zero-measure sets. We lastly prove that $\SQiSW$  has an advantage over the $\CNOT$ and $\iSWAP$  gates in the task of preparing W-like states. 

\subsection{Basic mathematical properties}
We summarize some useful mathematical properties of the $\SQiSW$  gate. Besides being the square root of the $\iSWAP$ gate, $\SQiSW$  satisfies the following properties:
\begin{itemize}
\item $\SQiSW$  lies in the third level of the Clifford hierarchy: just like the $T$ gate and the Controlled-$S$ gate, the $\SQiSW$  gate conjugates Pauli matrices to Clifford matrices. Also, it is not in the second level of the Clifford hierarchy, meaning that it itself is not a Clifford gate.
    \item $\SQiSW$  is a perfect entangler, that is, it maps a product state into a maximally entangled state. Explicitly, 
    $$\SQiSW |01\rangle = \frac{1}{\sqrt{2}}(|01\rangle + i|10\rangle).$$
    \item $\SQiSW$  is an excitation number preserving gate, meaning that for all $\theta$, $[\SQiSW, Z_\theta\otimes Z_\theta]=0$.
\end{itemize}
To explore further properties, we first introduce some mathematics.

\subsubsection{KAK decomposition and the Weyl chamber}
The KAK decomposition and Weyl chamber provide mathematical tools to characterize two-qubit gates up to single qubit gates. That is, they give the ``non-local'' information of a two-qubit gate. In particular, the KAK decomposition characterizes equivalence classes of two-qubit unitaries, or elements in the group $SU(4)$, under actions by single-qubit rotations in $SU(2) \otimes SU(2)$ before and after. This perspective is particularly useful when we can experimentally regard single qubit local operations as free resources that introduce little error compared to two-qubit gates. We here directly state the results and refer the reader to~\cite{zhang2003geometric,tucci2005introduction} for more detailed expositions.
\begin{thm}[KAK decomposition~\cite{zhang2003geometric}]
For an arbitrary $U\in SU(4)$, there exists a unique $\Vec{\eta}=(x,y,z), \frac{\pi}4\geq x\geq y\geq |z|$, single qubit rotations $A_0,A_1,B_0,B_1\in SU(2)$ and a global phase $g\in\{1,i\}$ such that
$$U = g\cdot\left(A_1\otimes A_2\right)\exp\{i\Vec{\eta}\cdot\Vec{\Sigma}\}\left(B_1\otimes B_2\right),$$
where $\Vec{\Sigma} \equiv [\sigma_X\otimes \sigma_X, \sigma_Y\otimes \sigma_Y,\sigma_Z\otimes \sigma_Z]$. The tuple $(g,\Vec{\eta}, A_0, A_1, B_0, B_1)$ is called the \emph{KAK decomposition} of the unitary $U$.
\end{thm}

Define the magic basis change matrix $M\equiv \frac{1}{\sqrt{2}}\begin{bmatrix}1&0&0&i\\ 0&i&1&0\\0&i&-1&0\\ 1&0&0&-i\end{bmatrix}$. The KAK decomposition theorem can be equivalently stated as follows: $$M^\dag U M = g\cdot A\cdot K\cdot B,$$ where $A,B\in SO(4)$ and 
$$K=\begin{bmatrix}e^{i(x-y+z)}&0&0&0\\0&e^{i(x+y-z)}&0&0\\0&0&e^{i(-x-y-z)}&0\\0&0&0&e^{i(-x+y+z)}\end{bmatrix}$$
is a diagonal matrix.

\begin{figure}
    \centering
\begin{tikzpicture}[scale=.5]
\def \tta{ 90.000000000000 } 
\def \k{    0.30000000000000 } 
\def \l{     7.00000000000000 } 
\def \d{     5.50000000000000 } 
\def \h{     7.00000000000000 } 
one
\coordinate (I) at (0,0); 
\coordinate (Cnot) at (0,{-\h}); 
\coordinate (iSwap) at ({-\l*sin(\tta))},
                    {-\h+\l*cos(\tta)}); 
\coordinate (B) at ({-\l*sin(\tta)/2)},
                    {-\h+\l*cos(\tta)/2});
\coordinate (Swap) at ({-\l*sin(\tta)-\d*sin(\k*\tta)},{-\h+\l*cos(\tta)+\d*cos(\k*\tta)}); 
\coordinate (Swapm) at ({-\l*sin(\tta)+\d*sin(\k*\tta)},{-\h+\l*cos(\tta)-\d*cos(\k*\tta)}); 
\coordinate (SQiSW) at ({-\l*sin(\tta))/2},
                    {(-\h+\l*cos(\tta))/2});
\coordinate (b) at ({-\l*sin(\tta))/2-1.9},
                    {(-\h+\l*cos(\tta))/2+0.6});                    
                    
\draw[-,thick] (Cnot) --  (Swapm)
                        (I) --  (Swap)
                        (I) -- (Swapm)
                        (I) -- (Cnot)
                        (Swap) --  (Swapm)
                        (I) --  (iSwap);

\draw[dashed,thick] (iSwap) --  (Cnot)
                        (Swap)  -- (Cnot);

\fill[black]  (I) circle [radius=2pt]; 
\fill[black]    (iSwap) circle [radius=2pt]; 
\fill[black]  (Cnot) circle [radius=2pt]; 
\fill[black] (Swap) circle [radius=2pt];
\fill[black] (Swapm) circle [radius=2pt];
\fill[black] (B) circle [radius=2pt];
\fill[red] (SQiSW) circle [radius=2pt];

\draw (I) node [right]           {$I (0,0,0)$}
      (Cnot) node [right]     {$\CNOT (\frac{\pi}{4},0,0)$}
      (Swap) node [left]           {$\SWAP^\dag (\frac\pi4,\frac\pi4,\frac{-\pi}4)$}
      (Swapm)  node [left]       {$\SWAP (\frac\pi4,\frac\pi4,\frac\pi4)$}
      (iSwap)  node [left]            {$\iSWAP (\frac\pi4, \frac\pi4, 0)$}
      (B) node [below] {$B$}
      (b) node    {$\SQiSW (\frac{\pi}8,\frac\pi8,0)$};
      
\end{tikzpicture}
\caption{An illustration of the Weyl chamber and the positions of common gates. Note that $\SQiSW$ lies in the midpoint of the identity and $\iSWAP$. The point $\SWAP^\dag$ is to be identified with the point $\SWAP$ but is drawn separately for easier visualization.}
\label{fig:weyl}
\end{figure}
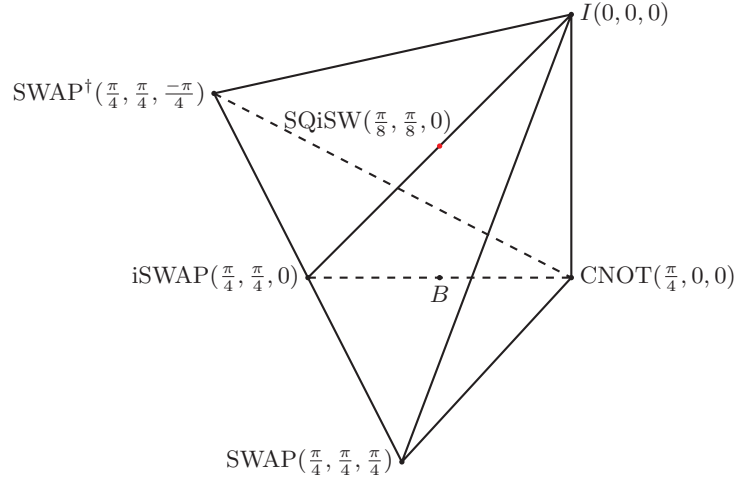

The equivalence class of a unitary $U$ under local unitaries is characterized by the \emph{interaction coefficients} $\eta(U)$, which lives in a 3-dimensional tetrahedron called the \emph{Weyl chamber}\footnote{Note that there are many ways one can choose the canonicalization. Our convention follows the one in \cite{cirq_developers_2021_4586899} and \cite{cross2019validating}.} $$W\equiv\{\pi/4\geq x\geq y\geq |z|\ \textrm{and}\ z\geq 0 \text{ if }x=\frac{\pi}{4}\mid(x,y,z)\in\mathbb{R}^3\}.$$ We say that two unitaries $U,V\in SU(4)$ are locally equivalent, or $U\sim V$, if $\eta(U)=\eta(V)$. We give the interaction coefficients for some common gates: 
\begin{itemize}
    \item $I:\eta(I)=(0,0,0)$;
    \item $\SWAP: \eta(\SWAP)=(\frac{\pi}4,\frac{\pi}4,\frac{\pi}4)$;
    \item $\CNOT$, CZ$: \eta(\CNOT)=\eta(\textrm{CZ})=(\frac{\pi}4,0,0)$. Note that $\CNOT\sim$ CZ by a local Hadamard conjugation on the target qubit;
    \item $\iSWAP:\eta(\iSWAP) = (\frac{\pi}4,\frac{\pi}4,0)$;
    \item $B$~\cite{zhang2004minimum}$:\eta(B)=(\frac{\pi}4,\frac{\pi}8,0)$;
    \item $\SQiSW: \eta(\SQiSW) = (\frac{\pi}8,\frac{\pi}8,0)$.
\end{itemize}
These gates and their positions in the Weyl chamber is given in~Fig.~\ref{fig:weyl}.
\begin{dfn}
Let $L(x,y,z) \equiv \exp\left(i [x,y,z]\cdot \Vec{\Sigma}\right)$ be the canonical element of the equivalence class.
\end{dfn} 
We have $U\sim L(\eta(U))$ for all $U\in SU(4)$.

\subsubsection{Local invariants and the character polynomial} The KAK decomposition geometrically characterizes the equivalence class of a unitary $U\in SU(4)$; however, it requires diagonalization of matrices and thus can sometimes be difficult to study analytically. Local invariants~\cite{makhlin2002nonlocal} characterizes the equivalence class of the unitary $U$, while still being easy to solve. There are many different choices of local invariants. We choose ours to be the degree-4 polynomial
$$F_U(t)=\det[\Re[M^\dag U M] + t\cdot \Im[M^\dag U M]],$$
where $\Re[\cdot]$ and $\Im[\cdot]$ represent the (element-wise) real and imaginary part of a matrix. We call this the \emph{character polynomial}. To see that the polynomial is locally invariant, we first observe $U\sim V\Leftrightarrow \exists O_1, O_2\in SO(4), O_1M^\dag UMO_2=M^\dag VM$. Then
\begin{align*}
    F_U(t)&=\det[\Re[M^\dag U M] + t\cdot \Im[M^\dag U M]]\\
    &=\det[\Re[O_1M^\dag V MO_2] + t\cdot \Im[O_1M^\dag V MO_2]]\\
    &=\det[O_1\Re[M^\dag V M]O_2 + t\cdot O_1\Im[M^\dag V M]O_2]\\
    &=\det[O_1(\Re[M^\dag V M] + t\cdot \Im[M^\dag VM])O_2]\\
    &=\det[O_1]\det[\Re[M^\dag V M] + t\cdot \Im[M^\dag V M]]\det[O_2]\\
    &=\det[\Re[M^\dag V M] + t\cdot \Im[M^\dag V M]]=F_V(t).\\
\end{align*}

The polynomial is a complete characterization of the equivalence classes as the zeros of the polynomial are $-\cot(x-y+z), -\cot(x+y-z), -\cot(-x-y-z)$ and $-\cot(-x+y+z)$ by evaluating it on the canonical element. Hence, one only needs to check the corresponding character polynomial coefficients in order to determine whether two unitaries are locally equivalent. Furthermore,
$$F_U(i) = \det[M^\dag U M] = 1, F_U(-i) = \det[M^\dag U^* M]=1,$$
since $U,U^*\in SU(4)$, leaving the character polynomial with three free coefficients. For $U$ with interaction coefficients $(x, y,z)$, we have
$$F_U(t)=(t^2+1)(Ct^2+Bt+A)-t^2,$$
where 
\begin{align}
    A&=\cos(x+y-z)\cos(x-y+z)\cos(-x-y-z)\cos(-x+y+z),\label{eq:charpoly_A}\\
    B&=-\sin2x\sin2y\sin2z,\label{eq:charpoly_B}\\
    C&=\sin(x+y-z)\sin(x-y+z)\sin(-x-y-z)\sin(-x+y+z).\label{eq:charpoly_C}
\end{align}

\subsubsection{Effective target size}
An interesting way to quantify how easy it is to realize a two-qubit gate with quantum control is its effective target size, as put forth in~\cite{watts2013metric}. Intuitively, the effective target size is the invariant volume of the region around a two-qubit gate that corresponds to a small perturbation of its interaction coefficients. We show in this section that the effective target size of $\SQiSW$ is larger than that of $\CNOT$ and $\iSWAP$, having a target size that scales with the perturbation better than any other common two-qubit gate, save for the $B$ gate.

Let $U \in SU(4)$ and its interaction coefficients $\eta(U) = (x,y,z) \in W$. Furthermore, let $\mathcal{U}$ be the neighborhood of $\eta(U)$ given by a box with edge length $a$ centered on $\eta(U)$ and with sides parallel to the $x,y,z$ axes. Then, the effective target size of $U$ is defined as
\begin{align*}
T(U) \equiv \int_{(SU(2) \otimes SU(2)) \times \mathcal{U} \times (SU(2) \otimes SU(2))} d\mu = \int_{\mathcal{U}} d\mu_W
\end{align*}
where $d\mu$ is the Haar measure over $SU(4)$ and 
\begin{align}
    d\mu_W&\equiv M_W(x,y,z) dx\land dy\land dz \nonumber\\
    &=\frac{3}\pi[\cos 2x\cos 4y+\cos 2y\cos 4z+\cos 2z\cos 4x-\cos 2x\cos 4z-\cos 2y\cos 4x-\cos 2z\cos 4y ]dx\land dy\land dz \label{eq:haar_weyl}
\end{align}
is the normalized Haar measure over $W$.

We note that the effective target size is the same for mirror gates (differ by a $\SWAP$) since its definition is symmetric under exchanging which qubits we deem as the first and the second. Now, the Weyl coordinates of $\SQiSW$  is $(\pi/8, \pi/8, 0)$. Its mirror gate has coordinates (e.g. see~\cite{cross2019validating}) 
\begin{align*}
    (\pi/4 - 0, \pi/4-\pi/8, 1\times(\pi/8 - \pi/4)) & = (\pi/4, \pi/8, -\pi/8)\\
    & \sim (-\pi/4, \pi/8, -\pi/8)\\
    & \sim (\pi/4, \pi/8,\pi/8),
\end{align*}
where the first equivalence follows by subtracting $\pi/2$ from the first coordinate and the second follows from flipping the signs of the first and third coordinates~\cite{tucci2005introduction}. Hence, the effective target size of $\SQiSW$ is the same as that of its mirror gate $(\pi/4, \pi/8, \pi/8)$ which is given~\footnote{Note that the convention in~\cite{watts2013metric} for the Weyl coordinates is $2$ times that of ours. } in (47) of~\cite{watts2013metric}:
\begin{align*}
    T(\SQiSW)&  = \frac{1}{2\pi} [3 \cos (a) - 3 \cos( 3a)  - 4a \sin (3a)  ] \\
    & = 4 a^4/\pi + O(a^6) \quad \text{as $a \to 0$.}
\end{align*}
This is a larger area than that of $\CNOT$ and $\iSWAP$~\cite{watts2013metric}:
\begin{align*}
    T(\CNOT) & = T(\iSWAP) \\
    & =  \frac{1}{2\pi} [8a+7a \cos(3a) -15a \cos(a)-9\sin(3a) +12\sin(2a)+3\sin(a)]\\
    & = 4a^5/\pi + O(a^7) \quad \text{as $a \to 0$.} 
\end{align*}
Note that the effective target sizes of $\CNOT$ and $\iSWAP$ are the same since they are mirror gates up to local equivalence.

\subsection{Compiling two-qubit gates into $\SQiSW$ and single-qubit rotations} 
We first study the problem of compiling arbitrary two-qubit gates. In particular, we prove the following theorem.
\begin{thm}
Every two-qubit unitary can be expressed by at most 3 $\SQiSW$  gates interleaved by single qubit gates.
\end{thm}

The proof of the theorem will consist of two steps. We first completely characterize the set $W(S_2)$ of all two-qubit gates that can be generated using only 2 uses of the $\SQiSW$  gate. We second show how to decompose a gate outside of $W(S_2)$ into one use of $\SQiSW$  and one use of a gate in $W(S_2)$. This completes the proof. We end by providing an explicit decomposition algorithm.

\subsubsection{Weyl chamber region spanned by two $\SQiSW$  gates}

We now study the region in the Weyl chamber that can be generated by two $\SQiSW$  gates interleaved with single qubit rotations. This will later help us establish compilation schemes of arbitrary two-qubit gates using $\SQiSW$  and single qubit rotations.

\begin{lem}
The Weyl chamber region spanned by two $\SQiSW$  gates is the region described by the inequalities $\frac{\pi}{4}\geq x\geq y\geq |z| \land x\geq y+|z|  .$
\label{lem:p}
\end{lem}

\begin{proof}
Denote the subset of the Weyl chamber spanned by two $\SQiSW$  gates $W(S_2)$ and $$W'\equiv\left\{\frac{\pi}{4}\geq x\geq y\geq |z| \land x\geq y+|z| \mid (x,y,z)\in W\right\}.$$ The proof proceeds in two steps: We first prove that $W'\subset W(S_2)$ by giving an analytical solution to the interleaving single qubit rotations for a general element in $W'$, and then prove that $W(S_2)\subset W'$ by investigating the character polynomial coefficients associated to a general element in $W(S_2)$.
\begin{description}
\item[$W'\subset W(S_2)$:] We prove this statement constructively by giving explicit analytical forms for the interleaving single qubit rotations: For $(x,y,z)\in W'$, consider the following gate in $W(S_2)$:
$$U(\alpha, \beta, \gamma)\equiv S\cdot\left(\begin{bmatrix}e^{i\gamma}\cos\alpha/2 & i\sin\alpha/2 \\ i\sin\alpha/2 & e^{-i\gamma}\cos\alpha/2\end{bmatrix}\otimes \begin{bmatrix}\cos\beta/2 & i\sin\beta/2 \\ i\sin\beta/2 & \cos\beta/2 \end{bmatrix}\right)\cdot S,$$
where
\begin{align}
    \alpha &= \arccos\left(\cos 2x-\cos 2y+\cos 2z+2\sqrt{C}\right), \label{eq:alpha_xyz}\\
    \beta &= \arccos\left(\cos 2x-\cos 2y+\cos 2z-2\sqrt{C}\right),\\
    \gamma &= \arccos\left(\mathrm{sgn} z\cdot\sqrt{\frac{4\cos^2x\cos^2z\sin^2y}{4\cos^2x\cos^2z\sin^2y+\cos2x\cos2y\cos2z}}\right).
\end{align}
Here we define $\mathrm{sgn}(z)=1$ if $z\geq 0$ and is otherwise $-1$.
Note that $C$ in terms of $x,y,z$ was given in~\cref{eq:charpoly_C}. One can verify that all operations including the square root and the inverse cosine functions are legal when $(x,y,z)\in W'$, and one can also verify that the interaction coefficient associated to $U(\alpha, \beta, \gamma)$ is indeed $(x,y,z)$ by comparing the coefficients in the character polynomial. 
\item[$W(S_2)\subset W'$:] Up to local equivalence, a general element in $W(S_2)$ can be parameterized by six parameters:
$$
         U(\alpha, \beta,\gamma_1, \gamma_2, \delta_1,\delta_2)
        =  S\cdot \left(
        \begin{bmatrix}
        e^{i\gamma_1}\cos \alpha  & e^{i\delta_1}\sin\alpha \\
        -e^{-i\delta_1}\sin\alpha & e^{-i\gamma_1}\cos \alpha
        \end{bmatrix}\otimes 
        \begin{bmatrix}
        e^{i\gamma_2}\cos\beta  &  e^{i\delta_2}\sin\beta\\
        -e^{-i\delta_2}\sin\beta & e^{-i\gamma_2}\cos\beta
        \end{bmatrix}\right)\cdot S,$$
where $S$ is shorthand for SQiSW. The corresponding coefficient $C$ in the character polynomial associated to it is then
\begin{align}
    C & =\frac{1}{16}(\cos2\alpha-\cos2\beta)^2.
    \label{eq:c_positive}
\end{align}
Combining~\cref{eq:c_positive} with with~\cref{eq:charpoly_C}, we have
$$C = \sin(x+y-z)\sin(x-y+z)\sin(-x-y-z)\sin(-x+y+z)\geq 0,$$
where $(x,y,z)=\eta(U(\alpha, \beta,\gamma_1, \gamma_2, \delta_1,\delta_2))$. Since $(x,y,z)\in W$ ensures that $\sin(x+y-z),\sin(x+y+z)\geq 0$, we know that $$\sin(x-y+z)\sin(x-y-z)\geq 0\Rightarrow |z|\leq x-y$$
when $\frac{\pi}4\geq x\geq y\geq |z|\geq0$.
Combining this constraint with the ones from $W$ gives us $$\frac{\pi}{4}\geq x\geq y\geq |z| \land x\geq y+|z| \Rightarrow W(S_2)\subset W'.$$

\end{description}
\end{proof}

In~Fig.~\ref{fig:s2_region} we show the region $W' \subset W$. We also show a schematic of how an element of $W(S_2)$ can be decomposed in~Fig.~\ref{fig:red}.

\begin{figure}
    \centering
\begin{tikzpicture}[scale=.5]
\def \tta{ 90.000000000000 } 
\def \k{    0.30000000000000 } 
\def \l{     7.00000000000000 } 
\def \d{     5.00000000000000 } 
\def \h{     7.0000000000000 } 

\coordinate (I) at (0,0); 
\coordinate (Cnot) at (0,{-\h}); 
\coordinate (iSwap) at ({-\l*sin(\tta))},
                    {-\h+\l*cos(\tta)}); 
\coordinate (Swap) at ({-\l*sin(\tta)-\d*sin(\k*\tta)},{-\h+\l*cos(\tta)+\d*cos(\k*\tta)}); 
\coordinate (Swapm) at ({-\l*sin(\tta)+\d*sin(\k*\tta)},{-\h+\l*cos(\tta)-\d*cos(\k*\tta)}); 
\coordinate (SQiSW) at ({-\l*sin(\tta))/2},
                    {(-\h+\l*cos(\tta))/2});
\coordinate (mSQiSWm) at ({(-\l*sin(\tta)-\d*sin(\k*\tta))/2},{-\h+(\l*cos(\tta)+\d*cos(\k*\tta))/2});

\coordinate (m) at ({(-\l*sin(\tta)-\d*sin(\k*\tta))/2+0.4},{-\h+(\l*cos(\tta)+\d*cos(\k*\tta))/2-1.2});
\coordinate (mSQiSW) at ({(-\l*sin(\tta)+\d*sin(\k*\tta))/2},{-\h+(\l*cos(\tta)-\d*cos(\k*\tta))/2});

 \filldraw[color=red, fill opacity = 0.5]
  (I) -- (Cnot) -- (mSQiSWm);
  \filldraw[color=red, fill opacity = 0.5]
  (I) -- (iSwap) -- (mSQiSWm);
  \filldraw[color=red, fill opacity = 0.5]
  (I) -- (iSwap) -- (mSQiSW);
\filldraw[color=red, fill opacity = 0.5]
  (I) -- (Cnot) -- (mSQiSW);
 \filldraw[color=red, fill opacity = 0.5]
  (mSQiSWm) -- (Cnot) -- (mSQiSW) -- (iSwap);
                    
\draw[-,thick] (Cnot) --  (Swapm)
                        (I) --  (Swap)
                        (I) -- (Swapm)
                        (I) -- (Cnot)
                        (Swap) --  (Swapm)
                        (I) --  (iSwap);

\draw[dashed,thick] (iSwap) --  (Cnot)
                        (Swap)  -- (Cnot);

\draw[dotted, thick] (iSwap) --  (mSQiSW)
                        (iSwap)  -- (mSQiSWm)
                        (I) -- (mSQiSW)
                        (I) -- (mSQiSWm);

\fill[black]  (I) circle [radius=2pt]; 
\fill[black]    (iSwap) circle [radius=2pt]; 
\fill[black]  (Cnot) circle [radius=2pt]; 
\fill[black] (Swap) circle [radius=2pt];
\fill[black] (Swapm) circle [radius=2pt];
\fill[black] (mSQiSW) circle [radius=2pt];
\fill[black] (mSQiSWm) circle [radius=2pt];
\fill[red] (SQiSW) circle [radius=2pt];

\draw (I) node [right]           {$I$}
      (Cnot) node [right]     {$\CNOT$}
      (Swap) node [left]           {$\SWAP^\dag$}
      (Swapm)  node [left]       {$\SWAP$}
      (iSwap)  node [left]            {$\iSWAP$}
      (SQiSW) node [above left]           {$\SQiSW$}
      (mSQiSW) node [right] {$(\frac{\pi}4,\frac{\pi}8,\frac{\pi}8)$}
      (m) node {$(\frac{\pi}4,\frac{\pi}8,\frac{-\pi}8)$};
\end{tikzpicture}
\caption{The region $W'=W(S_2)$ spanned by 2 $\SQiSW$ gates. It is a pyramid with vertices $I$, $\CNOT$, $(\frac{\pi}{4},\frac{\pi}{8},\frac{\pi}{8})$, $\iSWAP$, and $(\frac{\pi}{4},\frac{\pi}{8},-\frac{\pi}{8})$.}
\label{fig:s2_region}
\end{figure}
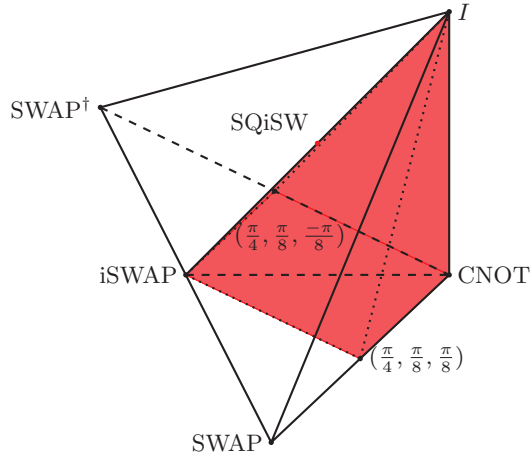

\begin{figure}
    \begin{tikzpicture}
    \centering
\node {
    \begin{quantikz}
             & \gate[style={fill=red!50}, 2]{U\in S_2} &  \qw \\
             & \qw & \qw
             \end{quantikz}
             $\sim$\begin{quantikz}
              & \gate[2]{\SQiSW} & \gate[1]{R_z(\gamma)R_x(\alpha)R_z(\gamma)}   &\gate[2]{\SQiSW}  &\qw  \\
              & \qw & \gate[1]{R_x(\beta)} & \qw&\qw
             \end{quantikz}};
             \end{tikzpicture}
    \caption{Illustration of decomposition of a two-qubit gate in $W(S_2)$ into two $\SQiSW$  gates up to local equivalence. The special form of the interleaving single qubit gates are due to the proof that the three parameters $\alpha, \beta, \gamma$ running over $[0,2\pi]$ are sufficient to generate the whole region $W'$.}
    \label{fig:red}
\end{figure}
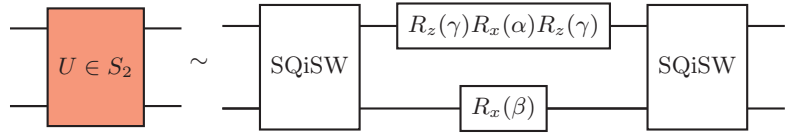

\subsubsection{Decomposing arbitrary two-qubit gate into $\leq 3$ $\SQiSW$  gates}

We now consider unitaries whose interaction coefficients lie outside of the region $W'$. Those gates include the SWAP family $(x,x,\pm x)$, the Sycamore fSim gates and so on. We show below that a third $\SQiSW$  gate is sufficient to span the whole Weyl chamber.

Given that all gates in the Weyl chamber region $W'$ can be generated using 2 $\SQiSW$  gates by Lemma~\ref{lem:p}, it suffices to prove the following.
\begin{lem}
For all $(x,y,z)\in W\setminus W'$, $L(x,y,z)$ can be generated with one use of some $L(x',y',z')$ and one $L(\frac{\pi}8,\frac{\pi}8,0)\sim \SQiSW$, where $(x',y',z')\in W'$.
\end{lem}
\begin{proof}

Before proceeding to the proof, we first visualize the constraints imposed by region $W$ and $W'$ in terms of the eigenphases $\{a_0,a_1,a_2,a_3\}$ of $L(x,y,z)$, where
\begin{align*}
    a_0 &= x+y-z,\\
    a_1 &= x-y+z,\\
    a_2 &= -x+y+z,\\
    a_3 &= -x-y-z.
\end{align*}
The constraint that $(x,y,z)\in W$ can be equivalently stated as
$$a_0\geq a_1\geq a_2\geq a_3, \sum_ia_i=0, a_0+a_1\leq \frac{\pi}2.$$
It can be deduced that $a_0\geq 0\geq a_3$. $(x,y,z)\in W'$ imposes an additional constraint:
$$a_0\geq a_1\geq{\color{red}0}\geq  a_2\geq a_3, \sum_ia_i=0, a_0+a_1\leq \frac{\pi}2.$$

Assuming that $z= \frac12(a_1+a_2) \geq  0$ (the other case can be reduced to this by observing that $\SQiSW\sim \SQiSW ^\dag$ and $L(x,y,z)\sim L^\dag(x,y,-z)$), $(x,y,z)\in W\setminus W'$ indicates that this additional constraint is violated via the sign violation $a_2>0$. We show that the following is true: we can always select $a_i, a_j, i\neq j$, append on them phases $\frac{\pi}4,-\frac{\pi}4$ such that $b_0\geq b_1\geq b_2\geq b_3$ being the sorted permutation of $(a_i+\frac{\pi}{4}, a_j-\frac{\pi}{4}, a_k, a_l), \{i,j,k,l\}\in\{0,1,2,3\}$ satisfies 
$$b_0\geq b_1\geq0\geq  b_2\geq b_3, \sum_ib_i=0, b_0+b_1\leq \frac{\pi}2.$$

This indicates that there is a way of decomposing $L(x,y,z)$ to $L(x',y',z')$ associated to the eigenphases $(b_0,b_1,b_2,b_3)$ and $L(\frac{\pi}8,\frac{\pi}8,0)$. Explicitly, we argue via the following two cases. We also give a visual argument in~Fig.~\ref{fig:visual_arg}. 

\begin{enumerate}
    \item $x\leq \frac{\pi}8$. Then $a_2\leq x\leq \frac{\pi}{8}$. One can take $$\{b_0,b_1\}=\left\{a_0+\frac{\pi}4, a_1\right\}, \{b_2,b_3\}=\textrm{sort}\left\{a_2-\frac{\pi}{4},a_3\right\},$$
    where ``$\textrm{sort}$'' means the set is sorted in descending order. One has $a_0+\frac{\pi}{4}\geq  a_1\geq 0, a_2-\frac{\pi}4, a_3\leq 0, a_0+a_1+\frac{\pi}4=2x+\frac{\pi}4\leq\frac{\pi}2$.
    \item $x>\frac{\pi}8.$ Then $a_3=-2x-a_2<-\frac{\pi}4$ and $a_2\leq x\leq \frac{\pi}{4}$. One can take
     $$\{b_0,b_1\}=\left\{a_0, a_1\right\}, \{b_2,b_3\}=\textrm{sort}\left\{a_2-\frac{\pi}4,a_3+\frac{\pi}4\right\}.$$
     One has $a_0\geq a_1\geq 0, a_2-\frac{\pi}4, a_3+\frac{\pi}4\leq 0, a_0+a_1\leq\frac{\pi}2$.
\end{enumerate}
Note that when an \emph{eigenphase crossing} happens, i.e.\ that $a_2-\frac\pi4<a_3$ in case 1 and $a_2-\frac\pi4<a_3+\frac\pi4$ in case 2, additional single qubit gates need to be applied to switch the two-qubit unitary to the canonical form for compilation purposes, see Algorithm~\ref{alg:decomp}.
\begin{figure}
\begin{subfigure}{\textwidth}
\centering
    \begin{tikzpicture}[scale=1,cap=round,>=latex]
    \def \r{3.0}
    \def \rr{2}
    \def \rl{4.0}
    \def \d{4.0}
    \def \x{20.0}
    \coordinate (x) at ({\r*cos(\x)}, {\r*sin(\x)});
    \coordinate (mx) at ({\r*cos(\x)}, {-\r*sin(\x)});
    \coordinate (a0) at ({\r*cos(35)}, {\r*sin(35)});
    \coordinate (a1) at ({\r*cos(5)}, {\r*sin(5)});
    \coordinate (a2) at ({\r*cos(10)}, {-\r*sin(10)});
    \coordinate (a3) at ({\r*cos(30)}, {-\r*sin(30)});
    \coordinate (pi) at ({\r*cos(45)}, {\r*sin(45)});
    \coordinate (mpi) at ({\r*cos(45)}, {-\r*sin(45)});
        \draw[->] (-\d,0) -- (\d,0) node[right,fill=white] {$0$};
        \draw[->] (0,-\d) -- (0,\d) node[above,fill=white] {$\frac\pi2$};

        \draw[thick] (0,0) circle(\r);
        \draw[gray, dotted, thick] (0,0) -- (\d,\d);
        \draw[gray, dotted, thick] (0,0) -- (\d,-\d);
        \draw (0,0) -- (a0);
        \draw (0,0) -- (a1);
        \draw (0,0) -- (a2);
        \draw (0,0) -- (a3);
        \draw[gray, dashed, thick] (0,0) -- (\d,{\d*tan(\x)});
        \draw[gray, dashed, thick] (0,0) -- (\d,{-\d*tan(\x)});
        \fill[black]  (x) circle [radius=1pt]; 
        
        \fill[black]  (mx) circle [radius=1pt]; 
        \fill[black]  (a0) circle [radius=2pt]; 
        \fill[black]  (a1) circle [radius=2pt]; 
        \fill[black]  (a2) circle [radius=2pt]; 
        \fill[black]  (a3) circle [radius=2pt]; 
        
        \draw (a0) node [right]     {$a_0$}
      (a1) node [right]           {$a_1$}
      (a2)  node [right]       {$a_2$}
      (a3)  node [right]            {$a_3$}
      ({\rl*cos(\x)},{\rl*sin(\x)}) node [fill=white]           {$x$}
      ({\rl*cos(\x)},{-\rl*sin(\x)}) node [fill=white] {$-x$}
      ({\rl*cos(45)},{\rl*sin(45)}) node [fill=white] {$\frac{\pi}4$}
      ({\rl*cos(45)},{-\rl*sin(45)}) node [ fill=white] {$-\frac\pi4$};

    \draw[very thick] ({cos(5)},{sin(5)}) arc [start angle=5, end angle=35, radius=1];
    \draw[very thick] ({0.95*cos(10)},{-0.95*sin(10)}) arc [start angle=-10, end angle=-30, radius=0.95];
    \draw[very thick] ({1.05*cos(10)},{-1.05*sin(10)}) arc [start angle=-10, end angle=-30, radius=1.05];
    \end{tikzpicture}
    \caption{Eigenphases without sign violation of $a_2$.}
    \label{fig:case0}
    \end{subfigure}
     \begin{subfigure}{0.45\textwidth}
    \centering
    \begin{tikzpicture}[cap=round,>=latex]
    \def \r{2}
    \def \rr{2}
    \def \rl{4.0}
    \def \d{2.7}
    \def \x{20.0}
    \coordinate (a0) at ({\r*cos(25)}, {\r*sin(25)});
    \coordinate (a0p) at ({\r*cos(70)}, {\r*sin(70)});
    \coordinate (a1) at ({\r*cos(15)}, {\r*sin(15)});
    \coordinate (a2) at ({\r*cos(10)}, {\r*sin(10)});
    \coordinate (a2p) at ({\r*cos(35)}, {-\r*sin(35)});
    \coordinate (a3) at ({\r*cos(50)}, {-\r*sin(50)});
    \coordinate (pi) at ({\r*cos(45)}, {\r*sin(45)});
    \coordinate (mpi) at ({\r*cos(45)}, {-\r*sin(45)});
        \draw[->] (-\d,0) -- (\d,0) node[right,fill=white] {$0$};
        \draw[->] (0,-\d) -- (0,\d) node[above,fill=white] {$\frac\pi2$};

        \draw[thick] (0,0) circle(\r);
        \draw[cyan] (0,0) -- (a0);
        \draw[cyan, dashed] (0,0) -- (a0p);
        \draw (0,0) -- (a1);
        \draw[red] (0,0) -- (a2);
        \draw[red, dashed] (0,0) -- (a2p);
        \draw (0,0) -- (a3);
        \fill[cyan]  (a0) circle [radius=2pt]; 
        \fill[cyan]  (a0p) circle [radius=2pt]; 
        \fill[black]  (a1) circle [radius=2pt]; 
        \fill[red]  (a2) circle [radius=2pt]; 
        \fill[red]  (a2p) circle [radius=2pt]; 
        \fill[black]  (a3) circle [radius=2pt]; 
        
        \draw (a0) node [right]     {\color{cyan} $a_0$}
      (a0p) node [right]     {\color{cyan} $a'_0$}
      (a1) node [right]           {$a_1$}
      (a2)  node [right]       {\color{red} $a_2$}
      (a2p)  node [right]       {\color{red} $a'_2$}
      (a3)  node [right]            {$a_3$};
    \draw[->, cyan, very thick] ({(\r+0.6)*cos(25)},{(\r+0.6)*sin(25)}) arc [start angle=25, end angle=70, radius={\r+0.5}];
    \draw[->, red, very thick] ({(\r+0.5)*cos(10)},{(\r+0.5)*sin(10)}) arc [start angle=10, end angle=-35, radius={\r+0.5}];
    \end{tikzpicture}
    \caption{$a_2$ has a sign violation and $x\leq \frac{\pi}8$.}
    \label{fig:case1}
    \end{subfigure}
    ~
        \begin{subfigure}{.45\textwidth}
    \centering
                \begin{tikzpicture}[cap=round,>=latex]
    \def \r{2.}
    \def \rr{2}
    \def \rl{4.0}
    \def \d{2.7}
    \def \x{25.0}
    \coordinate (x) at ({\r*cos(\x)}, {\r*sin(\x)});
    \coordinate (mx) at ({\r*cos(\x)}, {-\r*sin(\x)});
    \coordinate (a0) at ({\r*cos(30)}, {\r*sin(30)});
    \coordinate (a1) at ({\r*cos(20)}, {\r*sin(20)});
    \coordinate (a2) at ({\r*cos(10)}, {\r*sin(10)});
    \coordinate (a2p) at ({\r*cos(35)}, {-\r*sin(35)});
    \coordinate (a3) at ({\r*cos(60)}, {-\r*sin(60)});
    \coordinate (a3p) at ({\r*cos(15)}, {-\r*sin(15)});
    \coordinate (pi) at ({\r*cos(45)}, {\r*sin(45)});
    \coordinate (mpi) at ({\r*cos(45)}, {-\r*sin(45)});
        \draw[->] (-\d,0) -- (\d,0) node[right,fill=white] {$0$};
        \draw[->] (0,-\d) -- (0,\d) node[above,fill=white] {$\frac\pi2$};

        \draw[thick] (0,0) circle(\r);
        \draw (0,0) -- (a0);
        \draw[cyan, dashed] (0,0) -- (a3p);
        \draw (0,0) -- (a1);
        \draw[red] (0,0) -- (a2);
        \draw[red, dashed] (0,0) -- (a2p);
        \draw[cyan] (0,0) -- (a3);
        \fill[cyan]  (a3) circle [radius=2pt]; 
        \fill[cyan]  (a3p) circle [radius=2pt]; 
        \fill[black]  (a1) circle [radius=2pt]; 
        \fill[red]  (a2) circle [radius=2pt]; 
        \fill[red]  (a2p) circle [radius=2pt]; 
        \fill[black]  (a0) circle [radius=2pt]; 
        
        \draw (a0) node [right]     {$a_0$}
      (a3p) node [right]     {\color{cyan} $a'_3$}
      (a1) node [right]           {$a_1$}
      (a2)  node [right]       {\color{red} $a_2$}
      (a2p)  node [right]       {\color{red} $a'_2$}
      (a3)  node [right]            {\color{cyan} $a_3$};
    \draw[->, cyan, very thick] ({(\r+0.6)*cos(60)},{-(\r+0.6)*sin(60)}) arc [start angle=-60, end angle=-15, radius={\r+0.8}];
    \draw[->, red, very thick] ({(\r+0.5)*cos(10)},{(\r+0.5)*sin(10)}) arc [start angle=10, end angle=-35, radius={\r+0.5}];
    \end{tikzpicture}
    \caption{$a_2$ has a sign violation and $x> \frac{\pi}8$.}
    \label{fig:case2}
    \end{subfigure}
    \caption{Illustration of the eigenphases $a_0\geq a_1\geq a_2\geq a_3$. Being in $W$ requires that $x\leq\frac{\pi}{4}$, and $a_0, a_1$ and $a_2, a_3$ lie symmetrically with respect to $x$ and $-x$ respectively.~Fig.~\ref{fig:case0} corresponds to no sign violation of $a_2$ and hence can be generated using 2 $\SQiSW$  gates.~Fig.~\ref{fig:case1} and~Fig.~\ref{fig:case2} are possible value assignments corresponding to the two eigenphase modifications corresponding to case 1 and 2 in the proof, respectively. }
    \label{fig:visual_arg}
\end{figure}
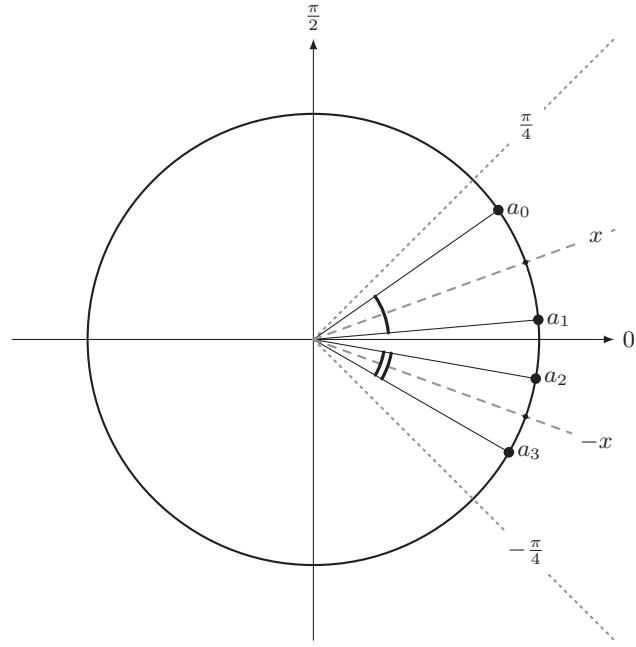
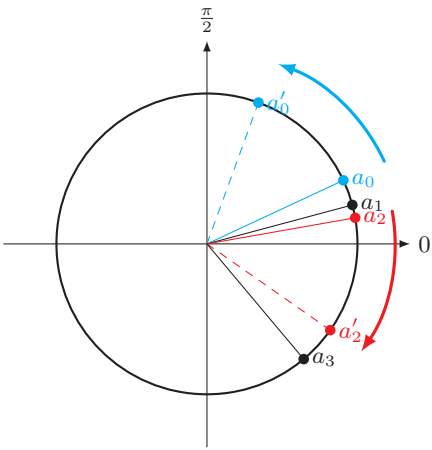
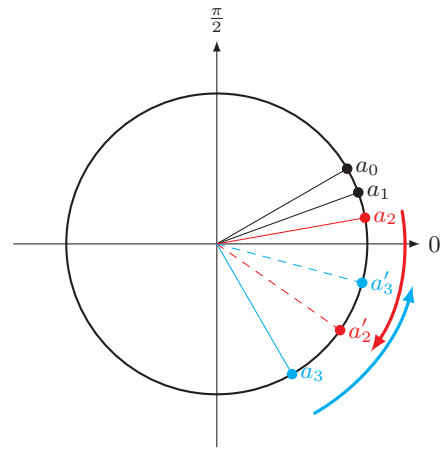

\begin{figure}
    \centering

        \begin{subfigure}{0.3\textwidth}
    \centering
\begin{tikzpicture}[scale=0.5]
\def \tta{ 90.000000000000 } 
\def \k{    0.30000000000000 } 
\def \l{     7.00000000000000 } 
\def \d{     5.00000000000000 } 
\def \h{     7.0000000000000 } 

\coordinate (I) at (0,0); 
\coordinate (Cnot) at (0,{-\h}); 
\coordinate (iSwap) at ({-\l*sin(\tta))},
                    {-\h+\l*cos(\tta)}); 
\coordinate (Swap) at ({-\l*sin(\tta)-\d*sin(\k*\tta)},{-\h+\l*cos(\tta)+\d*cos(\k*\tta)}); 
\coordinate (Swapm) at ({-\l*sin(\tta)+\d*sin(\k*\tta)},{-\h+\l*cos(\tta)-\d*cos(\k*\tta)}); 
\coordinate (SQiSW) at ({-\l*sin(\tta))/2},
                    {(-\h+\l*cos(\tta))/2});
\coordinate (mSQiSWm) at ({(-\l*sin(\tta)-\d*sin(\k*\tta))/2},{-\h+(\l*cos(\tta)+\d*cos(\k*\tta))/2});
\coordinate (mSQiSW) at ({(-\l*sin(\tta)+\d*sin(\k*\tta))/2},{-\h+(\l*cos(\tta)-\d*cos(\k*\tta))/2});
\coordinate (SQSW) at ({(-\l*sin(\tta)+\d*sin(\k*\tta))/2},{(-\h+\l*cos(\tta)-\d*cos(\k*\tta))/2});
\coordinate (SQSWm) at ({(-\l*sin(\tta)-\d*sin(\k*\tta))/2},{(-\h+\l*cos(\tta)+\d*cos(\k*\tta))/2});
\coordinate (mSQSW) at ({(-\l*sin(\tta)+\d*sin(\k*\tta))/4},{(-\h+\l*cos(\tta)-\d*cos(\k*\tta))/4});
\coordinate (mSQSWm) at ({(-\l*sin(\tta)-\d*sin(\k*\tta))/4},{(-\h+\l*cos(\tta)+\d*cos(\k*\tta))/4});
\coordinate (mmSQiSWm) at ({(-\l*sin(\tta)-\d*sin(\k*\tta))/4},{-\h/2+(\l*cos(\tta)+\d*cos(\k*\tta))/4});
\coordinate (mmSQiSW) at ({(-\l*sin(\tta)+\d*sin(\k*\tta))/4},{-\h/2+(\l*cos(\tta)-\d*cos(\k*\tta))/4});

\filldraw[color=green, fill opacity=0.5]
(SQSWm) -- (SQiSW) -- (mmSQiSWm);
\filldraw[color=green, fill opacity=0.5]
(SQSWm) -- (SQiSW) -- (iSwap) -- (Swap);
\filldraw[color=green, fill opacity=0.5]
(iSwap) -- (SQiSW) -- (mmSQiSWm) -- (mSQiSWm);
\filldraw[color=green, fill opacity=0.5]
(SQSWm) -- (Swap) -- (mSQiSWm) --  (mmSQiSWm);
\filldraw[color=green, fill opacity=0.5]
(Swap) -- (iSwap) -- (mSQiSWm);

 \filldraw[color=red, fill opacity = 0.5]
  (mSQiSWm) -- (Cnot) -- (mSQiSW) -- (iSwap);
 \filldraw[color=red, fill opacity = 0.5]
  (I) -- (Cnot) -- (mSQiSWm);
  \filldraw[color=red, fill opacity = 0.5]
  (I) -- (iSwap) -- (mSQiSWm);
  \filldraw[color=red, fill opacity = 0.5]
  (I) -- (iSwap) -- (mSQiSW);
\filldraw[color=red, fill opacity = 0.5]
  (I) -- (Cnot) -- (mSQiSW);
 
 \filldraw[color=green, fill opacity=0.5]
(SQSW) -- (SQiSW) -- (mmSQiSW);
\filldraw[color=green, fill opacity=0.5]
(SQSW) -- (SQiSW) -- (iSwap) -- (Swapm);
\filldraw[color=green, fill opacity=0.5]
(iSwap) -- (SQiSW) -- (mmSQiSW) -- (mSQiSW);
\filldraw[color=green, fill opacity=0.5]
(SQSW) -- (Swapm) -- (mSQiSW) --  (mmSQiSW);
\filldraw[color=green, fill opacity=0.5]
(Swapm) -- (iSwap) -- (mSQiSW);
                    
\draw[-,thick] (Cnot) --  (Swapm)
                        (I) --  (Swap)
                        (I) -- (Swapm)
                        (I) -- (Cnot)
                        (Swap) --  (Swapm)
                        (I) --  (iSwap);

\draw[dashed,thick] (iSwap) --  (Cnot)
                        (Swap)  -- (Cnot);

\draw[dotted, thick] (iSwap) --  (mSQiSW)
                        (iSwap)  -- (mSQiSWm)
                        (I) -- (mSQiSW)
                        (I) -- (mSQiSWm)
                        (SQiSW) -- (SQSW)
                        (SQiSW) -- (SQSWm)
                        (SQiSW) -- (mmSQiSW)
                        (SQiSW) -- (mmSQiSWm)
                        (SQSW) -- (mmSQiSW)
                        (SQSWm) -- (mmSQiSWm);

\fill[black]  (I) circle [radius=2pt]; 
\fill[black]    (iSwap) circle [radius=2pt]; 
\fill[black]  (Cnot) circle [radius=2pt]; 
\fill[black] (Swap) circle [radius=2pt];
\fill[black] (Swapm) circle [radius=2pt];
\fill[black] (mSQiSW) circle [radius=2pt];
\fill[black] (mSQiSWm) circle [radius=2pt];
\fill[black] (mmSQiSW) circle [radius=2pt];
\fill[black] (mmSQiSWm) circle [radius=2pt];
\fill[red] (SQiSW) circle [radius=2pt];

\end{tikzpicture}
\caption{$x>\frac\pi8$.}
    \end{subfigure}~
    \begin{subfigure}{0.3\textwidth}
    \centering
\begin{tikzpicture}[scale=0.5]
\def \tta{ 90.000000000000 } 
\def \k{    0.30000000000000 } 
\def \l{     7.00000000000000 } 
\def \d{     5.00000000000000 } 
\def \h{     7.0000000000000 } 

\coordinate (I) at (0,0); 
\coordinate (Cnot) at (0,{-\h}); 
\coordinate (iSwap) at ({-\l*sin(\tta))},
                    {-\h+\l*cos(\tta)}); 
\coordinate (Swap) at ({-\l*sin(\tta)-\d*sin(\k*\tta)},{-\h+\l*cos(\tta)+\d*cos(\k*\tta)}); 
\coordinate (Swapm) at ({-\l*sin(\tta)+\d*sin(\k*\tta)},{-\h+\l*cos(\tta)-\d*cos(\k*\tta)}); 
\coordinate (SQiSW) at ({-\l*sin(\tta))/2},
                    {(-\h+\l*cos(\tta))/2});
\coordinate (mSQiSWm) at ({(-\l*sin(\tta)-\d*sin(\k*\tta))/2},{-\h+(\l*cos(\tta)+\d*cos(\k*\tta))/2});
\coordinate (mSQiSW) at ({(-\l*sin(\tta)+\d*sin(\k*\tta))/2},{-\h+(\l*cos(\tta)-\d*cos(\k*\tta))/2});
\coordinate (SQSW) at ({(-\l*sin(\tta)+\d*sin(\k*\tta))/2},{(-\h+\l*cos(\tta)-\d*cos(\k*\tta))/2});
\coordinate (SQSWm) at ({(-\l*sin(\tta)-\d*sin(\k*\tta))/2},{(-\h+\l*cos(\tta)+\d*cos(\k*\tta))/2});
\coordinate (mSQSW) at ({(-\l*sin(\tta)+\d*sin(\k*\tta))/4},{(-\h+\l*cos(\tta)-\d*cos(\k*\tta))/4});
\coordinate (mSQSWm) at ({(-\l*sin(\tta)-\d*sin(\k*\tta))/4},{(-\h+\l*cos(\tta)+\d*cos(\k*\tta))/4});

  \filldraw[color=cyan, fill opacity = 0.5]
  (I) -- (iSwap) -- (Swap);
    \filldraw[color=cyan, fill opacity = 0.5]
  (I) -- (iSwap) -- (mSQiSWm);
    \filldraw[color=cyan, fill opacity = 0.5]
  (I) -- (mSQiSWm) -- (Swap);
    \filldraw[color=cyan, fill opacity = 0.8]
  (mSQiSWm) -- (iSwap) -- (Swap);

 \filldraw[color=red, fill opacity = 0.5]
  (mSQiSWm) -- (Cnot) -- (mSQiSW) -- (iSwap);
 \filldraw[color=red!60, fill opacity = 0.5]
  (I) -- (Cnot) -- (mSQiSWm);
  \filldraw[color=red, fill opacity = 0.5]
  (I) -- (iSwap) -- (mSQiSWm);
  \filldraw[color=red, fill opacity = 0.5]
  (I) -- (iSwap) -- (mSQiSW);
\filldraw[color=red, fill opacity = 0.5]
  (I) -- (Cnot) -- (mSQiSW);
                    
\draw[-,thick] (Cnot) --  (Swapm)
                        (I) --  (Swap)
                        (I) -- (Swapm)
                        (I) -- (Cnot)
                        (Swap) --  (Swapm)
                        (I) --  (iSwap);

\draw[dashed,thick] (iSwap) --  (Cnot)
                        (Swap)  -- (Cnot);

\draw[dotted, thick] 
                        (iSwap) --  (mSQiSW)
                        (iSwap)  -- (mSQiSWm)
                        (I) -- (mSQiSW)
                        (I) -- (mSQiSWm);

\fill[black]  (I) circle [radius=2pt]; 
\fill[black]    (iSwap) circle [radius=2pt]; 
\fill[black]  (Cnot) circle [radius=2pt]; 
\fill[black] (Swap) circle [radius=2pt];
\fill[black] (Swapm) circle [radius=2pt];
\fill[red] (SQiSW) circle [radius=2pt];

\end{tikzpicture}
\caption{$z<0$.}
    \end{subfigure}~
    \begin{subfigure}{0.3\textwidth}
    \centering
\begin{tikzpicture}[scale=0.5]
\def \tta{ 90.000000000000 } 
\def \k{    0.30000000000000 } 
\def \l{     7.00000000000000 } 
\def \d{     5.00000000000000 } 
\def \h{     7.0000000000000 } 

\coordinate (I) at (0,0); 
\coordinate (Cnot) at (0,{-\h}); 
\coordinate (iSwap) at ({-\l*sin(\tta))},
                    {-\h+\l*cos(\tta)}); 
\coordinate (Swap) at ({-\l*sin(\tta)-\d*sin(\k*\tta)},{-\h+\l*cos(\tta)+\d*cos(\k*\tta)}); 
\coordinate (Swapm) at ({-\l*sin(\tta)+\d*sin(\k*\tta)},{-\h+\l*cos(\tta)-\d*cos(\k*\tta)}); 
\coordinate (SQiSW) at ({-\l*sin(\tta))/2},
                    {(-\h+\l*cos(\tta))/2});
\coordinate (mSQiSWm) at ({(-\l*sin(\tta)-\d*sin(\k*\tta))/2},{-\h+(\l*cos(\tta)+\d*cos(\k*\tta))/2});
\coordinate (mSQiSW) at ({(-\l*sin(\tta)+\d*sin(\k*\tta))/2},{-\h+(\l*cos(\tta)-\d*cos(\k*\tta))/2});
\coordinate (SQSW) at ({(-\l*sin(\tta)+\d*sin(\k*\tta))/2},{(-\h+\l*cos(\tta)-\d*cos(\k*\tta))/2});
\coordinate (SQSWm) at ({(-\l*sin(\tta)-\d*sin(\k*\tta))/2},{(-\h+\l*cos(\tta)+\d*cos(\k*\tta))/2});
\coordinate (mSQSW) at ({(-\l*sin(\tta)+\d*sin(\k*\tta))/4},{(-\h+\l*cos(\tta)-\d*cos(\k*\tta))/4});
\coordinate (mSQSWm) at ({(-\l*sin(\tta)-\d*sin(\k*\tta))/4},{(-\h+\l*cos(\tta)+\d*cos(\k*\tta))/4});
\coordinate (mmSQiSWm) at ({(-\l*sin(\tta)-\d*sin(\k*\tta))/4},{-\h/2+(\l*cos(\tta)+\d*cos(\k*\tta))/4});
\coordinate (mmSQiSW) at ({(-\l*sin(\tta)+\d*sin(\k*\tta))/4},{-\h/2+(\l*cos(\tta)-\d*cos(\k*\tta))/4});

\filldraw[color=violet, fill opacity=0.5]
(I) -- (SQiSW) -- (mmSQiSWm);
\filldraw[color=violet, fill opacity=0.5]
(I) -- (mSQSWm) -- (mmSQiSWm);
\filldraw[color=violet, fill opacity=0.5]
(mSQSWm) -- (SQiSW) -- (mmSQiSWm);
\filldraw[color=violet, fill opacity=0.5]
(SQSWm) -- (SQiSW) -- (mmSQiSWm);
\filldraw[color=violet, fill opacity=0.5]
(SQSWm) -- (iSwap) -- (mSQiSWm);
\filldraw[color=violet, fill opacity=0.5]
(SQSWm) -- (mmSQiSWm) -- (mSQiSWm);
\filldraw[color=violet, fill opacity=0.5]
(SQSWm) -- (mSQiSWm) -- (mmSQiSWm);
\filldraw[color=violet, fill opacity=0.5]
(iSwap) -- (SQiSW) -- (mmSQiSWm) -- (mSQiSWm);

 \filldraw[color=red, fill opacity = 0.5]
  (mSQiSWm) -- (Cnot) -- (mSQiSW) -- (iSwap);
 \filldraw[color=red, fill opacity = 0.5]
  (I) -- (Cnot) -- (mSQiSWm);
  \filldraw[color=red, fill opacity = 0.5]
  (I) -- (iSwap) -- (mSQiSWm);
  \filldraw[color=red, fill opacity = 0.5]
  (I) -- (iSwap) -- (mSQiSW);
\filldraw[color=red, fill opacity = 0.5]
  (I) -- (Cnot) -- (mSQiSW);
 
 \filldraw[color=violet, fill opacity=0.5]
(I) -- (mSQSWm) -- (SQiSW) -- (mSQSW);
\filldraw[color=violet, fill opacity=0.5]
(SQSW) -- (SQSWm) -- (iSwap);
 \filldraw[color=violet, fill opacity=0.5]
(I) -- (SQiSW) -- (mmSQiSW);
\filldraw[color=violet, fill opacity=0.5]
(I) -- (mSQSW) -- (mmSQiSW);
\filldraw[color=violet, fill opacity=0.5]
(mSQSW) -- (SQiSW) -- (mmSQiSW);
\filldraw[color=violet, fill opacity=0.5]
(SQSW) -- (SQiSW) -- (mmSQiSW);
\filldraw[color=violet, fill opacity=0.5]
(SQSW) -- (iSwap) -- (mSQiSW);
\filldraw[color=violet, fill opacity=0.5]
(SQSW) -- (mmSQiSW) -- (mSQiSW);
\filldraw[color=violet, fill opacity=0.5]
(SQSW) -- (mSQiSW) -- (mmSQiSW);
\filldraw[color=violet, fill opacity=0.5]
(iSwap) -- (SQiSW) -- (mmSQiSW) -- (mSQiSW);

\draw[-,thick] (Cnot) --  (Swapm)
                        (I) --  (Swap)
                        (I) -- (Swapm)
                        (I) -- (Cnot)
                        (Swap) --  (Swapm)
                        (I) --  (iSwap);

\draw[dashed,thick] (iSwap) --  (Cnot)
                        (Swap)  -- (Cnot);

\draw[dotted, thick] (iSwap) --  (mSQiSW)
                        (iSwap)  -- (mSQiSWm)
                        (I) -- (mSQiSW)
                        (I) -- (mSQiSWm)
                        (SQiSW) -- (SQSW)
                        (SQiSW) -- (SQSWm)
                        (SQiSW) -- (mmSQiSW)
                        (SQiSW) -- (mmSQiSWm)
                        (SQSW) -- (mmSQiSW)
                        (SQSWm) -- (mmSQiSWm)
                        (SQSWm) -- (iSwap)
                        (SQSW) -- (iSwap)
                        (mSQiSW) -- (SQSW)
                        (mSQiSWm) -- (SQSWm)
                        (SQiSW) -- (mSQSW)
                        (mmSQiSW) -- (mSQSW)
                        (SQiSW) -- (mSQSWm)
                        (mmSQiSWm) -- (mSQSWm);

\fill[black]  (I) circle [radius=2pt]; 
\fill[black]    (iSwap) circle [radius=2pt]; 
\fill[black]  (Cnot) circle [radius=2pt]; 
\fill[black] (Swap) circle [radius=2pt];
\fill[black] (Swapm) circle [radius=2pt];
\fill[black] (mSQiSW) circle [radius=2pt];
\fill[black] (mSQiSWm) circle [radius=2pt];
\fill[black] (mmSQiSW) circle [radius=2pt];
\fill[black] (mmSQiSWm) circle [radius=2pt];
\fill[red] (SQiSW) circle [radius=2pt];

\end{tikzpicture}
\caption{Eigenphase crossing:\linebreak $y+|z|<\frac{\pi}{8}, x\leq\frac\pi8$ or \linebreak $ y+|z|<\frac{\pi}{4}, x>\frac\pi8 $.}
    \end{subfigure}\\
    \begin{subfigure}{\textwidth}
        \begin{tikzpicture}
\node {\begin{quantikz}
             &\gate{{\color{cyan}Z}\cdot {\color{violet}R_x(\frac\pi2)}\cdot {\color{cyan}Z}} &\gate[style={fill=red!50},2]{L(x',y',z')} &\gate{{\color{cyan}Z}\cdot {\color{violet}R_x(-\frac\pi2)}\cdot {\color{green}R_z(\frac{\pi}2)}\cdot R_y(-\frac{\pi}2)\cdot {\color{cyan}Z}} &\gate[2]{\SQiSW} & \gate{{\color{cyan}Z}\cdot R_y(\frac{\pi}2)\cdot {\color{green}R_z(-\frac\pi2)}\cdot {\color{cyan}Z}} &\qw  \\
             &\gate{{\color{violet}R_x(-\frac{\pi}2)}}&\qw &\gate{{\color{violet}R_x(\frac\pi2)}\cdot {\color{green}R_z(-\frac{\pi}2)}\cdot R_y(\frac{\pi}2)}&\qw&\gate{ R_y(-\frac{\pi}2)\cdot {\color{green}R_z(\frac\pi2)}}&\qw
             \end{quantikz}};
             \end{tikzpicture}
             \caption{}
    \end{subfigure}
\caption{Visualization of the full compilation scheme. When a gate is outside of the region $W'$, there are eight cases corresponding to different circuit compilations, indicated by three inequalities. (a) $x\stackrel{?}{>}\pi/8$, the $>$ case indicated in green. This corresponds to the two cases considered in the proof. (b) $z\stackrel{?}{<}0$, the $<$ case indicated in blue. The proof only deals with the case $z>0$, where the case $z<0$ follows similarly with appropriate inversions. (c) Whether there is \emph{eigenphase crossing}, i.e. whether the order of $a_0,\ldots, a_3$ is preserved after the phase modification. The purple region shows when it isn't, and a corresponding correction needs to be made in order to transform the gate to its canonical form. (d) The corresponding modifications are the green $R_z$-conjugations around the $\SQiSW$  gate, the violet $R_x$-conjugations around the $L(x',y',z')$ in red, and the cyan $Z$ gates on the first qubit.}
\label{fig:s3}
\end{figure}
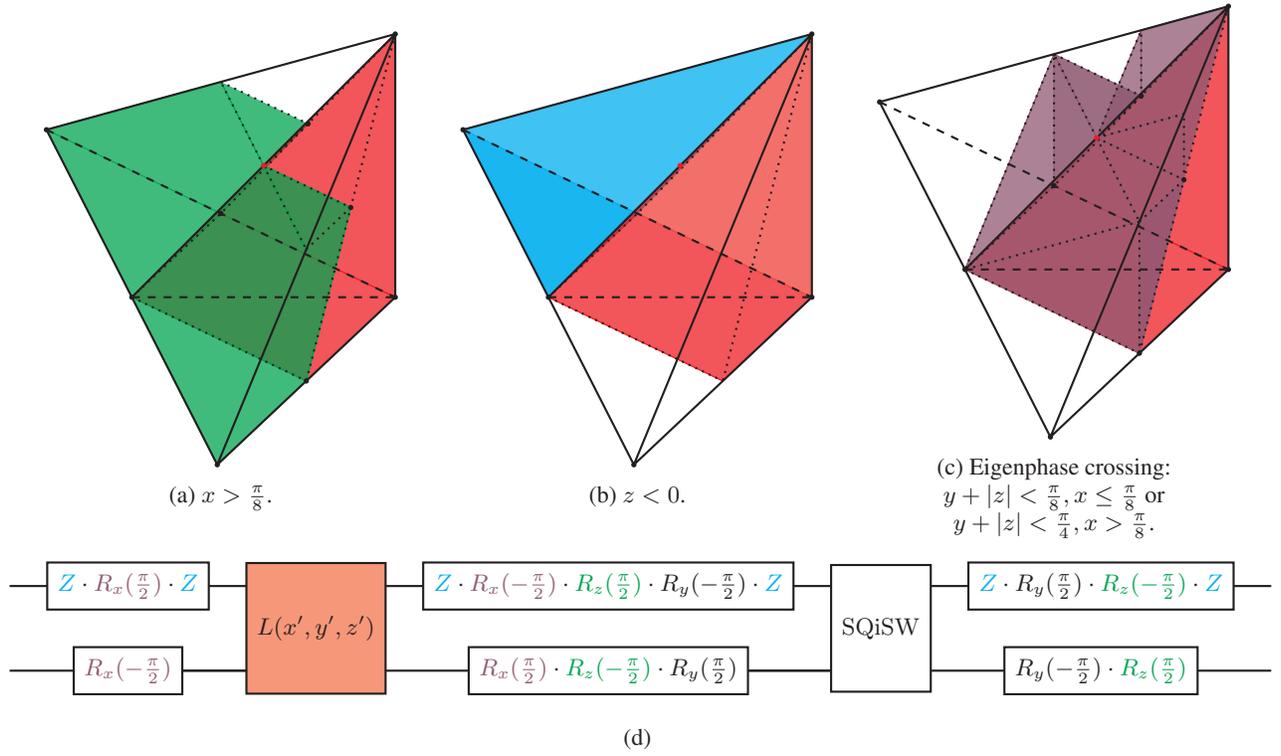

\end{proof}
\subsubsection{Decomposition algorithms for two-qubit gates into $\SQiSW$  gates}

The full decomposition algorithm for an arbitrary two-qubit gate into sequences of single qubit rotations and the $\SQiSW$  gate is summarized in~\cref{alg:decomp} and visualized in~Fig.~\ref{fig:s3}. We also list compilation schemes of some common two-qubit gates or gate families into $\SQiSW$  below and summarize the results in~Fig.~\ref{fig:special}. 

In this section and throughout the rest of the paper, we use $||$ to denote the concatenation of two quantum gates. For example, $A||B$ represents a composite quantum gates where $A$ is applied before $B$, resulting an overall operation of $B\cdot A$.

\begin{algorithm}[H]
  \caption{Decomposing an arbitrary two-qubit gate into a sequence of single qubit rotations and the $\SQiSW$  gate.}
  \label{alg:decomp}
   \begin{algorithmic}[1]
   \Procedure{Decomp}{U} \Comment{Decompose $U$ into single qubit gates and the $\SQiSW$  gate}
   \State $g,(x,y,z),A_1,A_2,B_1,B_2\leftarrow$\textsc{KAKDecomp}($U$)\Comment{$U = g\cdot (A_1\otimes A_2)L(x,y,z)(B_1\otimes B_2)$}
   \If{$|z|\leq x-y$} \Comment{2 $\SQiSW$  gates needed}
   \State $C_1,C_2\leftarrow $\textsc{InterleavingSIngleQubitRotations}($x,y,z$)
   \State $V\leftarrow \SQiSW (C_1\otimes C_2)\SQiSW$
   \State $g',(x,y,z),D_1,D_2,E_1,E_2\leftarrow$\textsc{KAKDecomp($V$)}\Comment{$L(x,y,z)=(1/g')(D_1^\dag\otimes D_2^\dag)\SQiSW(C_1\otimes C_2)\SQiSW(E_1^\dag\otimes E_2^\dag)$} 
   \State \Return $E_1^\dag B_1\otimes E_2^\dag B_2||\SQiSW||C_1\otimes C_2||\SQiSW||A_1D_1^\dag \otimes A_2D_2^\dag $ \Comment{Global phases $g,g'$ omitted} 
   \Else
   \State $(x',y',z'),F_1,F_2,G_1,G_2,H_1,H_2\leftarrow$\textsc{Canonicalize}($x,y,z$) \State\Comment{$L(x,y,z)=(F_1\otimes F_2)L(x',y',z')(G_1\otimes G_2)\SQiSW(H_1\otimes H_2), x',y',z'\in W'$}
   
   \State $C_1,C_2\leftarrow $\textsc{InterleavingSIngleQubitRotations}($x',y',z'$)
   \State $V\leftarrow \SQiSW (C_1\otimes C_2)\SQiSW$
   \State $g',(x',y',z'),D_1,D_2,E_1,E_2\leftarrow$\textsc{KAKDecomp($V$)}\Comment{$L(x',y',z')=(1/g')(D_1^\dag\otimes D_2^\dag)\SQiSW(C_1\otimes C_2)\SQiSW(E_1^\dag\otimes E_2^\dag)$}
   \State\Return $H_1B_1\otimes H_2B_2||\SQiSW||E_1^\dag G_1\otimes E_2^\dag G_2||\SQiSW||C_1\otimes C_2||\SQiSW||A_1F_1D_1^\dag\otimes A_2F_2D_2^\dag$
   \EndIf
   \EndProcedure
   \Procedure{InterleavingSingleQubitGates}{x,y,z} \Comment{Output the single qubit rotations given the interaction coefficients $(x,y,z)\in W'$ when sandwiched by two $\SQiSW$  gates}
   \State $C\leftarrow \sin(x+y-z)\sin(x-y+z)\sin(-x-y-z)\sin(-x+y+z)$
   \State $\alpha\leftarrow \arccos(\cos 2x-\cos 2y+\cos 2z+2\sqrt{C})$
   \State $\beta\leftarrow \arccos(\cos 2x-\cos 2y+\cos 2z-2\sqrt{C})$
   \State $\gamma\leftarrow \arccos(\mathrm{sgn} z\cdot \sqrt{\frac{4\cos^2x\cos^2z\sin^2y}{4\cos^2x\cos^2z\sin^2y+\cos2x\cos2y\cos2z}})$
   \State \Return $R_z(\gamma)R_x(\alpha)R_z(\gamma), R_x(\beta)$
   \EndProcedure
   \Procedure{Canonicalize}{x,y,z} \Comment{Decompose an arbitrary gate into one $\SQiSW$  and one $L(x',y',z')$ where $(x',y',z')\in W'$ and output the coefficients $(x',y',z')$ and the interleaving single qubit rotations}
   \State$A_0\leftarrow I, A_1\leftarrow I, B_1\leftarrow R_y(-\frac\pi2), B_2\leftarrow R_y(\frac\pi2), C_1\leftarrow R_y(\frac\pi2), C_2\leftarrow R_y(-\frac\pi2), s\leftarrow \mathrm{sgn}(z), x'\leftarrow x, y'\leftarrow y, z'\leftarrow |z|$
   \If{$x>\frac\pi8$} 
   \State $y'\leftarrow y'-\frac\pi8,z'\leftarrow z'-\frac\pi8,B_1\leftarrow R_z(\frac\pi2)B_1, B_2\leftarrow R_z(-\frac\pi2)B_2, C_1\leftarrow C_1R_z(-\frac\pi2), C_2\leftarrow C_2R_z(\frac\pi2)$
   \Else
   \State $x'\leftarrow x'+\frac\pi8,z'\leftarrow z'-\frac\pi8$
   \EndIf
   \If{$|y'|<|z'|$}\Comment{Eigenphase crossing}
   \State$y',z'\leftarrow -z',-y',A_1\leftarrow R_x(\frac\pi2), A_2\leftarrow R_x(-\frac\pi2), B_1\leftarrow R_x(-\frac{\pi}2)B_1, B_2\leftarrow R_x(\frac\pi2)B_2$
   \EndIf
   \If {$s<0$}
   \State$z'\leftarrow -z',A_1\leftarrow ZA_1Z, B_1\leftarrow ZB_1Z, C_1\leftarrow ZC_1Z$
   \EndIf
   \State \Return $(x',y',z'), A_1, A_2, B_1, B_2, C_1, C_2$
   \EndProcedure
   \end{algorithmic}
\end{algorithm}

Before proceeding, note that $[R_z(\alpha)\otimes R_z(\alpha), \SQiSW ]=0$ for all $\alpha$. This introduces gauge freedom in compilation of the circuit and enables us to choose the single qubit gates with the simplest form in our compilation.

\begin{description}
\item[Special orthogonal gates $(x,y,0)$] All gates locally equivalent to special orthogonal gates in $SO(4)$, i.e. gates that lie in the I--$\CNOT$--$\iSWAP$~plane can be generated with two $\SQiSW$  gates. Moreover, the expressions of $\alpha, \beta, \gamma$ can be simplified as
\begin{align*}
    \alpha&=0,\\
    \beta&=2\arccos \sqrt{\cos 2x+2\sin^2y},\\
    \gamma &= \arccos\sqrt{\frac{4\cos^2x\sin^2y}{\cos2x+2\sin^2 y}}.\\
\end{align*}
Therefore, one can check by applying the gauge freedom that
$$L(x,y,0)\sim \SQiSW \cdot (I\otimes B)\cdot \SQiSW,$$
where
$$B=R_z(\gamma)R_x(\beta)R_z(\gamma)=\begin{bmatrix}2\cos x\sin y-i\sqrt{\cos 2x\cos 2y}&i\sqrt{\cos 2y-\cos 2x}\\ i\sqrt{\cos 2y-\cos 2x}& 2\cos x\sin y+i\sqrt{\cos 2x\cos 2y}\end{bmatrix}.$$

In the case of special orthogonal gates, the single qubit corrections can be solved analytically. Let
\begin{align*}
    \xi &\equiv -\arcsin\left(\sin y\cdot \sqrt{\frac{2\cos 2x}{\cos2x+\cos2y}}\right),\\
    \phi &\equiv \arccos\left(-\cos y\cdot \sqrt{\frac{2\cos 2x}{\cos2x+\cos2y}}\right),\\
    \psi &\equiv-\arccos\left(\cot(x)\tan(y)\right). 
\end{align*}

Then
\begin{align*}
    L(x,y,0)=&(R_z(\xi)\otimes R_z(\phi)R_x(\psi))\cdot \SQiSW \cdot(I\otimes B)\cdot \SQiSW \cdot(R_z(\xi)\otimes R_x(\psi)R_z(\phi-\pi)).
\end{align*}
Specific examples of special orthogonal gates include: 
\begin{itemize}
    \item The $\CPhase$  ~family $(x,0,0)$, where $B=ZR_y(2\arcsin(2\sqrt{2}\sin x))$;
    \item The super-controlled gate family\cite{ye2004super} $(\frac{\pi}4, y, 0)$, where $B=R_x(2\arccos(2\sqrt{2}\sin y))$;
    \item The $\iSWAP$ ~family $(x,x,0)$, where $B=R_z( 4x-\frac\pi)$.
\end{itemize}

\item [Improper orthogonal gates $(\frac{\pi}4,y,z)$] Half of the improper orthogonal gate family can be generated by 2 $\SQiSW$  gates when $y+|z|\leq \frac{\pi}4$. In this case, one has
\begin{align*}
    \alpha&=\arccos\left(\cos 2z-\cos 2y+\sqrt{\frac{\cos 4z+\cos 4y}{2}}\right),\\
    \beta&=\arccos \left(\cos 2z-\cos 2y-\sqrt{\frac{\cos 4z+\cos 4y}{2}}\right),\\
    \gamma &= 0.\\
\end{align*}

Therefore,
$$L(\frac{\pi}4,y,z)\sim \SQiSW \cdot (R_x(\alpha)\otimes R_x(\beta))\cdot \SQiSW .$$
In this case we can also explicitly solve for the single qubit corrections. Let 
$$\phi = -\arccos\sqrt{\frac{1+\tan (y-z)}2},\psi = \arccos\sqrt{\frac{1+\tan(y+z)}2},$$
then
$$L(\frac{\pi}4,y,z)=(R_x(\phi+\psi)\otimes R_z(\frac\pi2)R_x(\phi-\psi))\cdot \SQiSW \cdot (R_x(\alpha)\otimes R_x(\beta))\cdot \SQiSW \cdot (R_x(\phi+\psi)\otimes R_x(\phi-\psi)R_z(-\frac\pi2)).$$

Decomposition for the other half of the improper orthogonal gates can be obtained by first decomposing it into one $\SQiSW$ {} and a gate in $W(S_2)$, then decomposing the gate in $W(S_2)$ by observing that it is an improper orthogonal gate.

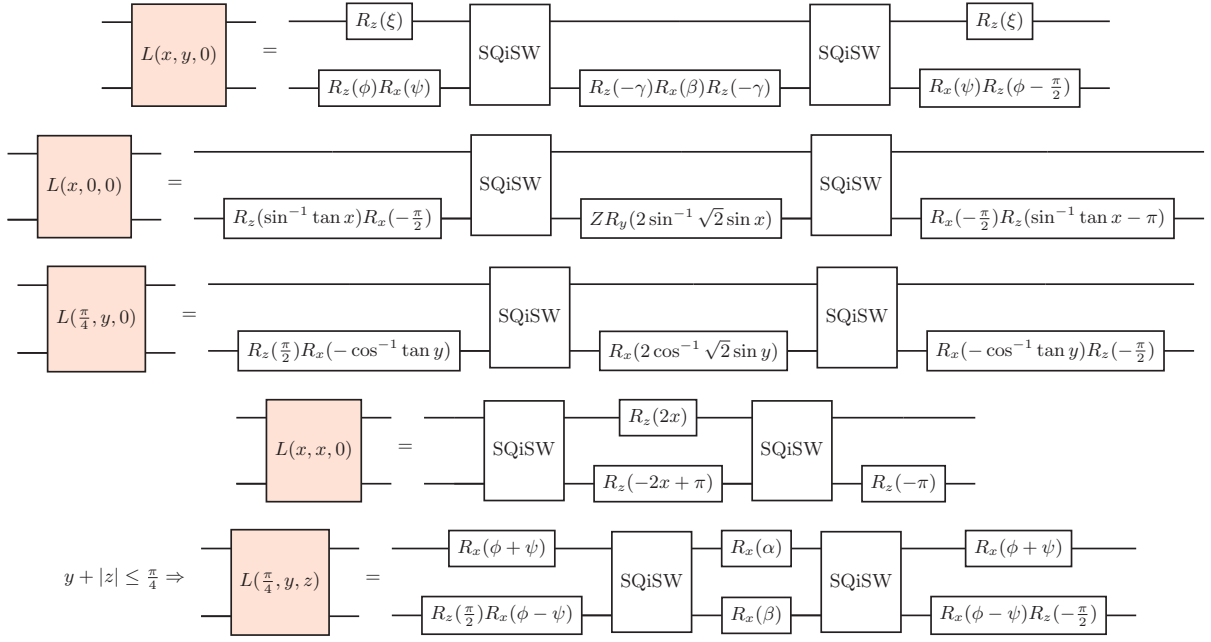
\begin{figure}
    \centering
    \begin{tikzpicture}
\node[scale=0.8] {
    \begin{quantikz}
             & \gate[style={fill=pink!50}, 2]{L(x,y,0)} &  \qw \\
             & \qw & \qw
             \end{quantikz}
             $=$\begin{quantikz}
             &\gate{R_z(\xi)}&\gate[2]{\SQiSW}  & \qw&\gate[2]{\SQiSW} &\gate{R_z(\xi)}&\qw  \\
             &\gate{R_z(\phi)R_x(\psi)}&\qw &\gate[1]{R_z(-\gamma) R_x(\beta) R_z(-\gamma)}& \qw& \gate{R_x(\psi)R_z(\phi-\frac\pi2)}&\qw
             \end{quantikz}};
             \end{tikzpicture}
    \begin{tikzpicture}
\node[scale=0.8]{
    \begin{quantikz}
             & \gate[style={fill=pink!50}, 2]{L(x,0,0)} &  \qw \\
             & \qw & \qw
             \end{quantikz}
             $=$\begin{quantikz}
             &\qw&\gate[2]{\SQiSW}  & \qw&\gate[2]{\SQiSW} &\qw&\qw  \\
             &\gate{R_z(\sin^{-1}\tan x)R_x(-\frac\pi2)}&\qw &\gate[1]{ZR_y(2\sin^{-1}\sqrt{2}\sin x)} & \qw& \gate{R_x(-\frac\pi2)R_z(\sin^{-1}\tan x-\pi)}&\qw
             \end{quantikz}};
             \end{tikzpicture}
                 \begin{tikzpicture}
\node[scale=0.8] {
    \begin{quantikz}
             &\gate[style={fill=pink!50}, 2]{L(\frac{\pi}4,y,0)}&\qw \\
             & \qw & \qw
             \end{quantikz}
             $=$\begin{quantikz}
             & \qw&\gate[2]{\SQiSW}  & \qw&\gate[2]{\SQiSW} &   \qw&\qw  \\
             &\gate{R_z(\frac\pi2)R_x(-\cos^{-1}\tan y)}&\qw &\gate[1]{R_x(2\cos^{-1}\sqrt{2}\sin y)} & \qw&\gate{R_x(-\cos^{-1}\tan y)R_z(-\frac\pi2)}&\qw
             \end{quantikz}};
             \end{tikzpicture}
                 \begin{tikzpicture}
\node[scale=0.8] {
    \begin{quantikz}
             & \gate[style={fill=pink!50}, 2]{L(x,x,0)} &  \qw \\
             & \qw & \qw
             \end{quantikz}
             $=$\begin{quantikz}
             &\qw&\gate[2]{\SQiSW}  & \gate{R_z(2x)}&\gate[2]{\SQiSW} &\qw&\qw  \\
             &\qw&\qw &\gate{R_z(-2x+\pi)} & \qw&\gate{R_z(-\pi)}&\qw
             \end{quantikz}};
             \end{tikzpicture}
    \begin{tikzpicture}
\node[scale=0.8] {$y+|z|\leq\frac{\pi}4\Rightarrow$
    \begin{quantikz}
             & \gate[style={fill=pink!50}, 2]{L(\frac{\pi}4,y,z)} &  \qw \\
             & \qw & \qw
             \end{quantikz}
             $=$\begin{quantikz}
             &\gate{R_x(\phi+\psi)}&\gate[2]{\SQiSW}  & \gate{R_x(\alpha)}&\gate[2]{\SQiSW} &\gate{R_x(\phi+\psi)}&\qw  \\
             &\gate{R_z(\frac\pi2)R_x(\phi-\psi)}&\qw &\gate[1]{R_x(\beta)} & \qw&\gate{R_x(\phi-\psi)R_z(-\frac\pi2)}&\qw
             \end{quantikz}};
             \end{tikzpicture}
    \caption{Summary of compilations of common two-qubit gates into $\SQiSW$  gates}
    \label{fig:special}
\end{figure}

\end{description}
\subsubsection{Comparison with other two-qubit gates}
It can be observed from the visualization in~Fig.~\ref{fig:s2_region} that $W'$ takes up $1/2$ of the entire Weyl chamber, similar to the set of all perfect entanglers~\cite{zhang2003geometric}. However, the measure in the Weyl chamber does not reflect the Haar measure of the unitary group $SU(4)$~\cite{watts2013metric}. Indeed, the probability that a Haar random element in $SU(4)$ can be decomposed with two $\SQiSW$  gates can be calculated as $$\int_{(x,y,z)\in W'}d\mu_W=\frac78-\frac4{15\pi}\approx 79\%.$$ It is well known that the B gate with interaction coefficients $(\frac{\pi}{4},\frac{\pi}{8},0)$ spans the whole Weyl chamber with only two uses~\cite{zhang2004minimum}. It is also well known that many two-qubit gates, including the CNOT gate, the $\iSWAP{}$ gate and the other gates in the super controlling gate family, generates the whole Weyl chamber with three uses\cite{zhang2004optimal}. We find that $\SQiSW$ lies in-between: although it cannot generate the whole Weyl chamber, it generates a unitary subset of a \emph{nonzero measure}. Although this holds for general two-qubit gates\cite{Peterson2020fixeddepthtwoqubit}, we show that two uses of standard gates such as $\CNOT$, $\iSWAP$ or the $\SWAP$  family actually generate a subset of the Weyl chamber with zero measure, even though three uses of either gate span the whole Weyl chamber.

\begin{prp}
For any two gates $U_1,U_2\in SU(4)$ in the $\CPhase$  ~gate family $\eta(U_1)=(x_1,0,0), \eta(U_2)=(x_2,0,0)$ and any two single qubit gates $A_1,A_2\in SU(2)$, define the gate
$$V\equiv U_1(A_1\otimes A_2)U_2.$$
Then the last element in $\eta(V)$ is always zero. Equivalently, $V$ must lie in the I-$\CNOT$-$\iSWAP$ ~plane in the Weyl chamber.
\end{prp}

\begin{proof}
Let $\eta(V)=(x',y',z')$. It can be checked from the characteristic polynomial $F_V(t)$ that the corresponding polynomial coefficient
$$B=\sin 2x'\sin2y'\sin2z'=0,$$
regardless of how $U_1,U_2,A_1,A_2$ are chosen. This indicates that $z'=0$ given $\frac\pi4\geq x\geq y\geq |z|$.
\end{proof}

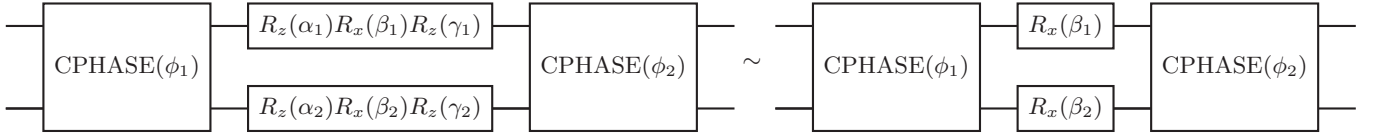
\begin{figure}
    \centering
    \begin{tikzpicture}
\node {
    \begin{quantikz}
             & \gate[2]{\CPhase(\phi_1)} & \gate[1]{R_z(\alpha_1)R_x(\beta_1)R_z(\gamma_1)}   &\gate[2]{\CPhase(\phi_2)}  &\qw  \\
              & \qw & \gate[1]{R_z(\alpha_2)R_x(\beta_2)R_z(\gamma_2)} & \qw&\qw
             \end{quantikz}
             $\sim$\begin{quantikz}
             & \gate[2]{\CPhase(\phi_1)} & \gate[1]{R_x(\beta_1)}   &\gate[2]{\CPhase(\phi_2)}  &\qw  \\
              & \qw & \gate[1]{R_x(\beta_2)} & \qw&\qw
             \end{quantikz}};
             \end{tikzpicture}
    \caption{Decomposing each single qubit gate as a $Z-X-Z$ sequence of rotations and commuting the Z-rotations with the $\CPhase$ gates, it can be seen that two $\CPhase$  gates only generate a two dimensional submanifold of the Weyl chamber.}
    \label{fig:cphase}
\end{figure}

Two $\CPhase$ family gates only generates a two dimensional submanifold  because they are $U(1)$-covariant, or ``leaky''\cite{koponen2006discrete,Peterson2020fixeddepthtwoqubit}; we have
$$[R_Z(\theta_1)\otimes R_Z(\theta_2)] \mathrm{diag}(1,1,e^{-i\phi},e^{i\phi}) = \mathrm{diag}(1,1,e^{-i\phi},e^{i\phi})[R_Z(\theta_1)\otimes R_Z(\theta_2) ]$$
for all $\theta_1,\theta_2$ and $\phi$. By commuting the Z-rotations, the interleaving single qubit gates, which can each be decomposed into a $Z-X-Z$ sequence of rotations, can only generate a two-dimensional manifold in the Weyl chamber, as illustrated in~Fig.~\ref{fig:cphase}.

By making use of the properties of mirror gates, we can extend this result to gates on the $\iSWAP-\SWAP$ line as well. The results are visualized in~Fig.~\ref{fig:c2}.

\begin{cor}
For any two gates $U_1,U_2\in SU(4)$ such that$$\eta(U_1)=(\frac{\pi}4,\frac{\pi}4,x_1), \eta(U_2)=(\frac{\pi}4,\frac{\pi}4,x_2)$$ and any two single qubit gates $A_1,A_2\in SU(2)$, define the gate
$$V\equiv U_1(A_1\otimes A_2)U_2.$$
Then the last element in $\eta(V)$ is always zero.
\end{cor}

\begin{cor}
For any two gates $U_1,U_2\in SU(4)$ such that $$\eta(U_1)=(\frac{\pi}4,\frac{\pi}4,x_1), \eta(U_2)=(x_2,0,0)$$ and any two single qubit gates $A_1,A_2\in SU(2)$, define the gate
$$V\equiv U_1(A_1\otimes A_2)U_2.$$
Then the first element in $\eta(V)$ is always $\frac\pi4$, i.e.\ the gate $V$ always lies inside the $\CNOT-\iSWAP-\SWAP$ plane of the Weyl chamber.
\end{cor}
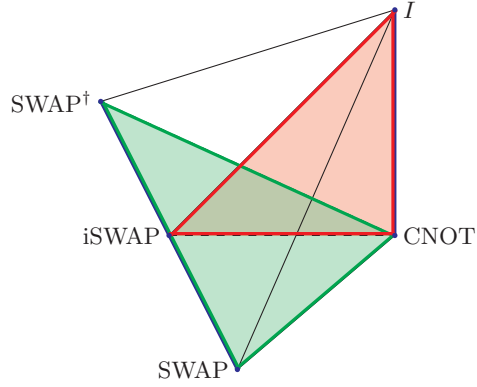
\begin{figure}
    \centering
\begin{tikzpicture} [scale=.5]
\def \tta{ 90.000000000000 } 
\def \k{    0.30000000000000 } 
\def \l{     6.00000000000000 } 
\def \d{     4.00000000000000 } 
\def \h{     6.0000000000000 } 

\coordinate (I) at (0,0); 
\coordinate (Cnot) at (0,{-\h}); 
\coordinate (iSwap) at ({-\l*sin(\tta))},
                    {-\h+\l*cos(\tta)}); 
\coordinate (Swap) at ({-\l*sin(\tta)-\d*sin(\k*\tta)},{-\h+\l*cos(\tta)+\d*cos(\k*\tta)}); 
\coordinate (Swapm) at ({-\l*sin(\tta)+\d*sin(\k*\tta)},{-\h+\l*cos(\tta)-\d*cos(\k*\tta)}); 
\coordinate (SQiSW) at ({-\l*sin(\tta))/2},
                    {(-\h+\l*cos(\tta))/2});

 \filldraw[color=red!50, fill opacity = 0.5]
  (I) -- (Cnot) -- (iSwap);
  \filldraw[color=green!50, fill opacity = 0.5]
  (Cnot) -- (Swap) -- (Swapm);
                    
\draw[-] (Cnot) --  (Swapm)
                        (I) --  (Swap)
                        (I) -- (Swapm)
                        (I) --  (iSwap);

\draw[-,very thick, blue] (I)--(Cnot)
                    (Swap) -- (Swapm);

\draw[dashed] (iSwap) --  (Cnot)
                        (Swap)  -- (Cnot);

 \draw[-, very thick,green] let \p0=(Cnot),\p1=(Swap) in
         ([xshift=-.05cm]\x0,\y0) --  ([xshift=-.05cm]\x1,\y1);  
\draw[-, very thick,green] let \p0=(Swapm),\p1=(Swap) in
         ([xshift=.05cm]\x0,\y0) --  ([xshift=.05cm]\x1,\y1);  
 \draw[-, very thick,green] let \p0=(Cnot),\p1=(Swapm) in
 ([xshift=-.05cm]\x0,\y0) --  ([xshift=-.05cm]\x1,\y1);  

\draw[-, very thick,red] let \p0=(I),\p1=(iSwap) in
         ([xshift=.05cm]\x0,\y0) --  ([xshift=.05cm]\x1,\y1);  
\draw[-, very thick,red] let \p0=(iSwap),\p1=(Cnot) in
         ([yshift=.05cm]\x0,\y0) --  ([yshift=.05cm]\x1,\y1);  
 \draw[-, very thick,red] let \p0=(I),\p1=(Cnot) in
 ([xshift=-.05cm]\x0,\y0) --  ([xshift=-.05cm]\x1,\y1); 

\fill[blue]  (I) circle [radius=2pt]; 
\fill[blue]    (iSwap) circle [radius=2pt]; 
\fill[blue]  (Cnot) circle [radius=2pt]; 
\fill[blue] (Swap) circle [radius=2pt];
\fill[blue] (Swapm) circle [radius=2pt];

\draw (I) node [right]           {$I$}
      (Cnot) node [right]     {$\CNOT$}
      (Swap) node [left]           {$\SWAP^\dag$}
      (Swapm)  node [left]       {$\SWAP$}
      (iSwap)  node [left]            {$\iSWAP$};
\end{tikzpicture}
\caption{The area spanned by 2 $\CPhase$  ~family gates or their mirror gates. Two gates on the I--$\CNOT$ line, or the $\SWAP^\dag$--$\SWAP$ line, spans the red area, whereas one gate on each line spans the green area.}
\label{fig:c2}
\end{figure}

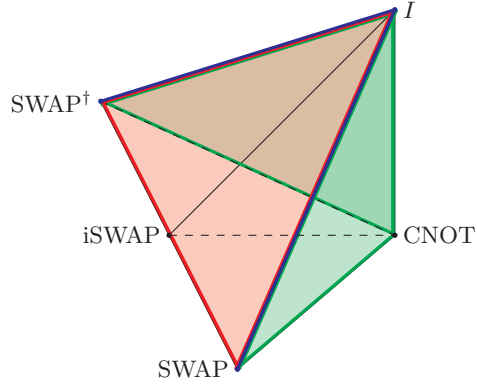
\begin{figure}
    \centering
\begin{tikzpicture} [scale=.5]
\def \tta{ 90.000000000000 } 
\def \k{    0.30000000000000 } 
\def \l{     6.00000000000000 } 
\def \d{     4.00000000000000 } 
\def \h{     6.0000000000000 } 

\coordinate (I) at (0,0); 
\coordinate (Cnot) at (0,{-\h}); 
\coordinate (iSwap) at ({-\l*sin(\tta))},
                    {-\h+\l*cos(\tta)}); 
\coordinate (Swap) at ({-\l*sin(\tta)-\d*sin(\k*\tta)},{-\h+\l*cos(\tta)+\d*cos(\k*\tta)}); 
\coordinate (Swapm) at ({-\l*sin(\tta)+\d*sin(\k*\tta)},{-\h+\l*cos(\tta)-\d*cos(\k*\tta)}); 
\coordinate (SQiSW) at ({-\l*sin(\tta))/2},
                    {(-\h+\l*cos(\tta))/2});

  \filldraw[color=green!50, fill opacity = 0.5]
  (I) -- (Swap) -- (Cnot);
\filldraw[color=green!50, fill opacity = 0.5]
  (I) -- (Cnot) -- (Swapm);
  \filldraw[color=red!50, fill opacity = 0.5]
  (I) -- (Swap) -- (Swapm);
                    
\draw[-] (Cnot) --  (Swapm)
                        (I) --  (Cnot)
                        (Swap) -- (Swapm)
                        (I) --  (iSwap);

 \draw[-, very thick,green] let \p0=(I),\p1=(Swap) in
         ([yshift=-.1cm]\x0,\y0) --  ([yshift=-.1cm]\x1,\y1);  
\draw[-, very thick,green] let \p0=(I),\p1=(Cnot) in
         ([xshift=-.05cm]\x0,\y0) --  ([xshift=-.05cm]\x1,\y1);  
 \draw[-, very thick,green] let \p0=(Cnot),\p1=(Swap) in
 ([xshift=.05cm]\x0,\y0) --  ([xshift=.05cm]\x1,\y1);  

 \draw[-, very thick,green] let \p0=(I),\p1=(Swapm) in
         ([xshift=.05cm]\x0,\y0) --  ([xshift=.05cm]\x1,\y1);  
\draw[-, very thick,green] let \p0=(I),\p1=(Cnot) in
         ([xshift=-.05cm]\x0,\y0) --  ([xshift=-.05cm]\x1,\y1);  
 \draw[-, very thick,green] let \p0=(Cnot),\p1=(Swapm) in
 ([xshift=-.05cm]\x0,\y0) --  ([xshift=-.05cm]\x1,\y1);  
\draw[dashed] (iSwap) --  (Cnot)
                        (Swap)  -- (Cnot);
\draw[-, very thick,red] let \p0=(I),\p1=(Swap) in
         ([yshift=-.05cm]\x0,\y0) --  ([yshift=-.05cm]\x1,\y1);  
\draw[-, very thick,red] let \p0=(Swapm),\p1=(Swap) in
         ([xshift=.05cm]\x0,\y0) --  ([xshift=.05cm]\x1,\y1);  
 \draw[-, very thick,red] let \p0=(I),\p1=(Swapm) in
 ([xshift=-.05cm]\x0,\y0) --  ([xshift=-.05cm]\x1,\y1);

\draw[-,very thick, blue] (I)--(Swap)
                    (I) -- (Swapm);

\fill[blue]  (I) circle [radius=2pt]; 
\fill[black]    (iSwap) circle [radius=2pt]; 
\fill[black]  (Cnot) circle [radius=2pt]; 
\fill[blue] (Swap) circle [radius=2pt];
\fill[blue] (Swapm) circle [radius=2pt];

\draw (I) node [right]           {$I$}
      (Cnot) node [right]     {$\CNOT$}
      (Swap) node [left]           {$\SWAP^\dag$}
      (Swapm)  node [left]       {$\SWAP$}
      (iSwap)  node [left]            {$\iSWAP$};
\end{tikzpicture}
\caption{The area spanned by 2 $\SWAP$  family gates. Two gates on the $\SWAP^\dag$--I--$\SWAP$ line spans either the red area (I--$\SWAP$--$\SWAP^\dag$) or the green area (I--$\SWAP$--$\CNOT$ or I--$\SWAP^\dag$--$\CNOT$).}
\label{fig:swap2}
\end{figure}

We can also prove a result for the case when we have two gates in the $\SWAP$  family. This result is visualized in~Fig.~\ref{fig:swap2}.

\begin{prp}
For any two gates $U_1,U_2\in SU(4)$ such that $$\eta(U_1)=(x_1,x_1,x_1), \eta(U_2)=(x_2,x_2,x_2)$$ and any two single qubit gates $A_1,A_2\in SU(2)$, define the gate
$$V\equiv U_1(A_1\otimes A_2)U_2.$$
Then $(x',y',z')\equiv \eta(V)$ must satisfy $y'=x'$ or $y'=|z'|$. Equivalently, $V$ must either lie in the $I-\CNOT-\SWAP$ plane, or the $I-\SWAP^\dag-\SWAP$ plane in the Weyl chamber.
\end{prp}

\begin{proof}
Let $\eta(V)=(x',y',z')$. It can be checked that the character polynomial $F_V(t)$ has a zero with multiplicity two:
$$(t\cdot \sin(x_1+x_2)+\cos(x_1+x_2))^2~\mid~F_V(t),$$
regardless of how $U_1,U_2,A_1,A_2$ are chosen. This indicates that there must be at least one equality in the inequalities $x'+y'-z'\geq x'-y'+z'\geq -x'+y'+z'\geq -x'-y'-z'$, or equivalently, one equality in $x'\geq y'\geq |z'|$.
\end{proof}

\section{Compilation: Circuits with more than two qubits}

\label{app:compilation_other}
To further demonstrate the information processing superiority of $\SQiSW$, we prove a linear separation between the number of gates needed to generate an $n$-qubit W-like state from the product state $\ket{0}^{\otimes n}$ using the $\CNOT$ gate and the $\SQiSW$ gate. Considering the corresponding family of circuits with the $\SQiSW$  gates, our proof extends to a linear separation between the gate counts in the task of compiling this family of circuits using the $\SQiSW$  gate and the $\CNOT$  gate. (Note that $\iSWAP$ is equivalent to $\CNOT$ for this purpose, since they are mirror gates up to local equivalence.) Throughout we consider single qubit rotations as free resources. 

\subsection{W state and W-like states}
The $n$-qubit W state is defined as
$$\ket{W_n} \equiv \frac{1}{\sqrt{n}}\sum_{i=1}^n \left(|0\rangle^{\otimes(i-1)}\otimes |1\rangle\otimes |0\rangle^{\otimes(n-i)}\right).$$
An interesting property of the W state is that it is robust against the disposal of qubits~\cite{dur2001multipartite}: even after tracing out any subset of $n-2$ qubits, the marginal state of the remaining two qubits is still an entangled state. In contrast, the most common multipartite generalization of maximally entangled states, the $n$-qubit GHZ state $\ket{GHZ_n} = \left(|0\rangle^{\otimes n}+|1\rangle^{\otimes n}\right)/{\sqrt{2}}$, does not satisfy this condition when $n\ge 3$. This special property of the W state can be abstracted as follows:
\begin{dfn}
An $n$-partite state $\ket{\Psi}$ is \emph{W-like} if the marginal state on any two subsystems is an entangled bipartite state. 
\end{dfn}

It is easy to see that $n$-qubit states of the form
\begin{align}
\sum_{i=1}^n \alpha_i \left(|0\rangle^{\otimes(i-1)}\otimes |1\rangle\otimes |0\rangle^{\otimes(n-i)}\right), \alpha_i > 0, \sum_{i=i}^n|\alpha_i|^2 = 1
\label{eq:w-like}
\end{align}
are special cases of  W-like states. Now, in order to generate any $n$-qubit W-like state from $\ket{0}^{\otimes n}$, it takes at least $n-1$ two-qubit gates, since the generating circuit as a graph needs to be at least connected. Surprisingly, $n-1$ $\SQiSW$  gates is also sufficient to generate a particular $W$-like state of the form in~\cref{eq:w-like}: it can be verified that
$$\SQiSW_{n-1, n}\cdots\SQiSW_{2, 3}\SQiSW_{1, 2}X_1\ket{0}^{\otimes n}$$
is a state of the above form where $\alpha_i = 2^{-i/2}$ for $i\le n-1$, and $\alpha_n = \alpha_{n-1}$.

In contrast, we show that two-qubit gates that are equivalent to a diagonal gate up to local unitaries, such as $\CNOT\sim$ CZ, are ill-suited for generating any W-like state. We have the following result.

\begin{thm}
\label{thm:cnot}
An $n$-qubit W-like state cannot be generated using single qubit gates and less than $\frac{15n-3}{14}$ $\CNOT$  gates.
\end{thm}
This is evidence that $\SQiSW$ has better information processing capabilities beyond just compiling two-qubit gates, but it is unclear how general the statement can be made.

In the rest of the section, we will provide a proof to~\cref{thm:cnot}. In~\cref{subsec:convert} we will provide a graph theoretical perspective towards quantum circuits. We will first establish the limitations on the entangling power of CNOT gate in~\cref{lem:non-robust}, which leads to nontrivial constraints on circuits that generate W-like states using CNOT gates and single-qubit gates. To be more precise, we will show in~\cref{lem:good-graph} that the graph given by a circuit generates a W-like state only if it is a ``good graph'' defined in~\cref{dfn:good}. Then in~\cref{subsec:graph} we will prove a lower bound for the number of edges in a good graph, which ultimately leads to~\cref{thm:cnot}.

\subsection{Conversion to graph problem}

\label{subsec:convert}
The key observation is that entanglement generated by a lone diagonal two-qubit gate is inherently non-robust against the disposal of qubits:
\begin{lem}
\label{lem:non-robust}
Consider a tripartite system $ABC$ where $AB$ is not entangled with $C$. If we apply a diagonal two-qubit gate between $B$ and $C$ and then immediately trace out $B$, then $A$ will still not be entangled with $C$ (i.e. the quantum state on system $AC$ is a separable state).
\end{lem}
\begin{proof}
Note that any diagonal two-qubit gate $U$ can be written in the block diagonal form
$$U = \ket{0}\bra{0}\otimes U_0 + \ket{1}\bra{1}\otimes U_1.$$
First we suppose that the initial state of the system $ABC$ is a product state $\rho_{AB}\otimes\rho_C$. Then after applying $U$ on $BC$ and tracing out $B$, the state of $AC$ becomes
\begin{align*}
&\Tr_B\left[(I_A\otimes U)(\rho_{AB}\otimes\rho_C)(I_A\otimes U^\dagger)\right]\\
={}&\Tr_B\left[(I_A\otimes \ket{0}\bra{0}\otimes U_0 + I_A\otimes \ket{1}\bra{1}\otimes U_1)(\rho_{AB}\otimes\rho_C)(I_A\otimes \ket{0}\bra{0}\otimes U_0^\dagger + I_A\otimes \ket{1}\bra{1}\otimes U_1^\dagger)\right]\\
={}&(I_A\otimes U_0)(\bra{0}_B\rho_{AB}\ket{0}_B\otimes\rho_C)(I_A\otimes U_0^\dagger) + (I_A\otimes U_1)(\bra{1}_B\rho_{AB}\ket{1}_B\otimes\rho_C)(I_A\otimes U_1^\dagger)\\
={}&(\bra{0}_B\rho_{AB}\ket{0}_B)\otimes(U_0\rho_C U_0^\dagger) + (\bra{1}_B\rho_{AB}\ket{1}_B)\otimes(U_1\rho_C U_1^\dagger),\\
\end{align*}
which is a mixture of two product states between $A$ and $C$ and thus is not entangled. If the initial state of the system $ABC$ is a mixture of multiple product states between $AB$ and $C$, then the final state will become a mixture of such mixtures, so $A$ and $C$ are still not entangled.
\end{proof}

\begin{cor} A $3$-qubit W-like state cannot be generated with only $2$ diagonal two-qubit gates.
\end{cor}
\begin{proof}
If such a generation scheme exists, then without loss of generality we can assume that the first two-qubit gate is between qubits $A$ and $B$ and the second is between qubits $B$ and $C$, but Lemma~\ref{lem:non-robust} shows that after the second two-qubit gate, $A$ is still not entangled with $C$, and single-qubit gates thereafter will not help.
\end{proof}

In fact, this argument can be generalized to give a non-trivial bound for any number of qubits.

Represent a circuit on $n$ qubits by an undirected graph on $n$ vertices with distinct edge weights. Larger edge weight means that an edge correspond to a two-qubit gate later in the circuit.
\begin{dfn}
\label{dfn:useful_edge}
An edge $C$--$D$ is considered a \emph{useful edge with respect to the vertices $A$ and $B$}, if there exists a path $C$--$D$--$D_1$--$D_2$--$\cdots$--$D_t$ such that all edges in the path have strictly increasing weights, and $D_t\in\{A, B\}$, and the same condition also holds for the other direction of the edge. 
\end{dfn}
Note that if $D\in\{A, B\}$ then the condition in one direction is automatically satisfied by $t=0$. Also, $D_1$ can be the same as $C$, but then by the strictly increasing condition there must be at least two edges between $C$ and $D$.
\begin{dfn}
\label{dfn:good}
A graph is \emph{good} if for all vertex pairs $A$ and $B$, there is a path between $A$ and $B$ that consists entirely of useful edges with respect to the vertices $A$ and $B$.
\end{dfn}
Note that this trivially implies a good graph must be connected.
\begin{lem}
Consider a circuit on $n$ qubits in which all two-qubit gates are diagonal. If that circuit can generate an $n$-qubit W-like state from $\ket{0}^{\otimes n}$, then the corresponding graph must be good.
\label{lem:good-graph}
\end{lem}
\begin{proof}
Consider any vertex pair $A$ and $B$ in the graph. Since the final state of the circuit is W-like, in the final state $A$ and $B$ must be entangled. We will show that this implies that $A$ and $B$ are connected by a path consisting entirely of useful edges (with respect to the vertices $A$ and $B$).

First, we trace out all qubits except $A$ and $B$ from the final state to get the marginal state $\rho_{AB}$. Then we remove some edges from the graph corresponding to the circuit in two sequential steps:
\begin{enumerate}
    \item Any single-qubit gates followed immediately by a trace operator, as well as two-qubit gates followed by two trace operators on both qubits involved, can be removed without affecting the marginal state. Repeat this step until there are no more gates to remove.
    \item Then, for each diagonal two-qubit gate $U$ followed immediately by a trace operator on \emph{one of} the qubits involved (say, the first qubit), we write $U$ as $\ket{0}\bra{0}\otimes U_0 + \ket{1}\bra{1}\otimes U_1$. Similar to in the proof of \cref{lem:non-robust}, after tracing out the qubit, the final state can be written as a mixture of two components, in each component the two-qubit gate is replaced by two single-qubit operations. Therefore, we remove all edges corresponding to such gates from the graph. (Note that this step cannot be repeated because this step removes the trace operator, too.)
\end{enumerate}

In what is left of the graph, there must still be a path connecting $A$ and $B$; otherwise, $A$ and $B$ will not be entangled in any of the components of the mixture, and thus they will not be entangled in $\rho_{AB}$. It suffices to show that this path consists entirely of useful edges in the beginning.

In fact, consider any edge $C$--$D$ left in the graph. Since this edge was not removed in the second step, either $D\in\{A, B\}$ or there was at least one other two-qubit gate on $D$ between $C$--$D$ and the final trace operator on $D$. In the second case, let the last of those gates be $D$--$D_1$. This gate must have been removed in the second step since it is followed immediately by a trace operator on $D$, but it must have not have been removed in the first step, either because $D_1\in\{A, B\}$ or because there was at least one other two-qubit gate on $D_1$ between it and the final trace operator on $D_1$. Repeating this argument, since there are a finite number of gates, we must end at some $D_t\in\{A, B\}$, which gives a path satisfying the condition in \cref{dfn:useful_edge}. This argument can be similarly applied to $C$. Therefore, $C$--$D$ was an useful edge in the original graph.
\end{proof}

\subsection{Bound of edge numbers in a good graph}
\label{subsec:graph}

We first consider the case where $G$ doesn't have any parallel edges and deal with the parallel edge case in the proof of~\cref{prp:edge-bound}. 
\begin{lem}
\label{lem:degree-1}
Consider a good graph without any parallel edges and with at least 3 vertices. Then, there can be at most one vertex with degree 1.
\end{lem}
\begin{proof}
Suppose there are at least 2 vertices with degree 1, and let $X$ and $Y$ be two of them. Let $e_X$ and $e_Y$ be the edges incident to $X$ and $Y$ respectively, and without loss of generality suppose $w(e_X) < w(e_Y)$. ($e_X$ and $e_Y$ cannot be the same edge, because otherwise $X$ and $Y$ will be disconnected from other vertices.) Then $e_Y$ cannot be a useful edge with respect to $X$ and $Y$, as the direction starting from $Y$ has to end on $X$ (it cannot end on $Y$ because there is no parallel edge) and have to go through $e_X$, but $e_X$ has a smaller weight. Then there cannot be any path of useful edges that connects $X$ and $Y$.
\end{proof}

We now prove the core result which with~\cref{lem:good-graph} leads directly to~\cref{thm:cnot}.

\begin{prp}
Any good graph with $n \geq 3$ vertices should have at least $\frac{15n-3}{14}$ edges.
\label{prp:edge-bound}
\end{prp}
\begin{proof}
\textbf{No parallel edges}:
We first consider graphs without any parallel edges. 

Given a connected graph $G$ without any parallel edges, we can consider the subgraph generated by its set of vertices with degree 2. Each connected component in this subgraph consists of vertices connected one after another. We call each component a \emph{chain}. We now consider two cases:

\noindent 1. In the degenerate case where a chain is a cycle, we can directly argue:
\begin{itemize}
\item
Suppose one component is a cycle and there are other connected components, then the cycle must be disconnected from other vertices in the original graph, and the graph cannot be good.
\item 
Otherwise suppose the cycle is the only connected component, then the original graph itself must be a cycle. This could be further divided into two cases based on the total number of vertices. 
\begin{itemize}
    \item A cycle with 3 vertices is a good graph and obeys $m \geq \frac{15n-3}{14}$. 
    \item A cycle with at least 4 vertices cannot be a good graph. Consider any 2 vertices that are not neighbors in the cycle. There are 2 paths connecting this pair of vertices, and each path contains at least 2 edges. The edge with largest weight in each path each cannot be useful with respect to this pair, so these 2 vertices are not connected by useful edges.
\end{itemize}
\end{itemize}

\noindent 2. No chains are cycles. Then, we first establish the following lemma:
\begin{lem}
\label{lem:chain}
For a good graph without any parallel edges or any cycle chains, each chain can have at most 4 vertices.
\end{lem}
\begin{proof}
Suppose a chain has at least 5 vertices. Denote the first 5 vertices starting from one end by $P_1, P_2, \ldots, P_5$. Let $M\not= P_2$ and $N\not=P_4$ be the vertices connected to $P_1$ and $P_5$ respectively. Then we define $e_k$ as the edge connecting $P_k$ and $P_{k+1}$, $1\le k \le 4$. The edges $(M,P_1)$ and $(P_5,N)$ are denoted by $e_0$ and $e_5$ respectively. Now we consider the useful edges with respect to $M$ and $P_2$. $e_0$ and $e_1$ cannot be both useful, so the path of useful edges between $M$ and $P_2$ should go through the edges $e_2$, $e_3$, \ldots, which implies
\[
    w(e_2) < w(e_3).
\]
Similar analysis could be done to $P_4$ and $N$ and yield $w(e_2) > w(e_3)$, which leads to a contradiction.
\end{proof}

Now, suppose that every vertex has degree at least 2. We can replace each chain by a single edge between the pair of vertices that the chain connects and remove all the disconnected vertices. In this new graph each vertex has degree at least 3, so we have
\[
    n' \le \frac{2}{3} m',
\]
where $n'$ and $m'$ are the number of vertices and edges respectively of the new graph. Letting $n_2$ be the number of vertices of degree 2, by Lemma~\ref{lem:chain},
\[
    n_2 \le 4m'.
\]
The numbers of vertices and edges in the original graph are $n'+n_2$ and $m'+n_2$, so we have
\[
    \frac{m'+n_2}{n'+n_2} \ge \frac{m' + 4m'}{n'+4m'} \ge \frac{m' + 4m'}{\frac{2}{3}m'+4m'} = \frac{15}{14}.
\]
Hence, $m \geq\frac{15}{14}n$.

By Lemma~\ref{lem:degree-1} there can only be at most 1 vertex with degree 1. In that case we can remove that vertex and apply the calculation above, which shows that $m \geq\frac{15}{14}(n-1) + 1 =\frac{15n-1}{14}$. 

In summary, if there are no parallel edges, we must have $m \geq \min\{\frac{15}{14}n, \frac{15n-1}{14}, \frac{15n-3}{14}\} = \frac{15n-3}{14}$. Note that we have this bound purely for the $n=3$ cycle case. For $n>3$, we can improve this to $\frac{15n-1}{14}$.
\\~\\
\noindent \textbf{Allowing parallel edges}: we prove the theorem by induction. As we have already known, the statement is true for $n=3$. For $n>3$, if there is no parallel edges, the theorem is proved true above. Otherwise, we can contract a pair of vertices connected by parallel edges and the resulting graph with $n-1$ vertices is still a good graph, thus by the inductive hypothesis should have at least $\frac{15(n-1)-3}{14}$ edges. Then the number of edges in the original graph is at least $\frac{15(n-1)-3}{14}+ 2 > \frac{15n-3}{14}$.
\end{proof}

\section{Control: Numerical simulation for the error rate}
\label{app:control}
In this section, we give numerical evidence that we can implement $\SQiSW$ on superconducting platforms with ultra-high fidelity. 

Two-qubit gates in the $\iSWAP$ family can be implemented by
tuning two superconducting qubits with transversal coupling into resonance.
This can be demonstrated in a two-qubit system such as two tunable transmons or tunable fluxonia capacitively coupled directly or through a bus resonator.
Such an implementation suffers from possible stray longitudinal coupling ($ZZ$ interaction),
inaccurate flux tuning and decoherence\cite{barends2019,ganzhorn2020,PhysRevX.11.021058}.  Here we show how these errors will affect the fidelity of implementation.

First we will investigate the coherent error in the gate.
Without loss of generality, we consider a two-qubit system with $YY$ coupling~\cite{krantz2019quantum},
with only the lowest two levels of each qubit included. The relevant Hamiltonians near resonance are given by 
\begin{align*}
    H_1 & =\left[\begin{matrix} 0 & 0 \\ 0 & \omega\end{matrix}\right]\otimes I_2 ;                                               \\
    H_2 & =I_1 \otimes \left[\begin{matrix} 0 & 0 \\ 0 & \omega+\Delta\end{matrix}\right]                                             ; \\
    H_c & =\frac{1}{4}\left(g_{yy}\sigma_{y1}\otimes\sigma_{y2}+g_{zz}\sigma_{z1}\otimes\sigma_{z2}\right) ;\\
    H   & = H_1 + H_2 + H_c = \left[\begin{matrix}
            g_{zz}/4  & 0                      & 0               & -g_{yy}/4               \\
            0         & \omega+\Delta-g_{zz}/4 & g_{yy}/4        & 0                       \\
            0         & g_{yy}/4               & \omega-g_{zz}/4 & 0                       \\
            -g_{yy}/4 & 0                      & 0               & 2\omega+\Delta+g_{zz}/4  \end{matrix}\right],
\end{align*}
where $H_1$ and $H_2$ are the single-qubit Hamiltonians, $H_c$ is
the coupling term, $H$ is the Hamiltonian of the whole two-qubit system, $\omega$ and $\omega+\Delta$ are frequencies of two qubits, and $g_{yy}$ and $g_{zz}$ are
the corresponding coupling strengths.
$g_{yy}$ is the term that contributes to the $\iSWAP$ family gate.
$\Delta$ and $g_{zz}$ are introduced to account for inaccurate flux tuning and stray longitudinal coupling, respectively.

We move into the frame rotating with both qubits at frequency $\omega$. The rotating frame transformation $R(t)$ and the rotated Hamiltonian $H^R(t)$ are given by
\begin{align*}
    R(t)   & =\left[\begin{matrix}
            0 & 0 \\ 0 & e^{-i\omega t}\end{matrix}\right]
    \otimes\left[\begin{matrix}0 & 0 \\ 0 & e^{-i\omega t}\end{matrix}\right],                                       \\
    H^R(t) & =R^{\dagger}(t)H(t)R(t) + i\frac{\partial R^{\dagger}(t)}{\partial t}R(t)=
    \left[\begin{matrix}
            g_{zz}/4               & 0               & 0         & e^{-i2\omega t}g_{yy}/4 \\
            0                      & \Delta-g_{zz}/4 & g_{yy}/4  & 0                       \\
            0                      & g_{yy}/4        & -g_{zz}/4 & 0                       \\
            e^{i2\omega t}g_{yy}/4 & 0               & 0         & \Delta+g_{zz}/4
        \end{matrix}\right].
\end{align*}
Assuming $\omega$
is much larger than any other frequency in the Hamiltonian, we can take the rotating wave approximation and eliminate
all fast-oscillation terms with
$e^{i2\omega t}$ since the time-average is approximately zero.
We thus have an approximate time-independent Hamiltonian of the system given by
\begin{align*}
    H_R & \approx
    \left[\begin{matrix}
            g_{zz}/4 & 0               & 0         & 0               \\
            0        & \Delta-g_{zz}/4 & g_{yy}/4  & 0               \\
            0        & g_{yy}/4        & -g_{zz}/4 & 0               \\
            0        & 0               & 0         & \Delta+g_{zz}/4
        \end{matrix}\right].
\end{align*}

The time evolution operator corresponding to this Hamiltonian is
\begin{align*}
    U(t) & = e^{-iH_Rt}.
\end{align*}
It is easy to verify that when all error terms are zero, that is $\Delta=0$ and $g_{zz}=0$,
$U(t)$ is given by the well-known form of $\iSWAP$ family gate:
\begin{align*}
    U(t)             & = \left[\begin{matrix}
            1 & 0              & 0              & 0 \\
            0 & \cos(\theta)   & -i\sin(\theta) & 0 \\
            0 & -i\sin(\theta) & \cos(\theta)   & 0 \\
            0 & 0              & 0              & 1
        \end{matrix}\right] ;\\
    U(t=\pi/g_{yy})  & = \SQiSW^\dag=\begin{bmatrix}1&0&0&0\\0&\frac1{\sqrt{2}}&-\frac{i}{\sqrt{2}}&0\\0&-\frac{i}{\sqrt{2}}&\frac1{\sqrt{2}}&0\\0&0&0&1\end{bmatrix}                                  ;\\
    U(t=2\pi/g_{yy}) & = \iSWAP^\dag=\begin{bmatrix}1&0&0&0\\0&0&-i&0\\0&-i&0&0\\0&0&0&1\end{bmatrix},
\end{align*}
where $\theta\equiv tg_{yy}/4$. Note that we assume above $g_{yy} >0$ since the gate time must be positive. When $g_{yy}<0$, we instead implement $U(t=-\pi/g_{yy})=\SQiSW$. Note that $\SQiSW$  is equivalent to $\SQiSW^\dag$ under local unitaries: we have $\SQiSW^\dag = (Z\otimes I)\SQiSW(Z\otimes I)$. The properties of the two gates are very similar despite minor sign differences in compilation. In the subsequent sections we will consider the $\SQiSW$  gate even though $\SQiSW^\dag$ is more common in physical implementations.

In realistic systems $g_{zz}$ and $\Delta$ are generally not zero.
This causes a nonzero error, which we can quantify using the average fidelity $F$ between two unitary matrices $U,V$~\cite{nielsen2002simple}: 
\begin{align*}
    F(U,V) & = \frac{\left|\Tr(V^{\dagger}U)\right|^2+d}{d(d+1)}
\end{align*}
where $d=4$ is the dimension of our matrices.
We can estimate the infidelity $E=1-F$ between the simulated physical implementation of $\SQiSW^\dag$ with $U(t=\pi/g_{yy})$
and the ideal $\SQiSW^\dag$ unitary induced by a specific error term by turning on the given error while turning off all others.

We find that the infidelity induced by the $ZZ$ interaction $g_{zz}$ and detuning $\Delta$
can be written as a power series of $g_{zz}/g_{yy}$ and $\Delta/g_{yy}$ respectively,
assuming each error term is small:
\begin{align}
    E_{ZZ}     & \approx \frac{\pi^2}{20}\left(\frac{g_{zz}}{g_{yy}}\right)^2 ;  \nonumber\\
    E_{\Delta} & \approx \frac{8+\pi^2}{10}\left(\frac{\Delta}{g_{yy}}\right)^2. \label{eq:quad_err}
\end{align}
Therefore, the infidelity of $\SQiSW$  scales quadratically with the ratio of $ZZ$ to $YY$ coupling and the ratio of $\Delta$ to $YY$ coupling. 
We also expect there to be a cross-interaction between these two errors, and this is investigated below.
To include the decoherence of the system, we perform numerical simulations
to estimate the effect of $T_1$ and $T_{\phi}$ on the fidelity.
The time evolution is based on the Lindblad master equations:
\begin{align*}
    \frac{d\rho(t)}{dt} & =-\frac{i}{\hbar}[H(t),\rho(t)]+\sum_{j=1,2}\left(
    \Gamma_{1,j}\mathcal{L}[\sigma_j^-] +
    \frac{1}{2}\Gamma_{\phi,j}\mathcal{L}[\sigma_j^z]
    \right)\rho(t)
\end{align*}
where $\rho(t)$ is the time-dependent density matrix, $H(t)$ is the time-dependent Hamiltonian, $\Gamma_{1,j}$ is the dissipation rate of $j$-th qubit, $\Gamma_{\phi,j}$ is the dephasing rate of $j$-th qubit, and $\mathcal{L}[c]\rho=c\rho c^{\dagger}-\frac{1}{2}(\rho c^{\dagger}c-c^{\dagger}c\rho)$ is the Lindblad superoperator. Note this model includes only the Markovian noise. The gate fidelity is affected little by the non-Markovian noise when the gate time is short~\cite{omalley2015}, so we neglected the non-Markovian noise in the discussion.
We can then compute the average gate fidelity via the process fidelity and $\chi$-matrix as follows~\cite{nielsen2002simple}:
\begin{align*}
    F_p(\chi_1,\chi_2) & =\Tr(\chi_1\chi_2);                                                     \\
    F(\chi_1, \chi_2)  & = \frac{dF_p(\chi_1,\chi_2)+1}{d+1} = \frac{\Tr(\chi_1\chi_2)d+1}{d+1}.
\end{align*}

We choose the following range of parameters, which can be experimentally realized in fabricated fluxonium qubits, to perform the numerical simulations.
For simplicity, we assume $T_1=T_{1,1}=T_{1,2}$ and $T_{\phi}=T_{\phi,1}=T_{\phi,2}$.
\begin{itemize}
    \item $g_{yy}/h=25\mathrm{MHz}$, corresponds to a $\SQiSW$  gate time of $20\mathrm{ns}$
    \item $g_{zz}/h\in [-0.6, +0.6]\ \mathrm{MHz}$
    \item $\Delta/h\in [-0.4, +0.4]\ \mathrm{MHz}$
    \item $T_1 \geq 25\mathrm{\mu s}$
    \item $T_{\phi}\geq 25\mathrm{\mu s}$
\end{itemize}
We check the effect of nonzero temperature as the thermal population of the first excited state $\ket{e}$
may be large. However, we find no major difference between the infidelity at zero temperature
and that of a typical fluxonium system ($T=50\mathrm{mK}$ and frequency $\omega_{ge}=2\pi\times1\mathrm{GHz}$).

\begin{figure}[ht]
    \centering
    \includegraphics[width=0.35\textwidth]{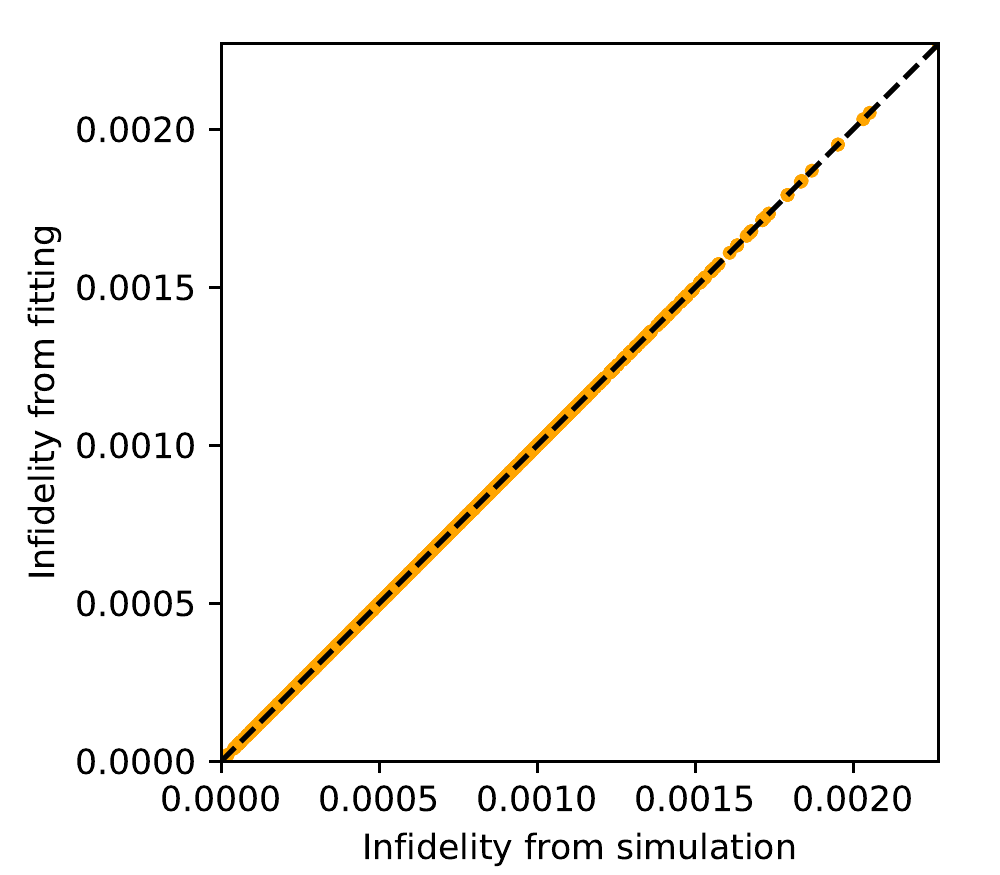}
    \caption{Infidelity via polynomial regression against simulated infidelity of $\SQiSW$  gates. The black dashed line is the $x=y$ line for visual effect. The root mean square error of the regression is on the order of $10^{-9}$. 
    }
    \label{fig:infidelity_regression}
\end{figure}

\begin{figure}[ht]
    \centering
    \includegraphics[width=0.35\textwidth]{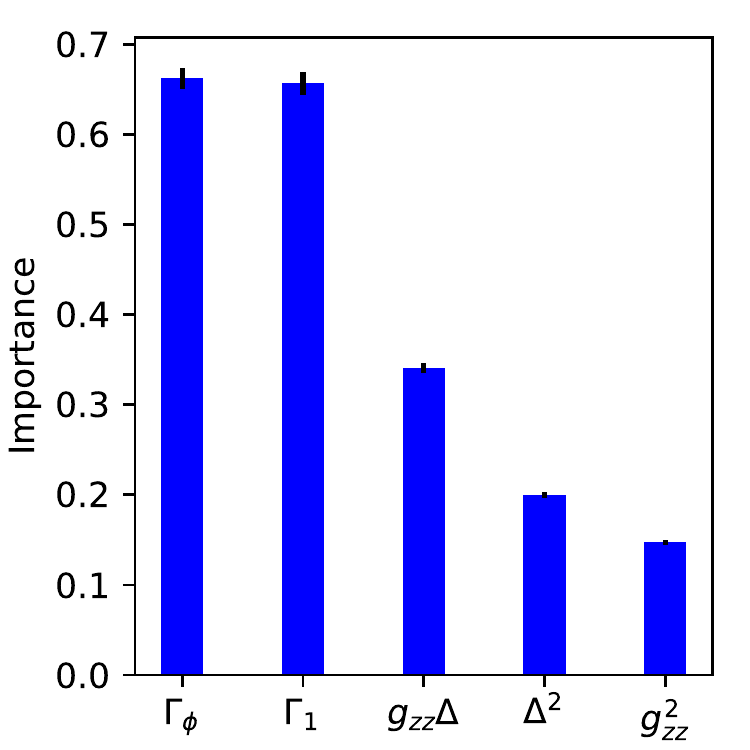}
    \includegraphics[width=0.40\textwidth]{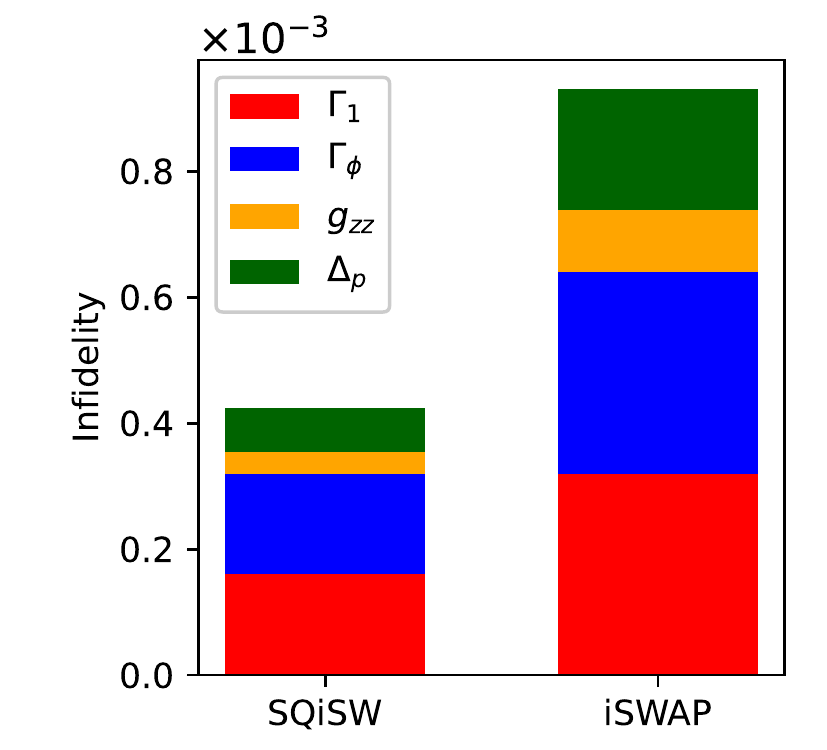}
    \caption{Left: The permutation feature importance for the polynomial regression in Fig.~\ref{fig:infidelity_regression}. Only features with importance $>0.01$ are included in the figure. Right: Comparison of features' contributions to the infidelity of $\SQiSW$  and $\iSWAP$ with the parameters $g_{yy}/h=25\mathrm{MHz}, T_1=100\mathrm{\mu s},T_{\phi}=100\mathrm{\mu s},
    g_{zz}/h=-0.3\mathrm{MHz},\Delta_p/h=0.18\mathrm{MHz}$.}
    \label{fig:importance_error}
\end{figure}

After computing the infidelity via numerical simulation, we perform
a polynomial regression up to 2nd order in the different parameters to identify the error sources (3rd order polynomial regression does not add any new important terms as shown in Fig.~\ref{fig:importance_error}). 
In particular, $g_{zz}$, $\Delta$, $\Gamma_1$ and $\Gamma_{\phi}$,
are features in the regression as they are expected to increase the infidelity linearly or quadratically. 
The regression accuracy is plotted in Fig. \ref{fig:infidelity_regression}.
The polynomial regression works very well in this case: the root mean square error of the regression is on the order of $10^{-9}$. 

To identify the key error sources contributing to the infidelity, we check the permutation feature importance as shown in Fig. \ref{fig:importance_error}.
There are 5 dominant features in the polynomial regression.
The infidelity depends linearly on $\Gamma_{\phi}$ and $\Gamma_1$, which is a general behavior of decoherence.
And the infidelity also depends quadratically on $g_{zz}$ and $\Delta$, which agrees with the symbolic results in~\cref{eq:quad_err}. 
Finally, an additional term $g_{zz}\Delta$ emerges, which is the cross-term between $g_{zz}$ and $\Delta$. This term makes it unclear what the error contributions are due to $g_{zz}$ and $\Delta$ individually. In our experiments, $g_{zz}$ is fixed at the design stage and $\Delta$ is microwave tunable, so we can choose an optimal $\Delta$ in the experiments to minimize the infidelity contributed by the terms $c_{\Delta^2}\Delta^2,c_{g_{zz}\Delta}g_{zz}\Delta, c_{g_{zz}^2} g_{zz}^2$. This isolates the contribution to the infidelity from $g_{zz}$. This optimal $\Delta$ will be our operating point. However, the precision of the frequency is limited by our instruments, and there is an error caused by the deviation from the optimal point which we call $\Delta_p$.  Overall, we define the infidelity from each error source as follows: 
\begin{itemize}
    \item $c_{\Delta^2}\Delta^2+c_{g_{zz}\Delta}g_{zz}\Delta+ c_{g_{zz}^2} g_{zz}^2$ at the optimal $\Delta$ as the error from the stray coupling $g_{zz}$;
    \item $c_{\Delta^2}\Delta_p^2$ as the error from the instrumental limitation of the frequency $\Delta_p$;
    \item $c_{\Gamma_1}\Gamma_1$ and $c_{\Gamma_{\phi}}\Gamma_{\phi}$ as the error from the decoherence processes.
\end{itemize}   
These errors are plotted for both $\SQiSW$  and $\iSWAP$ in Fig.~\ref{fig:importance_error}. Note that because we are working at the optimal $\Delta$, the above errors add up to the total error. 

Considering the recent progress on fluxonium fabrication~\cite{nguyen2019,Bao2021} and the precision of the arbitrary wave generator, we compute for a realistic set of parameters  $g_{yy}/h=25\mathrm{MHz}$, $T_1=100\mathrm{\mu s},T_{\phi}=100\mathrm{\mu s}, g_{zz}/h=-0.3\mathrm{MHz},\Delta_p/h=0.18\mathrm{MHz}$, which is what we take as input for the results in~Fig.~\ref{fig:importance_error}, $\SQiSW$  can be realized with about $5\times10^{-4}$ infidelity. Note in the fluxonium the $T_{\phi}$ at the operation point is limited by $1/f$ flux noise. Its error can be estimated by $t^2/3T_{\phi}^2$, which is $<10^{-4}$  when $T_{\phi}>2.5\mathrm{\mu s}$ for a 40 $\mathrm{ns}$ $\iSWAP{}$ gate and it is even smaller for a 20 $\mathrm{ns}$ $\SQiSW{}$ gate. So the error from $1/f$ flux noise can be neglected and it is not included in the discussion.

\section{Experimental Details}
\label{app:calibration}
\subsection{Calibration of an $i\mathrm{SWAP}$-like gate}
For an iSWAP-like gate, the matrix form of the unitary operator can be written as~\cite{arute2020observation}
$$
U(\theta, \zeta, \gamma,\chi-\Delta\cdot t, \phi) = \notag \\
\left[\begin{matrix}1 & 0 & 0 & 0\\0 & e^{-i(\gamma + \zeta)} \cos\theta & -ie^{-i(\gamma-\chi+\Delta \cdot t)} \sin\theta & 0\\0 & -ie^{-i(\gamma+\chi-\Delta \cdot t)} \sin\theta & e^{-i(\gamma-\zeta)} \cos\theta & 0\\0 & 0 & 0 & e^{-i(2\gamma+\phi)}\end{matrix}\right],
$$
where $\gamma$ is a common single-qubit phase induced by the flux modulation, $\chi$ is the relative phase between $|10\rangle$ and $|01\rangle$ states with population swapping, $\zeta$ is relative phase without population swapping, and $\phi$ is the controlled-phase induced by $ZZ$ coupling, and is believed to be negligible according to the analysis in \Cref{app:control}. We assume $\phi=0$ throughout the section. The swap angle $\theta$ equals to $\pi/2$ for $\iSWAP{}$ and equals to $\pi/4$ for $\SQiSW{}$. To activate the two-qubit interaction, a fast flux pulse brings the qubit with lower frequency into resonance with the other qubit. Technically, the iSWAP-like gate is realized when the qubits are kept in the same frequency. However in practice, the two qubits are biased at their own operation points with different frequencies $\omega_1$ and $\omega_2$ for individual single-qubit operations. As a result, the relative phase $\chi-\Delta\cdot t$ contains a time-dependent term, where $\Delta=\omega_1-\omega_2$ is the detuning of the two qubits, and $t$ is the start time of the gate. 

The swap angle $\theta$ determines the gate implemented up to local equivalence. All the other phase terms can be absorbed in single-qubit phase gates for corrections:
\begin{equation}
    U(\theta, \zeta, \gamma,\chi-\Delta\cdot t, 0) \propto (R_z(-\gamma + \beta)\otimes R_z( -\gamma -\beta))\cdot U(\theta, 0, 0, 0, 0)\cdot (R_z(\alpha)\otimes R_z(-\alpha)),
\end{equation}
where $\beta = (\zeta - \chi +\Delta\cdot  t)/2$, $\alpha=(\zeta + \chi -\Delta\cdot  t)/2$. Therefore, $\theta$ needs to be characterized and calibrated to $\pi/4$ with high accuracy, and the phase terms $\zeta, \gamma,\chi$ only needs to be characterized with high accuracy, and can be cancelled out in the compilation stage.

\subsection{Coarse calibration}
The calibrations of the bias-crosstalk, the flux-pulse correction and the single-qubit gates should be implemented before the coarse calibration~\cite{Bao2021}. 
Without need of special pulse-shaping for suppressing higher-energy-level leakage, we use an error-function shape pulse to modulate the external flux $\Phi_{\mathrm{ext}}$:
\begin{equation}
    \Phi_{\mathrm{ext}}(t) = \dfrac{\Phi_{\mathrm{amp}}}{2}\left[\mathrm{Erf}(\dfrac{t-4\sigma}{\sqrt{2}\sigma})-\mathrm{Erf}(\dfrac{t-t_g+4\sigma}{\sqrt{2}\sigma}) \right], \label{eq:pulse_shape}
\end{equation}
with $\sigma=0.5$~ns.

\begin{figure}
  \includegraphics{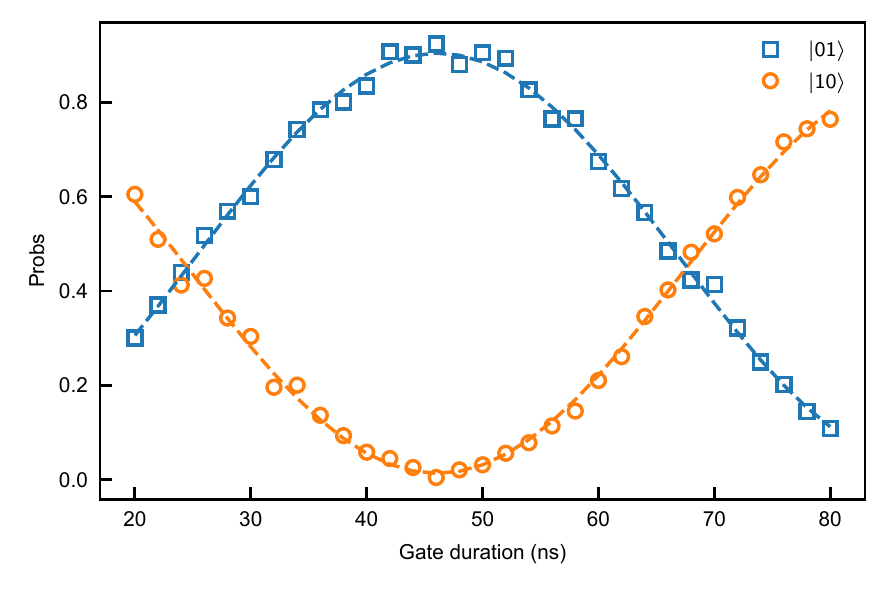}
  \caption{\label{fig:gate_duration} The population versus duration of iSWAP-like gate.}
\end{figure}

The first step is to measure the pulse amplitude $\Phi_{\mathrm{amp}}$ of the resonance point with both qubits initially biased at their respective sweet spots and prepared in the state $|10\rangle$. We measure $P_{10}$ as a function of the gate duration and amplitude. The measured $P_{10}$ oscillates versus the gate duration with an angular frequency $\omega = \sqrt{4g^2 + \left(\frac{\text{d} \omega_1}{\text{d}\Phi_{\mathrm{ext}}}\right)^2(\Phi_{\mathrm{amp}}-\Phi_{\mathrm{res}})^2}$, where $g$ is the coupling strength of the two qubits. The measurement results can be found in~\cite{Bao2021}. The resonance point $\Phi_{\mathrm{res}}$ 
corresponds to the minimal oscillation frequency. When pulse amplitude is fixed at $\Phi_{\mathrm{res}}$, the population as a function of gate duration is plotted in Fig.~\ref{fig:gate_duration}. The probability $P_{10}$ can be fitted with the function $P_{10}(t) = (1 + \cos(2 g t + \phi_0)) / 2$. The duration of the iSWAP-like gate is estimated as $T = (2\theta-\phi_0)/2g$. The detailed calibration process of $\iSWAP{}$ gate has been described in \cite{Bao2021}. Here we will focus on the calibration of $\SQiSW{}$.

\subsection{Fine calibration for $\theta$ and $\zeta$}
The gate duration of $\SQiSW{}$ is approximately 25~ns estimated from the previous coarse calibration. To minimize the control error, we use the following scheme to measure more accurate values of $\theta$ and $\zeta$. 

\begin{figure}
  \includegraphics{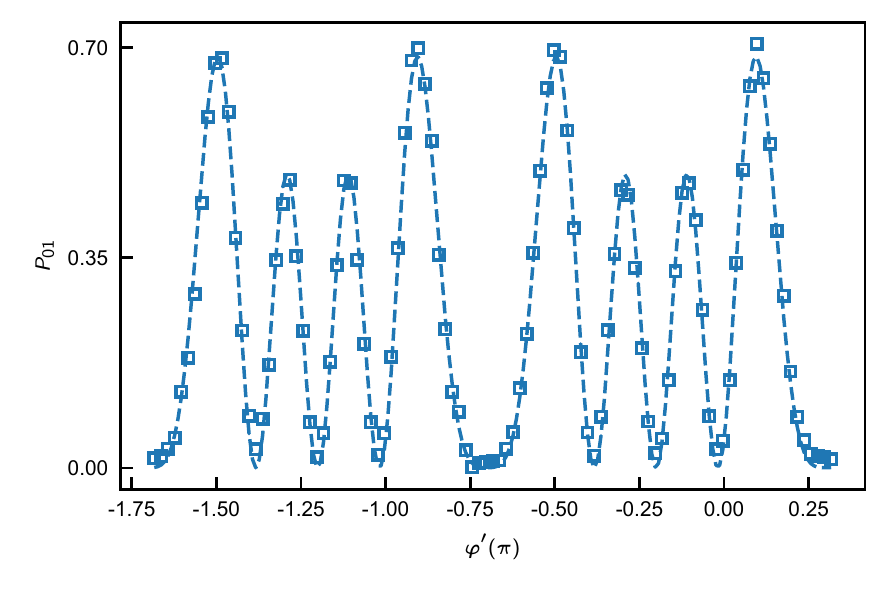}
  \caption{\label{fig:cal_theta_zeta} The $P_{01}$ as a function of phase of single-qubit phase gates with $N=8$.}
\end{figure}

\begin{figure}
  \includegraphics[scale=1]{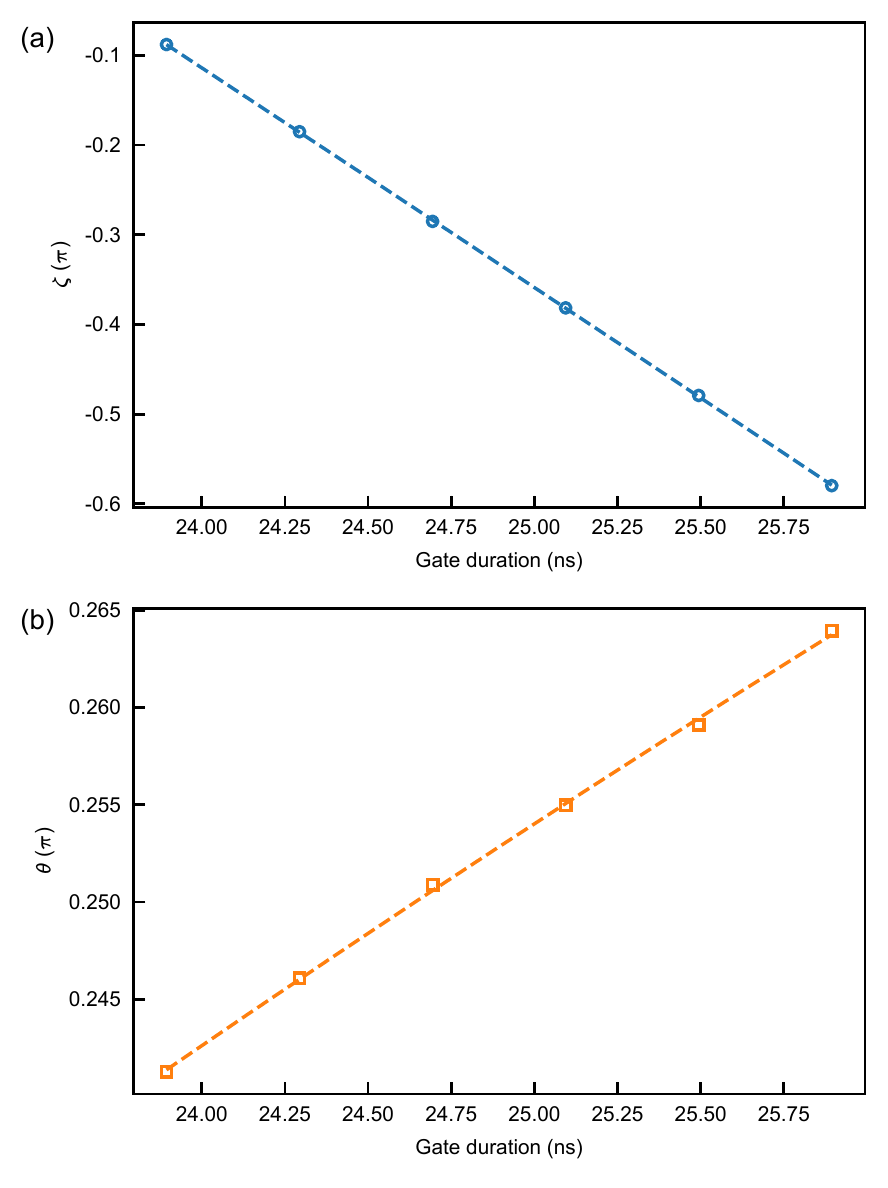}
  \caption{\label{fig:theta_zeta} $\theta$ (a) and $\zeta$ (b) as a function of gate duration.}
\end{figure}

A gate sequences $(U \cdot (R_z(\varphi)\otimes R_z(-\varphi)))^N$ is applied to the initial state $|10\rangle$. The repeat count $N$ is chosen to be $8$ to amplify the sensitivity of final $P_{01}$ to $\theta$ and $\zeta$. The final swapped probability can be written as 
\begin{equation}
    \begin{gathered}
        P_{01} = (\sin(N\Omega) \sin(\theta)/\sin(\Omega))^2,\\
        \Omega = \arccos(\cos(\theta)\cos(\zeta+\varphi-\Delta\cdot\delta t)).\label{eq:p01}
    \end{gathered}
\end{equation}
$\delta t = T + T_R$ is the separation of two $\SQiSW{}$ gate, where $T_R$ is the duration of the phase gates $R_z(\varphi)\otimes R_z(-\varphi)$. The single-qubit phase gates are realized with 
\begin{equation}
    R_z(\varphi)\otimes R_z(-\varphi) = (R_0(\pi)\cdot R_{\varphi/2}(\pi)) \otimes (R_0(\pi)\cdot R_{-\varphi/2}(\pi))
\end{equation}
where  $R_\theta(\pi) = i e^{i \theta} \cdot |0\rangle \langle 1| + i e^{-i \theta} \cdot |1\rangle \langle 0|$ is a $\pi$-rotation of single-qubit, and can be realized with a single microwave pulse. The duration of each pulse for single-qubit operation is set to 10~ns, which gives $T_R=20$~ns. We measure the $P_{01}$ as a function of $\varphi^\prime:= \varphi -\Delta \cdot \delta t$.
The experimental data and a fit to Eq.~(\ref{eq:p01}) yielding $\theta/\pi=0.2507$ and $\zeta/\pi=-0.2992$ are shown in Fig.~\ref{fig:cal_theta_zeta}.
To further reduce the error in $\theta$, we fix the pulse amplitude $\Phi_{\mathrm{amp}}$ and sweep gate duration over small range. We measure $\theta$ and $\zeta$ for each specified gate duration. The measured results versus different gate duration are plotted in Fig.~\ref{fig:theta_zeta}. The $\theta$ and $\zeta$ almost vary linearly with gate duration. By interpolating the data, we can extract a gate duration of $T=24.64$~ns and $\zeta/\pi=-0.2709$ when $\theta$ equals $\pi/4$. In order to facilitate the experiment and reduce the impacts from possible flux distortion, we add an idle length to each flux pulse to make the gate duration $T=30$~ns.

\subsection{Calibration for $\gamma$ and $\chi$}
\begin{figure}
  \includegraphics[scale=1]{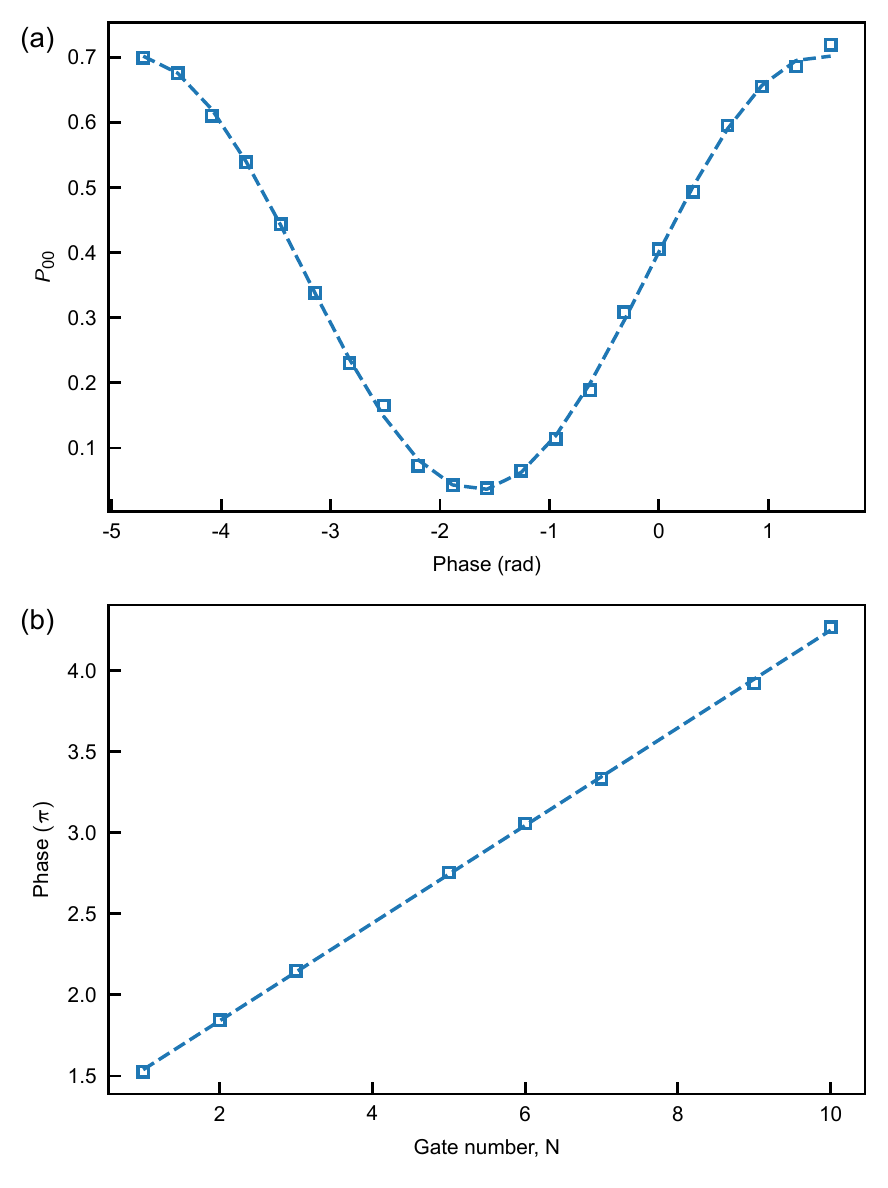}
  \caption{\label{fig:gamma_chi} (a) Ramsey-type experiment under $N=1$, the measured $P_{00}$ oscillates versus phase of the second half-$\pi$ pulse. (b) The extracted phase as a function of gate number $N$.}
\end{figure}

To further measure the phases $\gamma$ and $\chi$, we use the similar sequence applied to the initial state $|\psi_i\rangle = (|00\rangle - i |01\rangle)/\sqrt{2}$. To simplify the measurement, $\Omega$ is set to $\pi/4$ with $\varphi = \Delta \cdot \delta t -\zeta$. The increased phase of first qubit after multiple gates can be written as
\begin{align}
    \varphi_{sq} = \arg(\langle 10| (U \cdot R_z(\varphi)\otimes R_z( -\varphi))^N |\psi_i\rangle ) \notag, \\
     = N \gamma + \chi + \dfrac{\pi}{2}\cdot \mathrm{sgn}\left(\sin\left(\dfrac{N\pi}{4}\right)\right) + \Delta(t_1 + t_N) / 2,
\end{align}
where $t_1$ and $t_N$ are the start time of first and last $\SQiSW{}$ gate. We use a Ramsey-type experiment to extract the single-qubit phase $\varphi_{sq}$. After applying a second half-$\pi$ pulse to the first qubit, the $P_{00}$ relates to the $\varphi_{sq}$ and the varied phase of second half-$\pi$ pulse. Subtracting the known phase term $\dfrac{\pi}{2}\cdot \mathrm{sgn}\left(\sin\left(\dfrac{N\pi}{4}\right)\right) + \Delta(t_1 + t_N) / 2$, 
the residual phase of $N\gamma + \chi$ can be extracted by single-qubit tomography. We present an experiment of $N=1$ as an example in Fig.~\ref{fig:gamma_chi}(a). The measured $P_{00}$ can be well fitted with a cosine function, giving $(\gamma + \chi)/\pi\approx -0.4615$. We then increase the gate number $N$ and measure the corresponding phases $N\gamma + \chi$. A relatively accurate value of $\gamma/\pi=0.3012$ and $\chi/\pi=-0.7628$ can be extracted by fitting the measured phases presented in Fig.~\ref{fig:gamma_chi}(b) with a linear function.

\subsection{Drift of $T_1$ and $T_2$}
The coherence times $T_1$ and $T_2$ of our qubits fluctuated during our measurements. The highest number we measured are $T_1^A=75~\mathrm{\mu s},T_{2,R}^A=5~\mathrm{\mu s},T_1^B=67~\mathrm{\mu s},T_{2,R}^B=17.5~\mathrm{\mu s}$ at the sweet spots and $T_1^A=40~\mathrm{\mu s},T_{2,R}^A=1~\mathrm{\mu s}$ near the resonant point. Based on the estimation of error in Ref.~\cite{omalley2015}, we find that the decoherence error of $\iSWAP{}$ from the relaxation and the white noise dephasing is $4.7\times 10^{-3}$ and that from $1/f$ flux noise dephasing is $0.5\times 10^{-3}$. The coherent error is estimated to be $0.5\times 10^{-3}$~\cite{Bao2021}. This corresponds to the best $\iSWAP{}$ fidelity 0.993 in Figure~4 of the main text. Also, we observed that $T_1$ of $Q_A$ sometimes drops significantly, as low as $1~\mathrm{\mu s}$ at the sweet spot, possibly due to coupling with a fluctuating two-level system. In such case, the decoherence error from white noise dephasing and relaxation is as large as $2\times 10^{-2}$, which corresponds to the worst fidelity $0.98$. Overall, the decoherence error from white noise dephasing and relaxation dominate the infidelity of the $\iSWAP{}$ gate. As this error is proportional to the gate time, we expect the infidelity of the $\iSWAP{}$ gate and the $\SQiSW{}$ gate is approximately $2:1$. This roughly agrees with the measured fidelities of the two gates in Figure~1 of the main text. 

\subsection{Single-Qubit Gates}
The single-qubit Clifford gate is complied with a primary set of gate operations , denoted as $\{I, X_{\pi},Y_{\pi},X_{\pm \pi/2},Y_{\pm \pi/2}\}$. The measured average fidelities of this set operations of $Q_A$ and $Q_B$ are 99.90\% and 99.96\%, respectively.

\begin{figure}
  \includegraphics[scale=1]{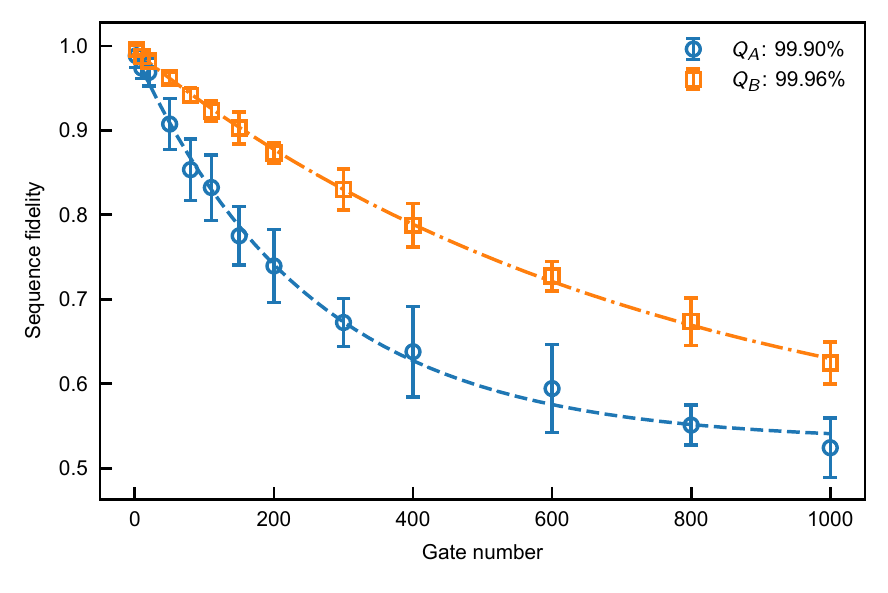}
  \caption{\label{fig:simultaneous_RB} Sequence fidelity of simultaneous RB versus gate number.}
\end{figure}

\section{Characterization: Interleaved fully randomized benchmarking}
\label{app:frb}
We now give a variant of the interleaved randomized benchmarking (iRB) framework~\cite{magesan2012efficient} for benchmarking the $\SQiSW$  gate. The iRB framework was first proposed to benchmark the average fidelity of a target gate given the ability to implement arbitrary Clifford gates with high fidelity. However, under the iRB framework, the target gate, i.e.\ the gate to be benchmarked, needs to be Clifford too. For this reason, the iRB framework is usually used on two-qubit gates such as the $\iSWAP$ gate or the $\CNOT$  gate, but not on non-Clifford gates such as the Controlled-$S$ gate , much of the fSim gate family, the matchgates\cite{helsen2020matchgate}, or $\SQiSW$. Our variant, called {\em interleaved fully randomized benchmarking (iFRB)}, relies on the efficient implementation of Haar random gates as the reference gate set. Compared to Clifford-based iRB, the iFRB scheme is readily applicable to benchmarking of arbitrary quantum gates (not necessarily Clifford) and especially useful when benchmarking on a small quantum system where efficiency of implementing Haar random gates is not an issue.

\subsection{FRB and iFRB}
\label{subsec:frb}
Before introducing iFRB, we first briefly recall the vanilla randomized benchmarking (RB) and the iRB frameworks. Randomized benchmarking~\cite{emerson2005scalable} was first proposed to study the amplitude of the gate-independent, time-independent, average noise level in a quantum system, while isolating out the errors caused by imperfect state preparation and measurement (SPAM error). The experimental protocol goes as follows: for a $d$-dimensional quantum system, choose an appropriate gate sequence length $m$, choose $m$ gates $U_1,\cdots, U_m$ i.i.d.\ from the Haar random distribution and compute $U_{m+1}$ the recovery gate such that $U_{m+1}U_m\cdots U_1=I$ assuming there are no errors. The gate sequence $U_1||U_2||\dots ||U_m||U_{m+1}$ is then performed on an initial state $|0\rangle$ and subsequently measured in the computational basis. A \emph{survival probability} $p_{m,j}$ of measuring $0$ is estimated from repeated experiments with sequence $j$ of length $m+1$. By averaging over many different sequences with the same length and collecting data over several different gate sequence lengths, one can extract the average gate fidelity  $r=1-(1-u)(d-1)/d$, where $d$ is the dimension of the system ($2^n$ for an $n$-qubit system) and $u$ is obtained from fitting the curve
$$\mathbb{E}_j[p_{m,j}]=A\cdot u^m+B.$$
Here, the parameters $A$ and $B$ are supposed to capture the SPAM error, leaving $u$ represent solely the imperfection in gate implementation.
The requirement for Haar randomness was subsequently found to be unnecessary and unscalable to larger sized quantum systems and is relaxed to a unitary 2-design, most commonly a uniform distribution on the Clifford gates\cite{knill2008randomized}. To avoid confusion, we refer to the Haar random randomized benchmarking as fully randomized benchmarking (FRB) and the Clifford-based one RB. 

After the first proposal of FRB, follow-up works flourished~\cite{magesan2012efficient, knill2008randomized, emerson2007symmetrized, gambetta2012characterization, wallman2015robust, wallman2016robust, garion2021experimental}, most of which were based on the Clifford gate set. In particular, interleaved randomized benchmarking~\cite{magesan2012efficient} was proposed to benchmark the average fidelity of a specific gate, referred to as the \emph{target gate}, with the hope to exclude not only the SPAM error, but the errors of other gates in a gate set. An iRB experiment consists of two parts. The first part is an ordinary RB protocol on a gate set, referred to as the \emph{reference gate set}. The second part performs RB with interleaved sequences. Given a target gate $T$, a random gate sequence of length $m$, $U_1,U_2,\cdots U_m$, is generated i.i.d.\ from the Clifford group, but the final recovery gate is chosen to be $U'_{m+1}=(T\cdot U_m\cdot T\cdot\dots\cdot T\cdot U_2\cdot T\cdot U_1)^\dag$, and the gate sequence $U_1||T||U_2||T||\dots||T||U_{m}||T||U'_{m+1}$ is then performed as in the RB experiment. A different error quantity $v$ can be calculated from the decay rate of the average survival probability with respect to the sequence length, similar to $u$ in the ordinary RB experiment. The average fidelity of the target gate can then be calculated as $$r_T=1-\frac{(1-v/u)(d-1)}{d}.$$

In order to be able to carry out the iRB experiment, it is crucial that the final recovery gate, $U'_{m+1}$, lies in the reference gate set, i.e.\ the Clifford group. Although this holds when the target gate itself lies in the Clifford group, this is not true for many common gates. That the iRB framework cannot be applied for non-Clifford gates is a serious outstanding issue as a non-Clifford gate is necessary for universal quantum computing, by the Gottesman-Knill Theorem~\cite{gottesman1998heisenberg}. Several alternatives have been proposed, including choosing different finite groups other than the Clifford group~\cite{garion2021experimental}. Altogether different benchmarking experiments were also proposed~\cite{arute2019quantum, reagor2018demonstration}, but these alternatives either rely on extensive algebraic studies of the target gate or lack a rigorous theoretical framework for analyzing their interpretation and applicability.

To resolve this issue thorugh iFRB, we simply apply iRB, except that instead of using random Clifford gates, we return to the original FRB proposal by using Haar random gates. As there are no restrictions on the recovery gate, iFRB applied to any gate. Since Haar random gates are trivially a unitary 2-design, iFRB carries the same theoretical guarantees of RB and iRB that the noise is gate-independent. For the more general Markovian and possibly gate-dependent noises, whether the iFRB framework works as expected requires further investigation.

\begin{figure}
    \centering
    \begin{subfigure}{\textwidth}
    \begin{tikzpicture}
\node {
    \begin{quantikz}
             \lstick{\ket{0}} &\gate[style={fill=green!20}]{C_1}&\gate[style={fill=green!20}]{C_2}  & \ \ldots \ \qw&\gate[style={fill=green!20}]{C_m}& \gate[style={fill=blue!20}]{C_{m+1}}& \meter{0}
             \end{quantikz}};\end{tikzpicture} \\\begin{tikzpicture}
             
             \node{
             
             \begin{quantikz}
             \lstick{\ket{0}} &\gate[style={fill=green!20}]{C'_1}&\gate[style={fill=yellow!20}]{T_1}&\gate[style={fill=green!20}]{C'_2}  &\gate[style={fill=yellow!20}]{T_2} &\ \ldots \ \qw &\gate[style={fill=green!20}]{C'_m}&\gate[style={fill=yellow!20}]{T_m}& \gate[style={fill=blue!20}]{C'_{m+1}}& \meter{0}
             \end{quantikz}};
             \end{tikzpicture}
             \caption{Clifford-based iRB. $C_1,\dots, C_m$ and $C'_1,\dots, C'_m$ are taken i.i.d. uniformly from the Clifford group $\mathcal{C}_d$.}
    \end{subfigure}
    \begin{subfigure}{\textwidth}
    \begin{tikzpicture}
\node {
    \begin{quantikz}
             \lstick{\ket{0}} &\gate[style={fill=orange!20}]{U_1}&\gate[style={fill=orange!20}]{U_2} &\ \ldots \ \qw &\gate[style={fill=orange!20}]{U_m}& \gate[style={fill=blue!20}]{U_{m+1}}& \meter{0}
             \end{quantikz}};\end{tikzpicture}\\\begin{tikzpicture}
\node{
             
             \begin{quantikz}
             \lstick{\ket{0}} &\gate[style={fill=orange!20}]{U'_1}&\gate[style={fill=yellow!20}]{T}&\gate[style={fill=orange!20}]{U'_2}  &\gate[style={fill=yellow!20}]{T}&\ \ldots \ \qw &\gate[style={fill=orange!20}]{U'_m}&\gate[style={fill=yellow!20}]{T}& \gate[style={fill=blue!20}]{U'_{m+1}}& \meter{0}
             \end{quantikz}};
             \end{tikzpicture}
             \caption{Interleaved Fully Randomized Benchmarking. $U_1,\cdots, U_m$ and $U'_1,\dots, U'_m$ are taken i.i.d. Haar randomly from $SU(d)$.}
    \end{subfigure}
    \caption{Illustration of the Clifford-based iRB and the iFRB.}
\end{figure}
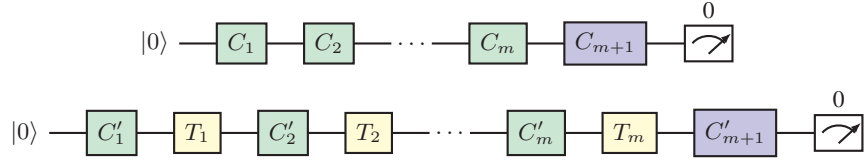
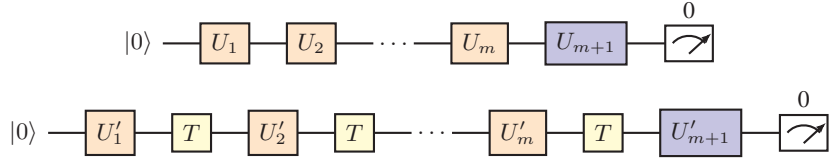

\subsection{iFRB on the $\SQiSW{}$ gate}

The biggest challenge of implementing such a fully randomized scheme is of course the efficient generation and implementation of arbitrary rotations. It is long known that the complexity of implementing Haar random gates grows exponentially with respect to the number of qubits\cite{knill1995approximation}. However, FRB/iFRB can still be a very useful tool to benchmark single qubit or two-qubit gates, or even unitaries acting on a small number of qubits. In superconducting systems, 1-qubit FRB/iFRB is readily realized via the virtual Z compilation scheme~\cite{mckay2017efficient}. On two-qubit systems, the FRB/iFRB framework requires the efficient generation of arbitrary two-qubit unitaries from native two-qubit gates. Luckily, for many families of gates, such as the super-controlling gates\cite{ye2004super} and the $\SQiSW$  gate, an efficient decomposition of arbitrary two-qubit gates into an optimal number of native two-qubit gate exists. Hence we can realize two-qubit FRB/iFRB in such cases.

\begin{algorithm}[H]
  \caption{Generating iFRB sequences}
  \label{alg:randgen}
   \begin{algorithmic}[1]
   \Procedure{GenRandSeq}{m,t} \Comment{Generate a random FRB sequence of length $m$, interleaved with $\SQiSW$ iff $t$ is True}
   \State $U\leftarrow I$
   \State $S\leftarrow \varepsilon$
   \For{$i=1,\dots, m-1$}
   \State $U_i, S_i\leftarrow$\textsc{GenRandGate}()
   \If{t == True}
   \State $U\leftarrow \SQiSW \cdot U_i\cdot U $
   \State $S\leftarrow S || S_i||\SQiSW$
   \Else
   \State $U\leftarrow U_i\cdot U $
   \State $S\leftarrow S || S_i$
   \EndIf
   \EndFor
   \State\Return  $S ||$ \textsc{Decomp}($U^\dag$)\Comment{Append the recovery gate}
   \EndProcedure
   \Procedure{GenRandGate}{} \Comment{Generate a Haar random $SU(4)$ gate and its corresponding decomposition into $\SQiSW$  sequence}
   \State $U\leftarrow$\textsc{GenHaarSU(4)} \Comment{Generate a Haar random element in $SU(4)$}
   \State \Return $U,$ \textsc{Decomp}($U$)
   \EndProcedure
   \end{algorithmic}
\end{algorithm}

\subsection{Experiment specs}

\section{Characterization: Numerical experiments of FRB and quantum volume}
The information processing capabilities of $\SQiSW$ we have proved all point to its superiority in actual experimental realizations. To strengthen this claim, in this section we conduct a series of numerical experiments comparing $\SQiSW$ to $\iSWAP$ with respect to different metrics under a noisy setting. For simplicity we assume a simple depolarizing noise model.

\subsection{Fidelity of Compiling Two-Qubit Gates}
In our first experiment we compare $\SQiSW$ to $\iSWAP$ by computing the fidelity of generating arbitrary two-qubit gates in a noisy setting. An arbitrary two-qubit gate is compiled using $\SQiSW$ according to~\cref{alg:decomp} and using $\iSWAP$ according to~\cite{ye2004super}. We consider a simple error model: each gate is followed by a depolarizing channel with error rate $p_{\text{iswap}} = 2p_{\text{sqisw}} = 0.005$, $p_{\text{single}}=0.0005$. For the two-qubit gates, each of the qubits undergo a depolarizing channel of the corresponding error rate.

\begin{figure}
    \centering
    \includegraphics[width=0.7\textwidth, trim=10cm 10cm 10cm 10cm]{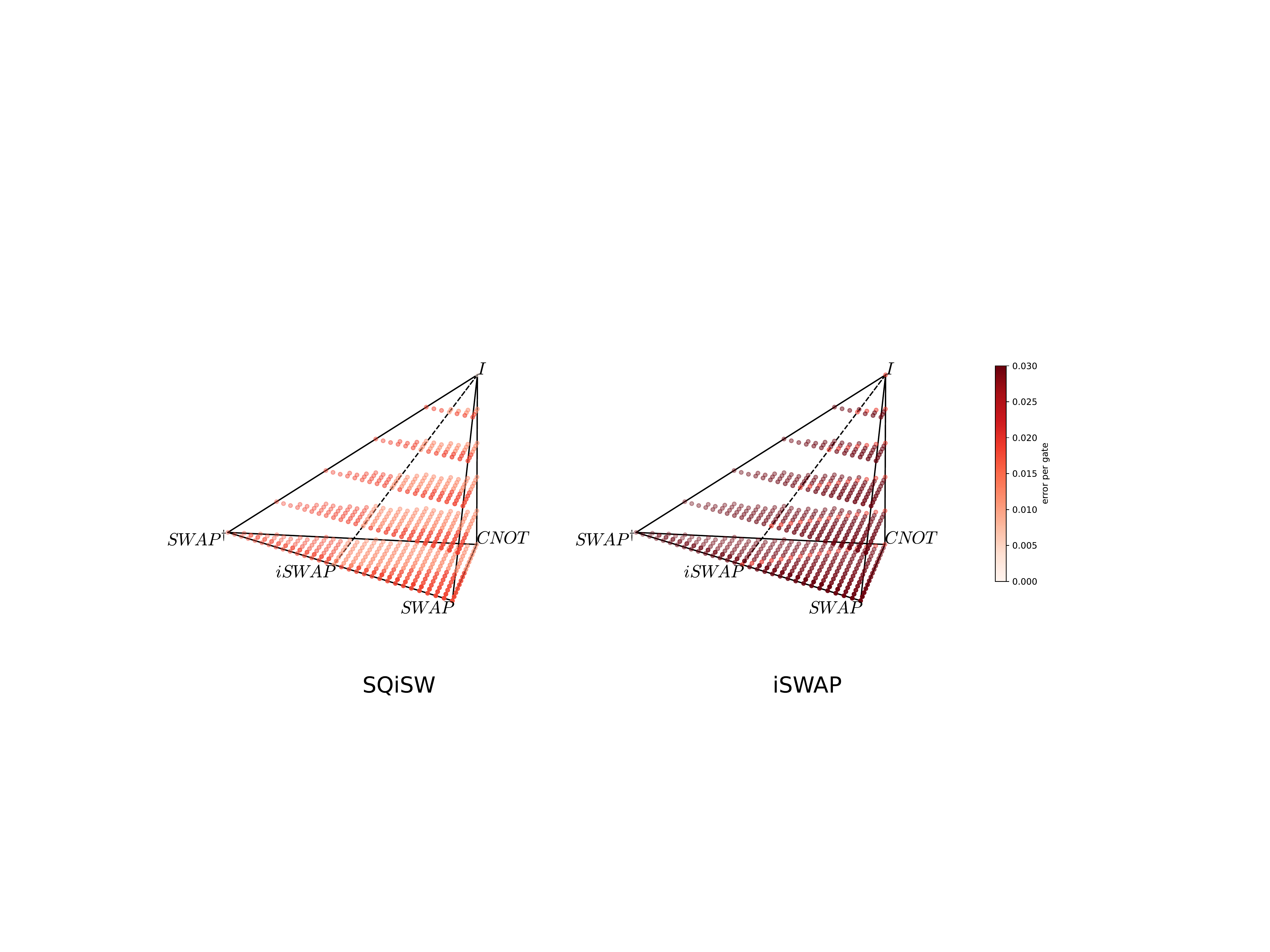}
    \caption{iFRB fidelity value projected onto the Weyl chamber. Data points are taken where $\eta_x$ are multiples of $\pi/20$, and $\eta_y$ and $\eta_z$ are multiples of $\pi/60$. Each data point is collected using iFRB on a gate with the Weyl chamber coordinate, with a randomly chosen set of single-qubit operations applied before and after. For demonstration, we consider a simple error model: each gate is followed by a depolarizing channel with error rate $p_{\text{iswap}} = 2p_{\text{sqisw}} = 0.005$, $p_{\text{single}}=0.0005$. It can be seen from the figure that the effects of the randomly chosen single-qubit operations are negligible as the predominant error sources are the two-qubit gates.}
    \label{fig:frb_wc}
\end{figure}

As the family of all two-qubit unitaries $\mathrm{SU}(4)$ has $15$ real degrees of freedom, we choose one element from each Weyl chamber coordinate, appending it with randomly chosen single-qubit gates. The results we find show that the errors are dominated by the two-qubit gates. We use an interleaved version of Fully Randomized Benchmarking, or iFRB~\cite{kong2021framework} to compute the fidelity value for each Weyl chamber coordinate (please refer to~\Cref{subsec:frb} for more details on iFRB). The corresponding results are shown in~\cref{fig:frb_wc}.

It can be seen from the figure that, under this particular noise model, all gates can be compiled using $\SQiSW$ with an error rate below $1.8\%$. Meanwhile, although gates in the $I$-$\CNOT$-$\iSWAP$ plane can be compiled with 2 applications of the $\iSWAP$ gate, reaching an error rate of about $2\%$, general gates requiring $3$ applications of the $\iSWAP$ gate has error rate about $3\%$. This significant difference indicates an appreciable advantage to using $\SQiSW$ for compiling quantum algorithms.

\subsection{Achievable Quantum Volume}
Quantum volume~\cite{cross2019validating} is a measure of the largest random quantum circuit of equal width and depth that a quantum computer can successfully implement. It is a all-around measure, taking into account gate fidelities, expressibility of native gate sets, quality of compilers, and even qubit connectivity. We conduct numerical experiments computing the quantum volume that directly compares using $\SQiSW$ to using $\iSWAP$ as the native two-qubit gate, ceteris paribus, under different noise levels and connectivities. Note that unlike compiled two-qubit gate fidelity, this compares the gates in a multi-qubit setting beyond just two qubits.

\begin{figure}
    \centering
    \includegraphics[width = 0.7\textwidth]{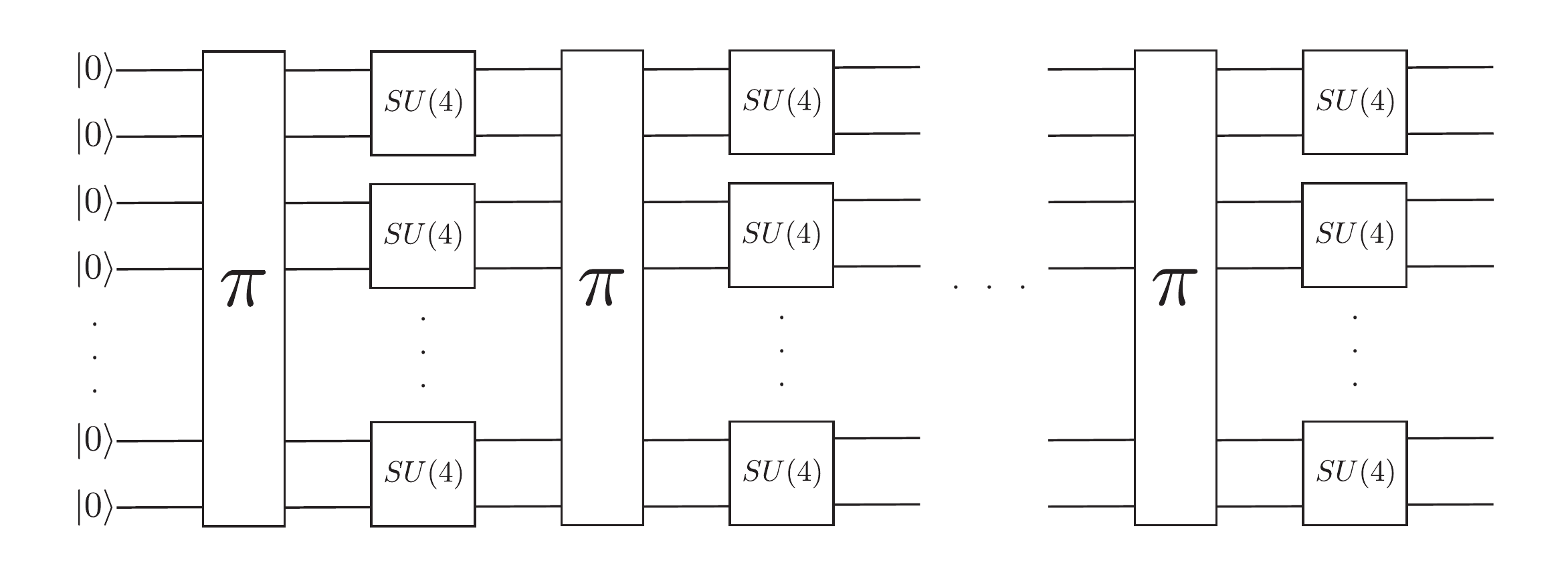}
    \caption{A random circuit used to evaluate quantum volume. The number of qubits and the number of cycles of permutation plus random two-qubit gates are both $d$.}
    \label{fig:qv_circuit}
\end{figure}

For the sake of being self-contained, we repeat here the definition of quantum volume. Given the number of qubits and depth $d$, we generate a random circuit of the form shown in~\cref{fig:qv_circuit}.
The $SU(4)$ box indicates a Haar-random two-qubit unitary, while the $\pi$ box indicates a uniformly randomly chosen permutation. The circuit as a whole defines an overall unitary $U \in SU(2^d)$. We first numerically compute the probability distribution over bit strings $x \in \{0,1\}^d$ measured if we implement $U$ on $\ket{0}^{\otimes d}$:
\begin{align*}
    p_U(x) \equiv \vert \bra{x} U \ket{0}^{\otimes d} \vert^2.
\end{align*}
Using this, we can define the \textit{heavy outputs} as the bit strings whose probability is higher than the median:
\begin{align*}
    H_U \equiv \{ x \in \{0,1\}^d \vert p_U(x) > p_\text{med} \},
\end{align*}
where $p_\text{med}$ is the median of the probabilities of the bit strings. Next, we compute the probability distribution obtained using imperfect gates, compilation and such: $q_U(x)$~\footnote{Note that this is possible for a numerical simulation, but in a real experimental settings a sampling approach is necessary. }. We then define
\begin{align*}
    h_U \equiv \sum_{x \in H_U} q_U(x).
\end{align*}
We average over unitaries $U$ according to the above distribution to obtain
\begin{align*}
    h_d \equiv \int_U dU h_U.
\end{align*}
The quantum volume is defined as
\begin{align*}
    V_Q \equiv 2^{\max\{ d  \vert h_d > \frac 2 3\}}.
\end{align*}

In our particular case, we generate $q_U(x)$ as follows. As before, an arbitrary two-qubit gate is compiled using $\SQiSW$ according to~\cref{alg:decomp} and using $\iSWAP$ according to~\cite{ye2004super}. The two-qubit gates are then subject to depolarizing noise with rate $p_\iSWAP = 2 p_\SQiSW$, and the single-qubit gates with rate $5\times 10^{-4}$. 

\begin{figure}
    \centering
    \includegraphics[width=0.7\textwidth]{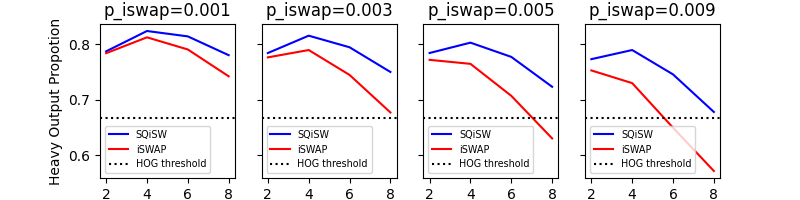}
    \caption{$h_d$ as a function of $d$ for different depolarizing noise rates assuming we use $\SQiSW$ or $\iSWAP$ as our native two-qubit gate. We assume here a complete connectivity graph.}
    \label{fig:qv_all}
\end{figure}

\begin{figure}
    \centering
    \includegraphics[width=0.7\textwidth]{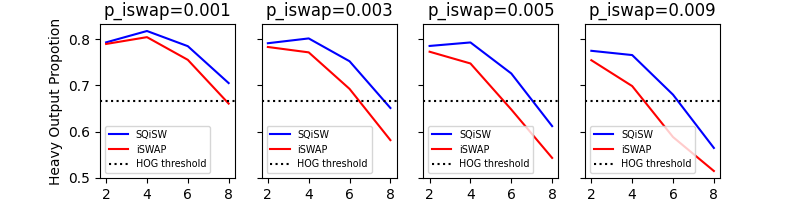}
    \caption{$h_d$ as a function of $d$ for different depolarizing noise rates assuming we use $\SQiSW$ or $\iSWAP$ as our native two-qubit gate. We assume here a 1-D chain connectivity graph.}
    \label{fig:qv_chain}
\end{figure}

We now show the results of our numerical experiments, conducted using the quantum volume module in Cirq~\cite{cirq_developers_2021_4586899}. $h_d$ is approximated by averaging over 1001 different $U$ for increasing $d$ (We find numerically that $h_U$ is approximately the same for all 1001 samples.), under different noise levels and connectivities. In~\cref{fig:qv_all}, a complete graph is assumed and~\cref{fig:qv_chain} assumes a 1-D chain graph. The chain case is done by computing a list of $\SWAP$ gates needed to implement the random permutations using the corresponding function in the quantum volume module of Cirq. Finally, in the figures we only compute even $d$ for simplicity --- the odd values show a similar trend. We see that for all the error rates we consider, $\SQiSW$ clearly outperforms $\iSWAP$, consistently achieving a higher quantum volume. This indicates that simply changing from $\iSWAP$ to $\SQiSW$ can appreciably change the quantum volume boasted by a quantum computer.

\bibliographystyle{unsrt}
\bibliography{ref}